\documentclass[prb,aps,twocolumn,10pt,floatfix,nofootinbib,superscriptaddress]{revtex4-2}

\usepackage{amsmath}
\usepackage{amssymb}
\usepackage{mathtools}

\usepackage{amsthm}

\usepackage{amsfonts}
\usepackage{txfonts}
\usepackage{dsfont}                     
\usepackage{bbold}                      
\usepackage{pifont}                     
\newcommand{\xmark}{\ding{55}}

\usepackage{array}

\usepackage{dcolumn}                    
\usepackage{makecell}                   
\usepackage{multirow}
\usepackage{booktabs}
\usepackage[flushleft]{threeparttable} 

\usepackage{enumerate}
\usepackage[shortlabels]{enumitem}

\usepackage[usenames,dvipsnames]{xcolor}
\usepackage{graphicx}
\graphicspath{{figs/}}                  

\usepackage[version=4]{mhchem}          
\usepackage{physics}                    
\usepackage{csquotes}                   
\usepackage[normalem]{ulem}             

\usepackage[colorlinks=true,citecolor=cyan,linkcolor=magenta,filecolor=magenta]{hyperref}
\usepackage[capitalize]{cleveref}

\usepackage{orcidlink}

\newcommand{\e}{\mathrm{e}}             

\renewcommand{\i}{\mathrm{i}}           

\DeclareMathOperator*{\argmin}{argmin}
\DeclareMathOperator{\sign}{sign}  	    

\def\abs#1{\left|{#1}\right|}      	    

\newcommand{\cconj}[1]{{#1}^*}

\newcommand{\subg}{<}

\renewcommand{\vec}[1]{{\vb{#1}}}       
\newcommand{\id}{\mathds{1}}            
\newcommand{\mat}[1]{\begin{pmatrix}#1\end{pmatrix}} 
\newcommand{\adjo}[1]{#1^\dagger}       

\newcommand{\nunit}[1]{\,\mathrm{#1}}   

\newcommand{\Ham}{\mathcal{H}}          

\def\mcT{\mathcal{T}}					
\def\mcP{\mathcal{P}}					
\def\mcK{\mathcal{K}}					

\newcommand{\sLG}{\mathcal{G}}          
\newcommand{\LG}[1]{\sLG^{\hspace{0.05em}#1}}
\newcommand{\sLcG}{\smash{\overline{\mathcal{G}}}}
\newcommand{\LcG}[1]{\sLcG^{\hspace{0.05em}#1}}
\newcommand{\TG}{\mathfrak{T}}          
\newcommand{\uG}{G}                     
\newcommand{\mG}{M}                     

\newcommand{\rep}{D}
\newcommand{\corep}{\overline{D}}

\newcommand{\projcorep}{\overline{\Delta}}

\DeclareMathAlphabet{\mathbbold}{U}{bbold}{m}{n}
\def\ztwo{\mathbbold{Z}_2}				
\def\cmplx{\mathbbold{C}}				

\newcommand{\SO}{\mathsf{SO}}           
\newcommand{\U}{\mathsf{U}}             

\newcommand{\td}{\text{--}}

\newcommand{\irrep}[3]{\prescript{#3\!}{}{#1}_{#2}} 

\newcommand{\kdotp}{\mbox{$\vec{k}\cdot\vec{p}$}} 
\newcommand{\Python}{\textsc{Python}}
\newcommand{\Mathematica}{\textsc{Mathematica}}

\newcommand{\matref}[1]{\ref{fig:TPmat:#1}, p.~\pageref{fig:TPmat:#1}}

\newcolumntype{x}[1]{>{\centering\let\newline\\\arraybackslash\hspace{0pt}}p{#1}}
\newcolumntype{C}{>{$}c<{$}} 
\newcolumntype{L}{>{$}l<{$}} 

\newcommand{\addextralinespace}[1]{\rule[#1\normalbaselineskip]{0pt}{0pt}}
	
\definecolor{MyOrange}{RGB}{255,180,0}
\definecolor{MyRed}{RGB}{204,0,0}
\definecolor{MyBlue}{RGB}{0,165.75,229.5}
\definecolor{MyIndigo}{RGB}{75,0,130}
\definecolor{MyTan}{RGB}{185,155,126}

\definecolor{MyGreen}{RGB}{0,216.75,0}

\crefname{section}{Sec.}{Secs.}
\Crefname{section}{Section}{Sections}

\begin{document}

\title{Triple nodal points characterized by their nodal-line structure in all magnetic space groups}

\author{Patrick M. Lenggenhager\,\orcidlink{0000-0001-6746-1387}}\email[corresponding author: ]{lenpatri@ethz.ch}
\affiliation{Condensed Matter Theory Group, Paul Scherrer Institute, 5232 Villigen PSI, Switzerland}
\affiliation{Institute for Theoretical Physics, ETH Zurich, 8093 Zurich, Switzerland}
\affiliation{Department of Physics, University of Zurich, Winterthurerstrasse 190, 8057 Zurich, Switzerland}

\author{Xiaoxiong Liu\,\orcidlink{0000-0002-2187-0035}}\email[corresponding author: ]{xxliu@physik.uzh.ch}
\affiliation{Department of Physics, University of Zurich, Winterthurerstrasse 190, 8057 Zurich, Switzerland}

\author{Titus Neupert\,\orcidlink{0000-0003-0604-041X}}
\affiliation{Department of Physics, University of Zurich, Winterthurerstrasse 190, 8057 Zurich, Switzerland}

\author{Tom\'{a}\v{s} Bzdu\v{s}ek\,\orcidlink{0000-0001-6904-5264}}
\affiliation{Condensed Matter Theory Group, Paul Scherrer Institute, 5232 Villigen PSI, Switzerland}
\affiliation{Department of Physics, University of Zurich, Winterthurerstrasse 190, 8057 Zurich, Switzerland}

\date{\today}

\begin{abstract}
We analyze triply degenerate nodal points [or triple points (TPs) for short] in energy bands of crystalline solids.
Specifically, we focus on spinless band structures, i.e., when spin-orbit coupling is negligible, and consider TPs formed along high-symmetry lines in the momentum space by a crossing of three bands transforming according to a 1D and a 2D irreducible corepresentation (ICR) of the little co-group.
The result is a complete classification of such TPs in all magnetic space groups, including the non-symmorphic ones, according to several characteristics of the nodal-line structure at and near the TP.
We show that the classification of the presently studied TPs is exhausted by $13$ magnetic point groups (MPGs) that can arise as the little co-group of a high-symmetry line and which support both 1D and 2D spinless ICRs.
For $10$ of the identified MPGs, the TP characteristics are uniquely determined without further information; in contrast, for the $3$ MPGs containing sixfold rotational symmetry, two types of TPs are possible, depending on the choice of the crossing ICRs.
The classification result for each of the $13$ MPGs is illustrated with first-principles calculations of a concrete material candidate.
\end{abstract}

\maketitle


\section{Introduction}\label{Sec:Intro}

The discovery of Weyl and Dirac semimetals~\cite{Wan:2011,Xu:2015a,Lv:2015,Soluyanov:2015,Young:2012,Yang:2014,Liu:2014a,Son:2013,Huang:2015} has ignited fruitful research into novel types of quasiparticle dispersions in semimetals.
These emerge when energy bands become degenerate in the vicinity of the Fermi energy, giving rise to so-called \emph{band nodes}, and they can feature non-trivial topological invariants, boundary signatures, and transport properties.
Since the originally proposed Weyl and Dirac point degeneracies, various other types of band nodes have been proposed and studied:
from nodal lines (NLs)~\cite{Burkov:2011,Fang:2016,Yu:2017,Kim:2015,Chan:2016} forming intricate linked, knotted, and intersecting structures~\cite{Bzdusek:2016,Yan:2017,Bi:2017,Ezawa:2017,Park:2022}, over nodal surfaces~\cite{Liang:2016,Agterberg:2017,Bzdusek:2017,Wu:2018,Yu:2019}, to nodal points with degeneracies different from two (Weyl) and four (Dirac)~\cite{Bradlyn:2016,Wieder:2016,Schroeter:2020,Yao:2020}.
In particular, three-fold degenerate points~\cite{Zhu:2016,Heikkila:2015,Hyart:2016,Weng:2016,Lv:2017,Wang:2017,Chang:2017,Zhang:2017,Ma:2018,Kim:2018,Chen:2018,Das:2020} [also called triply-degenerate nodal points, triple nodal points, or (for brevity) just \emph{triple points} (TPs)] have been widely investigated, as they constitute a special intermediate between Weyl and Dirac semimetals.

Triple points appear in \emph{two flavours}: (1)~as three-dimensional (3D) irreducible corepresentations (ICRs) of the little group of high-symmetry points (HSPs) in the Brillouin zone (BZ)~\cite{Bradlyn:2016,Park:2021,Feng:2021,Tian:2021}, where the crystalline symmetry forces a triplet of bands to be energetically degenerate, and (2)~as the crossing of a symmetry-protected 2D ICR of the little group of a high-symmetry line (HSL) by a 1D ICR.
Initially, TPs were considered within the context of spin-orbit-coupled (SOC) systems, where flavour-(2) TPs were classified into type~$\mathsf{A}$ vs.~type~$\mathsf{B}$ according to the absence/presence of attached nodal-line (NL) arcs~\cite{Zhu:2016,Chen:2018}.
Using photoemission spectroscopy, TPs were shown to exist in the band structure of various materials, including \ce{MoP}~\cite{Lv:2017}, \ce{WC}~\cite{Ma:2018} and ferroelectric \ce{GeTe}~\cite{Krempasky:2021}.

In contrast, TPs in \emph{spinless} band structures, which describe bosonic systems as well as electronic systems without magnetic order and negligible SOC, became the subject of a systematic analysis only recently~\cite{Lenggenhager:2021:MBNLs,Lenggenhager:2021:TPHOT,Park:2021,Feng:2021,Tian:2021} and have been reported in several compounds~\cite{Mikitik:2006,Zhang:2017,Zhang:2017b,Xie:2019,Jin:2019,Jin:2020}.
While it is difficult to find magnetic materials that are well described by spinless representations, classical metamaterials are naturally spinless and setups where time-reversal symmetry is broken are therefore expected to be described by spinless representations of magnetic groups.
Very recent works have systematically searched for TPs in spinless band structures of all magnetic space groups~\cite{Park:2021,Feng:2021,Tian:2021,Yu:2021,Liu:2022,Zhang:2021}, studied TPs at HSPs in more detail~\cite{Park:2021,Feng:2021,Tian:2021}, or classified TPs on HSLs according to their dispersion as linear or quadratic~\cite{Feng:2021}.
However, a systematic classification of the NL structure appended to TPs on HSLs (i.e., whether they are type~$\mathsf{A}$ vs.~type~$\mathsf{B}$~\cite{Zhu:2016}) in spinless band structures, as well as a discussion of their topology, is missing.

In the present work, we complete the missing aspects of the TP classification by considering the case of TPs lying on HSLs in spinless band structures.
In Ref.~\onlinecite{Lenggenhager:2021:MBNLs} we have classified all possible TPs in a subset of spinless systems, namely those in systems with space-time-inversion ($\mcP\mcT$) symmetry and symmorphic space group, according to a similar scheme as Ref.~\onlinecite{Zhu:2016}, and we revealed valuable connections of such TPs to non-Abelian band topology, monopole charges, and NL links~\cite{Lenggenhager:2021:MBNLs,Wu:2019,Tiwari:2020,Ahn:2018}.
Here, we describe the full derivation of the classification result shown in Ref.~\onlinecite{Lenggenhager:2021:MBNLs} and extend it to include TPs in spinless systems without $\mcP\mcT$ symmetry and in non-symmorphic space groups.
The result is a complete classification of TPs in spinless systems for \emph{all} magnetic space groups according to the NL structure appearing in the vicinity of the TPs.
Note that by \enquote{magnetic space groups} we mean all four types of Shubnikov groups, which include the ordinary space groups (type I) and space groups describing non-magnetic materials with time reversal symmetry (type II).
Let us also point out that this manuscript appears in parallel with another work~\cite{Lenggenhager:2021:TPHOT}, in which we reveal universal higher-order-topological signatures associated with \emph{pairs} of TPs (i.e, when a semimetallic band structure is formed by a 2D ICR being sequentially crossed by two 1D ICRs) classified here, thus filling the other major gap in the previous characterization of TPs.

The manuscript is structured as follows.
We start in \cref{Sec:Results} by setting the terminology necessary for describing the nodal structure near TPs.
Importantly, we report that even type-$\mathsf{A}$ TPs are often associated with a nexus of NLs that lies on the HSL in the vicinity of the TP, and which can collapse onto the TP after fine-tuning some model parameters.
Our main classification result, summarized by \cref{tab:MPGs,tab:Classification}, therefore includes not only the TP type, but also the \emph{codimension} for merging type-$\mathsf{A}$ TPs with a nexus of nodal lines.
The summary of results is followed by an outline of the actual derivation of the TP classification, starting with MPGs \emph{without} $\mcP\mcT$ symmetry in \cref{Sec:Classif:no_PT}, and followed by the MPGs \emph{with} $\mcP\mcT$ symmetry in \cref{Sec:Classif:PT}.
In the latter case, we briefly touch upon the non-Abelian topology~\cite{Ahn:2018,Wu:2019,Ahn:2019,Bouhon:2020,Bouhon:2020b,Guo:2021,Jiang:2021b} as described in Ref.~\onlinecite{Lenggenhager:2021:MBNLs}.
We proceed by demonstrating the classification result by showcasing TPs and their associated NL structures in concrete material examples in \cref{Sec:Materials}.
In particular, \cref{tab:TPmaterials} lists representative non-magnetic material examples with time-reversal symmetry (described by type-II magnetic space groups) hosting TPs on HSLs for each of the little groups tabulated in \cref{tab:MPGs}.
Using first principles calculations based on density functional theory (DFT), we compute the size of the relevant band gaps, identify the nodal structure, e.g., NLs and TPs, and infer the type of each TP to verify the predictions based upon our classification.
Finally, we conclude in \cref{Sec:Conclusion}.

The manuscript is supplemented by several appendices and supplementary data and code~\cite{Lenggenhager:2022:TPClassif:SDC}.
Since we discuss not only the 230 crystallographic space groups but all magnetic space groups, we have to deal with antiunitary symmetries and their representation theory.
For this reason, \cref{App:antiunitary_symmetries} provides a short review of the concepts, notation and properties that we use.
Second, in \cref{App:non-symm_SGs} we discuss the effect of possible non-symmorphic symmetries of the space group.
In particular, we show that non-symmorphicity does not alter the characterization of TPs along a HSL with a given little co-group whenever 1D ICRs are available along the HSL (information that is tabulated, e.g., in Ref.~\onlinecite{Bradley:1972} or on the Bilbao crystallographic server~\cite{BCS,Xu:2020,Elcoro:2021,Aroyo:2006a,Aroyo:2006b,Elcoro:2017}).
In the subsequent \cref{App:Classif:no_PT_mirror,App:Classif:PT} we provide detailed derivations and proofs for the classification results; additionally, all the \kdotp{} expansions are made available in the supplementary data and code~\cite{Lenggenhager:2022:TPClassif:SDC}.
Finally, in \cref{App:Materials} we present the band structure data supporting the observations summarized in \cref{tab:TPmaterials}.

\section{Terminology and classification results}\label{Sec:Results}

In this section, we summarize the key terminology adopted throughout the manuscript, and summarize the obtained classification of TPs in spinless band structures. 
We begin in \cref{Sec:Notions} by defining the notion of triple points (TPs) and nexus points.
This allows us to classify TPs according to their type ($\mathsf{A}$ vs.~$\mathsf{B}$) and certain additional characteristics (codimension for a nexus point coinciding with a TP for type-$\mathsf{A}$ TPs, number of NL arcs attached to a nexus and linear/quadratic attachment).
In \cref{Sec:Lines}, we discuss which high-symmetry lines (HSLs) in various space groups can potentially harbor TPs (of some type). 
Here, the key criterion is that the little group ($\LG{\textrm{HSL}}$) along the HSL should harbor both 1D and 2D irreducible corepresentations (ICRs). 
This happens only if the little co-group (the quotient of little group by the translation group, $\LcG{\textrm{HSL}} = \LG{\textrm{HSL}}/\TG$) is one of the 13 magnetic point groups (MPGs) listed in \cref{tab:MPGs}.
In the last paragraph of \cref{Sec:Lines}, we briefly discuss how to determine in which space groups this happens and provide references that have compiled such tables.

Finally, \cref{Sec:Classification} presents and briefly discusses the result of our classification, summarized in \cref{tab:Classification}.
Specifically, we find that the characteristics of the TP are uniquely fixed by the little co-group of the HSL; the sole exception being HSLs with sixfold rotational symmetry where one additionally needs to specify which 1D and 2D ICRs of the little co-group are crossing at the TP.
Crucially, this result applies irrespective of the \mbox{(non-)symmorphicity} of the space group: we find that non-symmorphicity can only forbid the existence of TPs (if 1D ICRs of the little group do not exist), but cannot alter the characteristics of the TPs (if 1D ICRs do exist).

\subsection{Basic notions}\label{Sec:Notions}

\begin{figure}[t]
    \centering
    \includegraphics{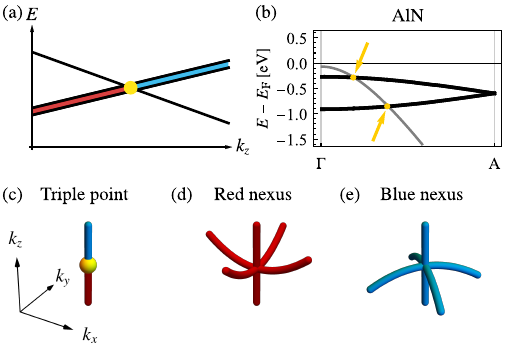}
    \caption{
        Triple points (TPs) and their characteristic nodal features.
        (a) Schematic band structure along a high-symmetry line (HSL) $k_z$ that can stabilize TPs (cf.~\cref{tab:MPGs}). Note the non-degenerate crossing the doubly-degenerate band. Nodal lines (NLs) in the lower (upper) band gap of a three-band model are shown in red (blue), and the TP is indicated by a yellow disk or sphere.
        (b) Band structure of the material \ce{AlN} along the $\Delta=\Gamma A$ line, which we predict to host triple points (cf.~\cref{tab:TPmaterials} and \cref{fig:TPmat:AlN_1,fig:TPmat:AlN_2}).
        Non-degenerate (doubly-degenerate) bands are shown as thin gray (thick black) lines and TPs are indicated by yellow arrows.
        (c) In the NL structure, a TP can be recognized by the change of a NL color from red to blue.
        (d,e) The point where additional NLs (called \emph{NL arcs}) coalesce with the central NL (along $k_z$) is called \emph{nexus}. All NLs involved in the nexus are in the same gap, i.e., a red (blue) nexus involves NLs in the lower (upper) gap. The number $n_a^\mathrm{nexus}$ of NL arcs that coalesce with the central NL depends on the little co-group of the HSL (cf.~\cref{tab:Classification}).
    }
    \label{fig:TPElements}
\end{figure}

Before introducing the classification scheme in more detail, we set some terminology.
Consider a HSL with little group $\LG{\textrm{HSL}}$; the little group $\LG{\vec{k}}$ of $\vec{k}$ is defined as the subgroup of the space group that leaves the momentum $\vec{k}$ invariant modulo translations by reciprocal lattice vectors.
Note that $\LG{\vec{k}}$ always contains as a subgroup the (infinite) group $\TG$ of translations by Bravais vectors.
The existence of a TP along the HSL means that a non-degenerate band (1D ICR of $\LG{\textrm{HSL}}$) crosses a doubly degenerate band (2D ICR of $\LG{\textrm{HSL}}$), such that the nodal line formed by the latter (which we call the \emph{central} NL) changes the energy gap [cf.~\cref{fig:TPElements}(a)].
In such a three-band system, let us denote the two gaps (and the NLs in the corresponding gaps) by the colors used in the illustrations: the lower gap is red and the upper gap is blue.
Then, the defining feature of a TP is the change of the central NL from red to blue, see \cref{fig:TPElements}(c)~\cite{Lenggenhager:2021:MBNLs}.

\begin{figure}[t]
    \centering
    \includegraphics{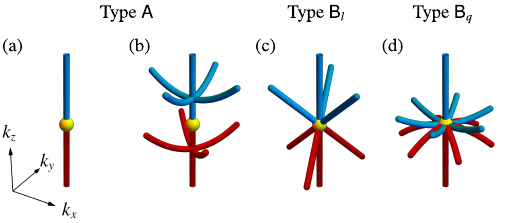}
    \caption{
        Types and subtypes of triple points (TPs) in spinless systems. Nodal lines (NLs) are shown in red (blue) if they are in the lower (upper) band gap of the three-band model and the TP is indicated in yellow.
        TPs are conventionally classified as type $\mathsf{A}$ vs.\ type $\mathsf{B}$ according to the absence vs.\ presence of NL arcs \emph{attached to} the TP besides the central NL along the high-symmetry line.
        Type-$\mathsf{A}$ TPs arise (a) without nexus points or (b) with nexus points that do generically (i.e., without fine tuning) \emph{not} coincide with the TP.
        For type-$\mathsf{B}$ TPs a red and blue nexus generally coincide with the TP. We differentiate between (c) type $\mathsf{B}_l$ with three linearly attached NLs and (d) type $\mathsf{B}_q$ with six quadratically attached NLs.
    }
    \label{fig:TPTypes}
\end{figure}

Often, TPs are accompanied by additional NLs that coalesce with the central NL of the same color at some point on the HSL.
We call such a point a red or blue \emph{nexus} (point), depending on the gap in which it appears -- see \cref{fig:TPElements}(d) and (e), respectively.
In some cases, the little group and the corepresentations of the bands can force nexus points to coincide with the TP [cf.~\cref{fig:TPTypes}(c,d)], while in other cases they generically do not coincide with the TP [cf.~\cref{fig:TPTypes}(b)] or are completely absent [cf.~\cref{fig:TPTypes}(a)].
Following the terminology of Ref.~\onlinecite{Zhu:2016} that introduced the classification of triple points in spinful systems, we call TPs that coincide with both a red and a blue nexus \emph{type $\mathsf{B}$} and the others \emph{type $\mathsf{A}$}.
For brevity, we term the additional NLs involved in the nexus points \emph{NL arcs}.

Going beyond the terminology of Ref.~\onlinecite{Zhu:2016}, we further subdivide type-$\mathsf{B}$ TPs into type-$\mathsf{B}_l$ TPs with linearly attached NL arcs [\cref{fig:TPTypes}(c)] and type-$\mathsf{B}_q$ TPs with quadratically attached NL arcs [\cref{fig:TPTypes}(d)].
For type-$\mathsf{A}$ TPs in symmetry groups that admit nexus points along the rotation axis, we additionally define the number $n_a^\mathrm{nexus}$ of NL arcs attached to a nexus point and specify the codimension for colliding at least one nexus point with the TP, i.e., quantify how many parameters need to be fine-tuned to achieve the coincidence of the TP with the nexus of NL~arcs.
Sometimes, because of reasons rooted in symmetry, the smallest number of nexus points that can collide with the TP is larger than one, i.e., several nexus points necessarily collide with the TP simultaneously.
We remark that understanding these codimensions is not just an abstract academic problem, but it may become important for the analysis of real materials' band structures when the codimension is small.
To illustrate this aspect, we show a particular example of a material that exhibits such a fine-tuned type-$\mathsf{A}$ TP in \cref{Sec:TPMaterials:examples}.

\subsection{High-symmetry lines that admit TPs}\label{Sec:Lines}
The presence of stable TPs requires the crossing of 1D and 2D ICRs on the HSL.
Thus, a necessary condition is the existence of both 1D and 2D ICRs in the little-group of the HSL.
To analyze which HSLs obey these conditions, it is necessary to distinguish the case of symmorphic vs.~non-symmorphic space groups.

In symmorphic space groups, the ICRs of any little group $\LG{\vec{k}}$ are readily deduced from the ICRs of the corresponding little \emph{co-group} $\LcG{\vec{k}}$, which is defined as the quotient group $\LG{\vec{k}}/\TG$ and equal to one of the 122 MPGs. 
By screening through the irreducible representations of MPGs characterizing HSLs~\cite{Bradley:1972,BCS,Xu:2020,Elcoro:2021}, we find that this condition is satisfied (1)~if a rotational symmetry $C_n$ of order $n\in\{3,4,6\}$ with rotation axis along the HSL is supplemented with $\mcP\mcT$ or a \emph{vertical} mirror symmetry $m_v$ (i.e., one containing the rotation axis), or both; 
alternatively, (2)~the combined symmetry $C_n\mcP\mcT$ (which we call \emph{antiunitary rotation})
of order $n\in\{4,6\}$ can stabilize TPs with or without the $\mcP\mcT$ and $m_v$ symmetry. 
We summarize all these options in \cref{tab:MPGs}.
The particular choice of the little co-group and of its 1D and 2D ICRs that describe the triplet of crossing bands constrain the form of the Hamiltonian close to the TP, thus determining the type and other characteristics of the TP.

In non-symmorphic space groups, on the other hand, the ICRs of the little group $\LG{\vec{k}}$ are generically not related to the ordinary ICRs of the little co-group $\LcG{\vec{k}}$, but they are instead obtained from the \emph{projective} ICRs of $\LcG{\vec{k}}$.
More precisely, one needs to identify representations $\Delta:\LcG{\vec{k}}\to \mathsf{U}(n)$ such that $\forall g_i,g_j\in\LcG{\vec{k}}: \Delta(g_i)\Delta(g_j) = \mu(g_i,g_j)\Delta(g_{i}g_{j})$, where the \emph{factor system} $\mu: \LcG{\vec{k}}\times \LcG{\vec{k}} \to \mathsf{U}(1)$ is fixed by the space group symmetry~\cite{Bradley:1972}. 
However, as we show in \cref{App:non-symm_SGs}, if the little group supports a 1D ICR (a necessary condition to admit TPs), it must hold that the factor system $\mu$ constraining the admissible projective ICRs of the little co-group $\LcG{\vec{k}}$ belongs to a \emph{trivial equivalence class} -- meaning that an appropriate transformation brings all values of $\mu$ to $1$, and that one in fact studies the ordinary representations of $\LcG{\vec{k}}$.
We thus find, \emph{even for non-symmorphic space groups}, that if the little group $\LG{\textrm{HSL}}$ at a HSL supports both 1D and 2D ICRs, then (1)~the corresponding little co-group $\LcG{\textrm{HSL}}$ is one of the 13 MPGs listed in \cref{tab:MPGs}, and (2)~the symmetry constraints on the Hamiltonian close to the TP (and therefore its type and the other characteristics) are fully determined by the corresponding \emph{ordinary} ICRs of the little co-group.

In this work, we classify triple points based on their properties (notably according to their associated NL structure) given a concrete little group of a HSL, while we leave the systematic identification of the space groups that host TPs to other works.
Given \cref{tab:MPGs} and databases such as the Bilbao crystallographic server~\cite{BCS} this is, in principle, a straightforward task.
Let us point out that, in parallel to our work, space groups that support variety of band nodes (\enquote{quasiparticles}), including TPs, were tabulated in Refs.~\onlinecite{Yu:2021,Liu:2022,Zhang:2021}.
In the corresponding supplementary materials, the authors list for each type-II, type-III and type-IV magnetic SG, respectively, at each HSL the generators of the little co-group and the admissible nodal quasiparticles.
Searching for TPs in those tables, one can find all SGs and HSLs supporting TPs as well as the relevant little co-group such that our classification can be easily applied.

\subsection{Characterization of TPs for each admissible HSL}\label{Sec:Classification}

\begin{table}[t]
    \centering
    \caption{
        Symmetry conditions for triple points (TPs) along high-symmetry lines (HSLs) in momentum space of spinless systems.
        The table displays all magnetic point groups (MPGs) in Hermann-Mauguin notation~\cite{Bradley:1972} that can stabilize TPs if they occur as the little co-group of a HSL.
        These are exactly the MPGs that (1) preserve one momentum component (such that the MPG corresponds to a little co-group of some HSL), and that (2) support both 1D and 2D irreducible corepresentations.
        The columns and the rows indicate generators of the MPG, where $C_n$ is rotational symmetry of order $n$, $\mcP\mcT$ is space-time inversion symmetry, and $m_v$ is mirror symmetry with respect to a plane containing the rotation axis.
        Entries marked by {\xmark} violate condition (2) and can therefore not stabilize TPs.
    }
    \begin{ruledtabular}
        \begin{tabular}{cCCCCC}
            Generators & C_3 & C_4 & C_4\mcP\mcT & C_6 & C_6\mcP\mcT\\\hline\addlinespace
            $\varnothing$ & \text{\xmark} & \text{\xmark} & \bar{4}' & \text{\xmark} & \bar{6}'\\\addlinespace
            $\mcP\mcT$ & \bar{3}' & \multicolumn{2}{C}{4/m'} & \multicolumn{2}{C}{6/m'}\\\addlinespace
            $m_v$ & 3m & 4mm & \bar{4}'2'm & 6mm & \bar{6}'m2'\\\addlinespace
            $\{\mcP\mcT,m_v\}$ & \bar{3}'m & \multicolumn{2}{C}{4/m'mm} & \multicolumn{2}{C}{6/m'mm}
        \end{tabular}
    \end{ruledtabular}
    \label{tab:MPGs}
\end{table}

Having determined which HSLs of which space groups can potentially harbor TPs, one can proceed to analyze their type and the other characteristics.
This involves a detailed study of \kdotp{} expansions near the TP for the various choices of (1)~the little co-group and of (2) a pair of its 1D and 2D ICRs. 
Such a technical analysis constitutes the bulk of \cref{Sec:Classif:no_PT,Sec:Classif:PT} (further supplemented by \cref{App:antiunitary_symmetries,App:Classif:no_PT_mirror,App:Classif:PT}).
The classification result derived from our analysis is summarized in \cref{tab:Classification}.
We find that for almost all MPGs, the little co-group uniquely determines the type and the codimension of the TP.
The only exception are the MPGs that contain $C_6$ symmetry, where either type-$\mathsf{A}$ or type-$\mathsf{B}$ TPs can arise, depending on the ICRs of the bands forming the TP.

Frequently, even type-$\mathsf{A}$ TPs are accompanied by one or more nexus points (in the same or in different gaps) that attach to the central NL near the TP.
We therefore determine the number $n_a^\mathrm{nexus}$ of NL arcs attached to the nexus point and the codimension (i.e., the number of parameters that have to be fine-tuned) to collide at least one of the nexus points with the TP.
The MPGs $\bar{4}'$ and $\bar{6}'$ are exceptions to this feature, because they do not support NL arcs away from 
the central line; this is denoted in \cref{tab:Classification} by a codimension `$\infty$'.
For type-$\mathsf{B}$ TPs, on the other hand, the codimension can be interpreted as being zero.
Note that in contrast to nexus points coinciding with a type-$\mathsf{B}$ TP, nexus points \emph{near} a type-$\mathsf{A}$ TP are not enforced by symmetry and therefore not a parameter-independent consequence of the TP.

Finally, we observe that the subtype of type-$\mathsf{B}$ TPs is determined by the order of rotational symmetry. 
Writing $k^2=k_x^2+k_y^2$, we find that three-fold rotational symmetry results in three NL arcs attaching linearly to the TP $k_z\propto k$ [cf.~\cref{fig:TPTypes}(c)] in each (i.e., both red \emph{and} blue) energy gap, while six-fold rotation gives six quadratically attaching NL arcs $k_z\propto k^2$ [cf.~\cref{fig:TPTypes}(d)] in each gap.
Analogously, we characterize how NL arcs attach to nexus points in the vicinity of type-$\mathsf{A}$ TPs.
We find that they always attach quadratically, $k_z\propto k^2$ [cf.~\cref{fig:TPTypes}(b)].

\begin{table}[t]
    \centering
    \begin{threeparttable}
    \caption{
        Classification of triple points (TPs) on high-symmetry lines in spinless systems as type $\mathsf{A}$ vs.\ type $\mathsf{B}_q$ vs.\ type $\mathsf{B}_l$ based on the little co-group of the high-symmetry line.
        In the last three columns we list the number $n_a^\mathrm{nexus}$ of nodal-line (NL) arcs attached to a nexus point occurring near or at the TP (for $\bar{4}'$ and $\bar{6}'$ no NL arcs are possible, which is denoted by the entry `\td'), the scaling $\mu$ of those NL arcs at the nexus point [$k_z\propto(k_x^2+k_y^2)^{\mu/2}$], and the codimension for colliding at least one nexus point with the TP (for type-$\mathsf{B}$ this is naturally $0$ and if nexus points cannot be stabilized we write `$\infty$').
        For point groups with $C_6$ rotational symmetry, type and codimension of the TP additionally depend on the irreducible corepresentations (ICRs) $\rho^\mathrm{2D}\oplus\rho^\mathrm{1D}$ of the bands involved in the formation of the TP.
    }
    \begin{ruledtabular}
    \begin{tabular}{CCCCCCC}
    	     \text{Little co-group}      & \text{ICRs}^\textrm{a} & \text{Type}  &  n_a^\mathrm{nexus}   & \mu & \text{Codimension} &  \\ \hline\addlinespace
    	   3m,\,\bar{3}',\, \bar{3}'m    &       \text{any}       & \mathsf{B}_l &           6           &  1  &         0          &  \\ \addlinespace
    	      \bar{4}',\,\bar{6}'        &       \text{any}       &  \mathsf{A}  &          \td          & \td &       \infty       &  \\
    	       \bar{4}'2'm,\, 4mm        &       \text{any}       &  \mathsf{A}  &           4           &  2  &         2          &  \\
    	         4/m',\, 4/m'mm          &       \text{any}       &  \mathsf{A}  &           4           &  2  &         1          &  \\ \addlinespace
    	          \bar{6}'m2'            &       \text{any}       &  \mathsf{A}  &           6           &  2  &         1          &  \\ \addlinespace
    	     \multirow{2}{*}{$6mm$}      &   (E_1;A),\,(E_2;B)    &  \mathsf{A}  & \multirow{2}{*}{$12$} &  2  &         2          &  \\
    	                                 &   (E_1;B),\,(E_2;A)    & \mathsf{B}_q &                       &  2  &         0          &  \\ \addlinespace[0.5\defaultaddspace]
    	\multirow{2}{*}{$6/m',\,6/m'mm$} &   (E_1;A),\,(E_2;B)    &  \mathsf{A}  & \multirow{2}{*}{$12$} &  2  &         1          &  \\
    	                                 &   (E_1;B),\,(E_2;A)    & \mathsf{B}_q &                       &  2  &         0          &
    \end{tabular}
    \end{ruledtabular}
    \begin{tablenotes}[flushleft]
        \item[$\textrm{a}$] The notation for the ICRs follows Ref.~\onlinecite{Bradley:1972}, where we drop the subscripts of the 1D ICRs if they do not affect the result. 
        Note that for the 2D ICRs of $6/m'$ we define: $^2E_2\!\,^1E_2 \mapsto E_1$, and $^2E_1\!\,^1E_1 \mapsto E_2$ to get labels consistent with those of $6mm$ and $6/m'mm$.
    \end{tablenotes}
    \label{tab:Classification}
    \end{threeparttable}
\end{table}

In the next two sections, we present the derivation of these results by constructing minimal \kdotp{} models for each possible combination of a 2D and a 1D ICR of each of the 13 MPGs shown in \cref{tab:MPGs}.
The type, subtype, and codimension of the TP, i.e., the absence vs.~presence of nexus points of NL arcs at or near the TP, is governed by the absence vs.~presence of NLs lying off the rotation axis and connecting to the rotation axis at some point.
These NLs are protected either by vertical mirror symmetry $m_v$ (in which case they are constrained to lie in the corresponding vertical mirror planes) or by $\mcP\mcT$ symmetry (in which case they can curve arbitrarily inside momentum space)~\cite{Fang:2016}.
To reflect this dichotomy, we divide the 13 MPGs into those \emph{without} $\mcP\mcT$ symmetry, discussed in \cref{Sec:Classif:no_PT}, and those \emph{with} $\mcP\mcT$ symmetry, discussed in \cref{Sec:Classif:PT}.

\section{Derivation in the absence of \texorpdfstring{$\mcP\mcT$}{PT} symmetry}\label{Sec:Classif:no_PT}

\subsection{MPGs without mirror symmetry}\label{Sec:Classif:no_PT_no_mirror}

We begin the derivation of the classification of TPs in \cref{tab:Classification} by swiftly considering those MPGs in \cref{tab:MPGs} that contain neither space-time-inversion symmetry $\mcP\mcT$ nor mirror symmetry $m_v$.
This includes MPGs $\bar{4}'$ and $\bar{6}'$ that are generated by a single element: $C_{4z}\mcP\mcT$ and $C_{6z}\mcP\mcT$, respectively.

These symmetries act like rotations inside the $\vec{k}$-space. 
Therefore, the little co-group of all points lying off the rotation axis contains only the identity element, in which case the codimension for node formation is three (i.e., point nodes in 3D)~\cite{Bzdusek:2017}.
It follows that NLs can only be stabilized along the corresponding rotation axis, preventing the existence of any stable NL arcs~\cite{Fang:2016}.
This implies that all TPs on lines with the little co-group $\bar{4}'$ or $\bar{6}'$ are type $\mathsf{A}$ and that nexus points are absent, i.e., the nodal structure is always the one shown in \cref{fig:TPTypes}(a).

\subsection{MPGs with mirror symmetry }\label{Sec:Classif:no_PT_mirror}

We continue with the characterization of TPs in those MPGs in \cref{tab:MPGs} that contain vertical mirror symmetry $m_v$ but not $\mcP\mcT$.
This includes: $3m$, $4mm$, $\bar{4}'2'm$, $6mm$ and $\bar{6}'m2'$. 
The presence of vertical mirror symmetries implies that there can be stable NLs in the corresponding mirror planes, which may connect as NL arcs either to type-$\mathsf{B}$ TPs or to a nexus point in the vicinity of a type-$\mathsf{A}$ TP.

For each of the listed MPGs we first derive a \kdotp{} expansion near the TP within the two-dimensional mirror planes.
More precisely, we perform the expansion only in the distance from the rotation axis ($k_x$ coordinate inside the mirror plane), whereas we keep the full dependence on the coordinate along the rotation axis ($k_z$).
The derived \kdotp{} expansions allow us to determine the TP type and the number of NL arcs attached to a nexus point.
For the cases where the TPs are identified as type-$\mathsf{A}$, we derive the codimension for colliding the TP with a nexus point.
We also determine the minimal number of nexus points that can simultaneously collide with the TP, which requires us to relate the parameters in the \kdotp{} model within symmetry-unrelated sets of mirror planes.
The discussion is subdivided into four parts: \cref{Sec:Classif:mirror_no_PT:C4vC6v,Sec:Classif:mirror_no_PT:C4vC6v-part2} about MPGs $4mm$, $\bar{4}'2'm$ and $6mm$, \cref{Sec:Classif:mirror_no_PT:C6PT} about MPG $\bar{6}'m2'$, and finally \cref{Sec:Classif:mirror_no_PT:C3v} about MPG $3m$.

\subsubsection{TP type derivation for MPGs \texorpdfstring{$4mm$, $\bar{4}'2'm$ and $6mm$}{4mm, -4'2'm and 6mm}}\label{Sec:Classif:mirror_no_PT:C4vC6v}

\begin{figure}[t]
    \centering
    \includegraphics{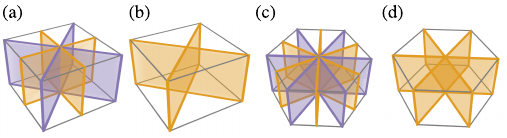}
    \caption{
        Vertical mirror planes, i.e., the ones containing the rotation axis, of the magnetic point groups (a) $4mm$ and $4/m'mm$, (b) $\bar{4}'2'm$, (c) $6mm$ and $6/m'mm$, (d) $\bar{6}'m2'$, $3m$, and $\bar{3}'m$.
        Mirror planes related by \mbox{(rotation-)symmetry} are shown in the same color, such that in (a,b) the different symmetry-unrelated sets of mirror planes can be distinguished by their color.
        The grey boxes are guides to the~eye.
    }
    \label{fig:mirror-planes}
\end{figure}

The MPGs $4mm$, $\bar{4}'2'm$ and $6mm$, are all generated by two elements: a rotational symmetry $C_n$ (or antiunitary rotational symmetry $C_n\mcP\mcT$) with $n=4,6$ and a vertical mirror symmetry $m$ with mirror plane containing the rotation axis.
For each of these MPGs, the vertical mirror planes come in pairs that are orthogonal to each other:
$4mm$ has two mirror planes along the $x$- and $y$-axes (for suitably chosen coordinates) and two diagonal mirror planes [cf.~\cref{fig:mirror-planes}(a)], $\bar{4}'2'm$ has only the two diagonal mirror planes  [cf.~\cref{fig:mirror-planes}(b)], and $6mm$ has a total of six mirror planes at angles differing by $\tfrac{\pi}{6}$  [cf.~\cref{fig:mirror-planes}(c)].
For any vertical mirror symmetry $m$, we label the associated orthogonal vertical mirror symmetry by $m_\perp$. 
For brevity, we call the set of $\vec{k}$-points invariant under $m_{(\perp)}$ the $m_{(\perp)}$-plane.

The convenience of considering the pair of symmetries $(m,m_\perp)$ is that the composition of $m$ and $m_\perp$ is simply the $C_{2}$ rotational symmetry around the HSL, and that these symmetries commute and thus can be diagonalized simultaneously.
Additionally, the subgroup of the MPG that maps points on the $m$-plane back to the $m$-plane is exactly the Abelian group $\{\id,m,m_\perp,C_2\}$ (with $\id$ the identity element), which is fully generated by $m$ and $m_\perp$.
Therefore, to derive a symmetry-compatible \kdotp{} Hamiltonian inside the $m$-plane, it is sufficient to study the constraints from $m$ and $m_\perp$.

To introduce concrete labels, say that a TP is formed along the HSL by the crossing of a 1D ICR $\rho^\mathrm{1D}$ with a 2D ICR $\rho^\mathrm{2D}$.
The common feature of $\rho^\textrm{2D}$ in all the present cases is that any mirror symmetry $m\in\textrm{MPG}$ satisfies $\tr[\rho^\textrm{2D}(m)]=0$.
Combined with the fact that the possible eigenvalues of $m$ are just $+1$ and $-1$, the vanishing trace implies that the 2D ICR is spanned by two states with opposite eigenvalues of $m$.
Additionally, $m$ and $m_\perp$ commute, meaning that they can be diagonalized simultaneously. 
Adopting a basis in which the two symmetry operators are diagonal, the possible mirror eigenvalues of the ICRs $\rho^\textrm{2D}$ and $\rho^\textrm{1D}$ (which appear on the operator diagonals) are shown in \cref{tab:mirror_eigvals}, where the only adjustable parameters are the four signs~$\pm_{p,q,r,t}$.

\begin{table}
    \centering
    \caption{
        Possible mirror eigenvalues for the irreducible representations involved in a triple point in one of the following magnetic point groups: $4mm$, $\bar{4}'2'm$ and $6mm$.
        The two rows of $\rho^\mathrm{2D}$ correspond to the eigenvalues in the order defined by a particular choice of basis of the 2D representation.
        This is well-defined, because $m$ and $m_\perp$ commute, such that the corresponding matrix representations can be simultaneously diagonalized.
        Subscripts to $\pm$ indicate independent signs.
        In the analysis of \cref{Sec:Classif:mirror_no_PT:C4vC6v}, we always rotate the basis such that $\pm_r 1 = \mp_p 1$.
    }
    \begin{ruledtabular}
        \begin{tabular}{CCCC}
            & m & m_\perp & C_2=m\circ m_\perp \\\hline\addlinespace
            \multirow{2}{*}{$\rho^\mathrm{2D}$} & \pm_p 1 & \pm_q 1 & (\pm_p 1)(\pm_q 1)\\
            & \mp_p 1 & \mp_q 1 & (\pm_p 1)(\pm_q 1)\\\addlinespace
            \rho^\mathrm{1D} & \pm_r 1 & \pm_t 1 & (\pm_r 1)(\pm_t 1)
        \end{tabular}
    \end{ruledtabular}
    \label{tab:mirror_eigvals}
\end{table}

Let $\rep$ be the representation that captures the three bands involved in the triple point, $\rep=\rho^\mathrm{2D}\oplus\rho^\mathrm{1D}$.
We work in the basis in which $\rho^\textrm{2D}(m)$ and $\rho^\textrm{2D}(m_\perp)$ are diagonal, and we permute the basis of $\rho^\mathrm{2D}$ such that the second eigenvalue of $\rho^\mathrm{2D}(m)$ is equal to $\rho^\mathrm{1D}(m)$.
Referring to \cref{tab:mirror_eigvals}, this implies 
\begin{equation}
    \pm_r1=\mp_p1, \label{eq:signs-a-c}
\end{equation} 
i.e., it fixes one of the four signs.
In this basis, we then have the following matrix representations
\begin{equation}
    \rep(m) = \mqty(\dmat{\pm_p1,\mp_p1,\mp_p1})\,,\;\; \rep(m_\perp) = \mqty(\dmat{\pm_q1,\mp_q1,\pm_t1}).
\end{equation}
Recall that a unitary symmetry $g$ with representation $D(g)$ leads to the following constraint on the Hamiltonian
\begin{equation}
    \Ham(\vec{k}) = \rep(g)\Ham(g^{-1}\vec{k})\rep(g)^{-1}.
    \label{eq:sym_constraint}
\end{equation}
We define orthogonal coordinates $(k_x,k_z)$ in the $m$-plane such that $k_z$ runs along the rotation axis and such that the TP is located at $(k_x,k_z)=0$.
An expansion to leading order in $k_x$ around the TP then takes the following form (see \cref{App:Classif:no_PT_mirror:kdotp:C4vC6v} for the derivation):
\begin{equation}
    \Ham(k_x,k_z) = \mqty(ak_x^2 & 0 & 0\\0 & -ak_x^2 & Ak_x^{\frac{3+s}{2}}\\0 & \cconj{A}k_x^{\frac{3+s}{2}} & k_z b+ck_x^2),
    \label{eq:kdotp_C4vC6v}
\end{equation}
where $s=-(\pm_q)(\pm_t)$ is a sign corresponding to the sign of the product of the $C_2$-characters of $\rho^\mathrm{2D}$ and $\rho^\mathrm{1D}$, $a(k_z),b(k_z),c(k_z)$ are continuous real-valued functions, and $A(k_z)$ is a continuous complex-valued function.
We suppress the $k_z$ argument of the four functions in \cref{eq:kdotp_C4vC6v} to improve readability.
Note that in \cref{eq:kdotp_C4vC6v} we dropped some terms proportional to the identity matrix, which are irrelevant to the nodal structure, as they merely shift all energy bands equally. 
The block-diagonal structure (which does \emph{not} parallel the blocks of $\rho^\mathrm{2D}\oplus\rho^\mathrm{1D}$) arises from the two different eigenvalues of $D(m)$ in the $m$-plane, and therefore persists to all orders of the expansion.

There are two possibilities for $\Ham(k_x,k_z)$ to have degeneracies in the spectrum (corresponding to band nodes): either (1)~the bottom right $2\times 2$ block $h_{23}(k_x,k_z)$ has degenerate eigenvalues, or (2)~it has an eigenvalue $a(k_z) k_x^2$.
For (1), recall that spectral degeneracies of a matrix correspond to the roots of the characteristic polynomial, and that coinciding roots of a polynomial can be diagnosed by a vanishing discriminant.
Therefore, we compute the discriminant of the characteristic polynomial of $h_{23}(k_x,k_z)$,
\begin{equation}
    \left[(a+c)k_x^2+bk_z\right]^2+4\abs{A}^2k_x^{3+s} = 0.\label{eq:m-plane-polynomial}
\end{equation}
Because $s=\pm 1$, the left-hand side of the equation is a sum of two squares, and as such it has no real solutions except for $k_x=k_z=0$, which, by construction, is the TP.

On the other hand, condition (2) is satisfied if and only if
\begin{equation}
    \begin{split}
        0 &= \det\left[h_{23}(k_x,k_z) - a(k_z) k_x^2\id\right]\\
        &= k_x^2\left[-2abk_z + 2a(a-c)k_x^2 - \abs{A}^2k_x^{1+s}\right],
    \end{split}
    \label{eq:arcs-in-mirror-planes}
\end{equation}
where in the second line we dropped the functional dependences of $a,b,c$ and $A$.
\Cref{eq:arcs-in-mirror-planes} admits two types of solutions.
First, $k_x=0$ is always a solution that defines the central NL along the $k_z$-axis. 
The second type of solution corresponds to zeros of the expression in the square brackets. 
Assuming that this solution appears close to the TP (which is always the case for NL arcs of type-$\textsf{B}$ TPs), we approximate the variable functions by their values at $k_z=0$, i.e.,
\begin{equation}
    f(k_z) \approx f(k_z=0) \equiv f_0 \quad\textrm{for $f\in\{a,b,c,A\}$}
    \label{eq:parameters_kz_expansion}
\end{equation} 
and find the explicit root
\begin{equation}
    k_z^\text{arc}(k_x) = \frac{a_0-c_0}{b_0}k_x^2 - \frac{\abs{A_0}^2}{2a_0b_0}k_x^{1+s}
    \label{eq:NL_C4vC6v}
\end{equation}
that describes the NL arcs.

To analyze the result in \cref{eq:NL_C4vC6v}, first observe that for $s=+1$, the NL arcs attach to the TP [because $\lim_{k_x\to 0}k_z^\text{arc}(k_x) = 0$], while they generically do \emph{not} attach to the TP for $s=-1$ [because $\lim_{k_x\to 0}k_z^\text{arc}(k_x) = -{\abs{A_0}^2}/{2a_0b_0}$]. 
We further read from \cref{eq:NL_C4vC6v} that the NL arcs scale quadratically as a function of $k:=(k_x^2+k_y^2)^{1/2}$, i.e., $k_z^\text{arc}(k)-k_z^\text{arc}(0)\propto k^2$.
Thus, we conclude that the TP is type $\mathsf{A}$ (type $\mathsf{B}_q$) if the product of $C_2$-characters of $\rho^\mathrm{2D}$ and $\rho^\mathrm{1D}$ is negative (positive). 
Additionally, we see from the more general \cref{eq:arcs-in-mirror-planes} that a nexus of NL arcs would coincide with a type-$\mathsf{A}$ TP when $A_0 = 0$. 
Since $A(k_z)$ is a complex function, this is generically achieved by tuning two real parameters, i.e., the sought codimension equals $2$.

To finalize the type classification for the 3 MPGs discussed here, we determine the value of $s$ for each combination of 2D and 1D ICRs $\rho^\mathrm{2D}\oplus\rho^\mathrm{1D}$. 
This is achieved by looking up the ICRs and $C_2$-characters on the Bilbao crystallographic server~\cite{Aroyo:2011,Aroyo:2006a,Aroyo:2006b} using the program \textsc{Corepresentations PG}~\cite{Xu:2020,Elcoro:2021}. 
We find that for $4mm$ and $\bar{4}'2'm$ all combinations of ICRs have $s=-1$, such that any TPs in those groups are type $\mathsf{A}$.
The analysis is more subtle for $6mm$: here, type-$\mathsf{A}$ TPs ($s=-1$) are realized for the ICR combinations $(E_1;A_i)$ and $(E_2;B_i)$, while type-$\mathsf{B}_q$ TPs ($s=+1$) for ICR combinations $(E_1;B_i)$ and $(E_2;A_i)$.
In the latter case we can also immediately conclude that $n_a^\mathrm{nexus}=12$ (because there are two sets of three symmetry-related mirror planes, and each mirror plane contains two NL arcs starting at the TP).

\subsubsection{Nodal-line arc characterization in MPGs \texorpdfstring{$4mm$, $\bar{4}'2'm$ and $6mm$}{4mm,-4'2'm 
and 6mm}}\label{Sec:Classif:mirror_no_PT:C4vC6v-part2}

We next derive characteristics of the NL arcs that appear near the type-$\mathsf{A}$ and at the type-$\mathsf{B}$ TPs just identified.
Although the analysis in \cref{Sec:Classif:mirror_no_PT:C4vC6v} was performed for one particular choice of mirror-invariant $m$-plane, the arguments [including the results in \cref{eq:m-plane-polynomial,eq:arcs-in-mirror-planes,eq:parameters_kz_expansion,eq:NL_C4vC6v}] straightforwardly generalize.

To begin, note that the band structure respects the MPG symmetry, which readily implies that
\begin{itemize}
    \item[(\emph{i})] NL arcs described by the same \cref{eq:arcs-in-mirror-planes} and with the same functions $a,b,c,A$ also appear in all the symmetry-related $m$-planes [cf.~\cref{fig:mirror-planes}(a--c)].
\end{itemize}
Additionally, note that the sign $s$ is a characteristic of the two crossing ICRs at the HSL; especially, it does not depend on the particular choice of $m$. Therefore, the whole algebraic analysis can be repeated for vertical mirror planes $m'$ that are not symmetry-related to $m$ [colored differently in \cref{fig:mirror-planes}(a--c)].
Note that for $4mm$ the mirrors $m'$ and $m_\perp$ are different, for $6mm$ we have $m=m_\perp$, and for $\bar{4}'2'm$ there is no $m'$.
We therefore deduce that 
\begin{itemize}
    \item[(\emph{ii})] NL arcs in the $m'$-plane are defined by the same implicit \cref{eq:arcs-in-mirror-planes}, although with a potentially different set of functions $a'$, $b'$, $c'$, and $A'$.
\end{itemize}
In the following, we study the implications of observations (\emph{i}) and (\emph{ii}) for the NL structure near the discussed TPs.

Let us first analyze the implications when $s=+1$, i.e., when the NL arcs are attached to a type-$\mathsf{B}$ TP. 
It follows from (\emph{i}) and (\emph{ii}) that the TP must be connected to one NL arc in each vertical mirror plane~[such as shown in \cref{fig:TPTypes}(d) for $6mm$].
The NL arcs in the two sets of mirror planes are generally in different energy gaps, which can be seen as follows.
According to the derivation in \cref{Sec:Classif:mirror_no_PT:C4vC6v}, the energy of the two bands involved in the NL along the NL arc $k_z=k_z^\text{arc}(k_x)$ is $ak_x^2$.
To determine in which gap the NL lies, we need to compare it to the energy of the third band, which can be deduced from \cref{eq:kdotp_C4vC6v,eq:arcs-in-mirror-planes} to be\footnote{
As an intermediate step, note that \cref{eq:arcs-in-mirror-planes} with $s=+1$ and $k_x\neq 0$ implies that, near the TP, $bk_z + ck_x^2 = ak_x^2 - (\abs{A}^2/2a)k_x^2$, and that this is also equal to $\tr(\Ham)$ after substituting for the bottom-right element of the matrix in \cref{eq:kdotp_C4vC6v}.
Since the trace is the sum of all eigenvalues, and the other two eigenvalues were determined as $a k_x^2$, the result in \cref{eq:third-band-energy} follows.} 
\begin{equation}
    -\left(a+\frac{\abs{A}^2}{2a}\right)k_x^2.\label{eq:third-band-energy}
\end{equation}
It follows from comparing the three band energies that the NL arc is in the red (blue) gap if $a_0<0$ ($a_0>0$).

\begin{figure}[t]
    \centering
    \includegraphics{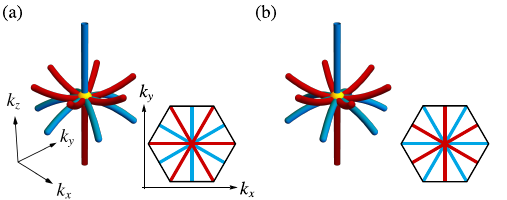}
    \caption{
        Nodal-line (NL) structure near a type-$\mathsf{B}_q$ triple point (TP) on a high-symmetry line (HSL) with little co-group $6mm$, which requires a specific combination (class II) of 2D and 1D irreducible representations (cf.~\cref{tab:Classification}).
        NLs are shown in red (blue) if they are in the lower (upper) band gap of the three-band model and the TP is indicated in yellow.
        Panels (a) and (b) show the two possibilities with red (blue) NL arcs in one set of symmetry-related mirror planes and blue (red) NL arcs in the other.
        The insets show the view on the nodal lines from the top to clarify the arrangement of the NL arcs.
        An analogous discussion applies to HSLs with little co-group $6/m'mm$ (with a class-II combination of irreducible corepresentations).
    }
    \label{fig:NLStructure:6mmII}
\end{figure}

Observation (\emph{ii}) seems to suggest that the NL arcs in symmetry-unrelated planes are independent of each other.
However, as revealed in \cref{App:Classif:no_PT_mirror:mirror_planes}, the four functions describing the expansions in $m$ and $m'$ obey certain constraints (explicitly derived in the supplementary data and code~\cite{Lenggenhager:2022:TPClassif:SDC}).
For the MPG $6mm$ and ICRs such that $s=+1$, i.e., such that the resulting TP is type $\mathsf{B}_q$, the derived constraints are 
\begin{equation}
    a'=-a,\quad b'=b, \quad c'=c, \quad \abs{A'}= \abs{A}.\label{eq:inequiv-m-relation-1}
\end{equation}
Therefore, the NL arcs in one set of planes are in the red gap, while the NL arcs in the other set of planes are in the blue gap.
Which set of planes contains NL arcs in which gap, depends on the parameter values as illustrated in \cref{fig:NLStructure:6mmII}.

In the remainder of this section, we analyze the implications of (\emph{i}) and (\emph{ii}) for $s=-1$, i.e., when the NL arcs connect to the rotation axis away from a type-$\mathsf{A}$ TP.
Here, note that a crossing of NLs in the two different gaps would automatically imply a threefold degeneracy (i.e., a TP) at the crossing; therefore, it follows that the nexus of NL arcs associated with type-$\mathsf{A}$ TPs are formed in the same gap (red/blue) as the central nodal line at which the arcs converge [cf.~\cref{fig:TPTypes}(b)].

\begin{figure}[t]
    \centering
    \includegraphics{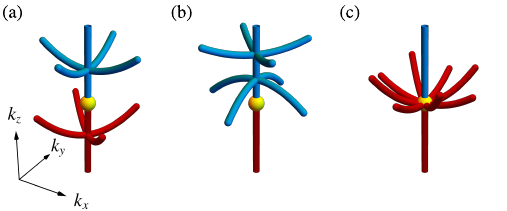}
    \caption{
        Nodal-line (NL) structure near a type-$\mathsf{A}$ triple point (TP) on a high-symmetry line (HSL) with little co-group $4mm$.
        NLs are shown in red (blue) if they are in the lower (upper) band gap of the three-band model and the TP is indicated in yellow.
        Type-$\mathsf{A}$ TPs can be accompanied by nearby nexus points on the central nodal line: (a) on opposite sides of the the TP if $\sign(a_0)=-\sign(a_0')$, or (b) on the same side if $\sign(a_0)=\sign(a_0')$.
        (c) If the complex parameter $A_0$ is fine-tuned to $0$, the two nexus points (independently of their color) simultaneously collide with the TP.
        An analogous discussion applies for HSLs with little co-group $4/m'mm$, with the only relevant change being the reduction of $A_0$ to be a real (rather than complex) parameter.
    }
    \label{fig:NLStructure:4mm}
\end{figure}

First, for MPG $4mm$, the derived constraints on the functions describing the expansion in $m$ and $m'$ are 
\begin{equation}
    b'=b, \quad c'=c, \quad \abs{A'}= \abs{A},\label{eq:inequiv-m-relation-2}
\end{equation}
while $a(k_z)$ and $a'(k_z)$ are unrelated.
When plugged into \cref{eq:arcs-in-mirror-planes}, we find that the functions $a$ and $a'$ provide enough freedom to realize nexus points of NLs in the two sets of mirror planes that attach to the rotation axis at arbitrary and different positions.  
In particular, the approximate result in \cref{eq:NL_C4vC6v} derived in the lowest order in $k_z$ suggests one nexus point for each set of symmetry-related $m$-planes (this can change when terms of higher order in $k_z$ are included).
The two nexus points can be on the opposite or on the same sides of the TP, depending on the relative sign of the coefficients $a_0$ and $a_0'$, see \cref{fig:NLStructure:4mm}(a,b).
Generically, the two nexus points do not coincide, such that $n_a^\mathrm{nexus}=4 = 2\times 2$ (each nexus point arises due to NL arcs in one set of two symmetry-related mirror planes, and each plane contains two NL arcs attached to the central NL.)

Curiously, the constraint $\abs{A}=\abs{A'}$ implies that the two nexus points in the two pairs of planes collide with the TP \emph{simultaneously}.
Notably, if $\sign(a_0)=\sign(a_0')$, such that the two nexus points are on the same side of the TP for $A\neq 0$, all the NL arcs converging at the fine-tuned type-$\mathsf{A}$ TP for $A_0=0$ would be in the \emph{same} gap, cf.~\cref{fig:NLStructure:4mm}(c).
This feature sharply contrasts to type-$\mathsf{B}$ TPs for which the number of attached NL arcs is always distributed equally over both gaps.
However, a situation similar to a type-$\mathsf{B}$ TP is also possible if $\sign(a_0)=-\sign(a_0')$ and $A_0=0$.

\begin{figure}[t]
    \centering
    \includegraphics{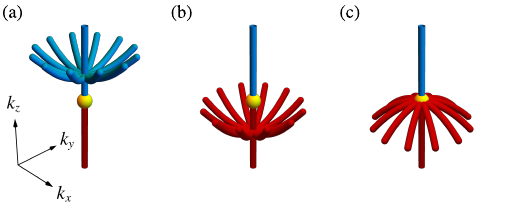}
    \caption{
        Nodal-line (NL) structure near a type-$\mathsf{A}$ triple point (TP) on a high-symmetry line (HSL) with little co-group $6mm$, which requires a specific combination (class I) of 2D and 1D irreducible representations (cf.~\cref{tab:Classification}).
        NLs are shown in red (blue) if they are in the lower (upper) band gap of the three-band model and the TP is indicated in yellow.
        As in the case of the little co-group $4mm$ (cf. \cref{fig:NLStructure:4mm}), type-$\mathsf{A}$ TPs in $6mm$ can be accompanied by nexus points nearby on the central nodal line (NL).
        However, the points of coalescence of NL arcs with the central NL in the inequivalent sets of mirror planes always coincide, such that we only distinguish (a) blue nexus points ($a_0>0$), (b) red nexus points ($a_0<0$) and (c) nexus points fine-tuned to collide with the TP (if the complex parameter $A_0$ is set to zero), all with $12$ NL arcs attached.
        An analogous discussion applies for HSLs with little co-group $6/m'mm$, with the only relevant change being the reduction of $A_0$ to real values.
    }
    \label{fig:NLStructure:6mmI}
\end{figure}

Next, we analyze type-$\mathsf{A}$ TPs in the MPG $6mm$.
It is derived in \cref{App:Classif:no_PT_mirror:mirror_planes} that the expansions in the two planes obey 
\begin{equation}
    a'=a,\quad b'=b,\quad c'=c,\quad \abs{A'}=  \abs{A}.\label{eq:inequiv-m-relation-3}
\end{equation}
Looking at \cref{eq:arcs-in-mirror-planes}, we find that NL arcs in both sets of symmetry-related planes always attach to the \emph{same} nexus point as illustrated in \cref{fig:NLStructure:6mmI}(a,b), with blue and red nexus points, respectively (distinguished by the sign of $a_0$).
Therefore, we find $n_a^\mathrm{nexus}=12$, i.e., the same as for type-$\mathsf{B}$ TPs in the same MPG.
However, one should bear in mind that \cref{eq:arcs-in-mirror-planes} was derived from a \kdotp{} expansion to finite order in $k_x$.
It can be shown that higher-order terms discriminate between the two sets of planes, meaning that the NL arcs in $m$-planes and $m'$-planes disperse differently for large enough distance $k$ from the rotation axis.
All these described features are clearly manifested by the type-$\mathsf{A}$ TP of the compound \ce{AlN}, see \cref{fig:TPmat:AlN_1} in the Appendix.
Finally, when the complex parameter $A_0$ is set to zero, the $12$ NL arcs attached to the fine-tuned type-$\mathsf{A}$ TP are all in the same gap, cf.~\cref{fig:NLStructure:6mmI}(c), such that the type-$\mathsf{A}$ TP can always be distinguished from the type-$\mathsf{B}$ TP.

Lastly, for $\bar{4}'2'm$ there is only one set of symmetry-related mirror planes; in this case, there is only a single nexus point with four NL arcs: $n_a^\mathrm{nexus}=4$.

\subsubsection{\texorpdfstring{$\bar{6}'m2'$}{-6'm2'}}\label{Sec:Classif:mirror_no_PT:C6PT}

The group $\bar{6}'m2'$ is similar to the cases discussed above in \cref{Sec:Classif:mirror_no_PT:C4vC6v,Sec:Classif:mirror_no_PT:C4vC6v-part2}, with one crucial difference: each of the three $C_3$-symmetry-related vertical mirrors $m$ [cf.~\cref{fig:mirror-planes}(d)] is paired with a \emph{pseudo-mirror} $m_\perp\mcP\mcT$ (here $m_\perp$ is again the mirror perpendicular to $m$; note that neither $m_\perp$ nor $\mcP\mcT$ is an element of the symmetry group).
We proceed analogously, but need to be careful because $m_\perp\mcP\mcT$ is antiunitary, such that we have to deal with corepresentations rather than representations.
For an \emph{antiunitary} symmetry $g$, the symmetry constraint in \cref{eq:sym_constraint} needs to be replaced by
\begin{equation}
    \Ham(\vec{k}) = \corep(g)\cconj{\Ham(g^{-1}\vec{k})}\corep(g)^{-1},
    \label{eq:sym_constraint:antiunitary}
\end{equation}
where $\corep(g)$ is the matrix \emph{corepresentation} of $g$.
(See \cref{App:antiunitary_symmetries} for a brief review of the representation theory of groups with antiunitary symmetries.)
For the two possible combinations of irreducible corepresentations of $\bar{6}'m2'$ $(E;A_1)$ [upper sign in \cref{eq:kdotp_C6PT}] and $(E;A_2)$ (lower sign), this implies the following form of the Hamiltonian in the mirror plane (see \cref{App:Classif:no_PT_mirror:kdotp:C6PT} for the derivation)
\begin{equation}
    \Ham(k_x,k_z) = \mqty(ak_x^2 & 0 & 0\\0 & -ak_x^2 & d(1\mp\i)k_x\\0 & d(1\pm\i)k_x & bk_z+ck_x^2)
    \label{eq:kdotp_C6PT}
\end{equation}
where $a,b,c,d$ are continuous real-valued functions $of k_z$.

We observe that \cref{eq:kdotp_C6PT} is identical to \cref{eq:kdotp_C4vC6v} with $A=d(1\mp\i)$ and $s=-1$, such that we can immediately obtain the implicit equation for NL arcs from \cref{eq:arcs-in-mirror-planes}, namely
\begin{equation}
    -2abk_z + 2a(a-c)k_x^2 - 2\abs{d}^2k_x^{1+s}=0.
\end{equation}
From here, assuming the functions $a,b,c,d$ are approximately constant near the TP, we obtain the result for the NL arcs
\begin{equation}
    k_z = k_z^\text{arc}(k_x) \equiv \frac{a_0-c_0}{b_0}k_x^2 - \frac{d_0^2}{a_0b_0},
    \label{eq:NL_C6PT}
\end{equation}
where $d_0 = d(k_z=0)$.
Therefore, we find that TPs in $\bar{6}'m2'$ are always type $\mathsf{A}$.
Since there is only one set of symmetry-related planes, only one nexus point is present in the \kdotp{} expansion, and $n_a^\mathrm{nexus}=6$, because there are three mirror planes with two NL arcs starting at the nexus point each.
Additionally, because $d$ is a real parameter, the codimension for colliding the nexus point with the TP is $1$.

\subsubsection{\texorpdfstring{$3m$}{3m}}\label{Sec:Classif:mirror_no_PT:C3v}

Finally, we discuss the remaining MPG $3m$, which has three-fold rotational symmetry with three vertical mirrors that are not associated with corresponding orthogonal \mbox{(pseudo-)mirror} planes [cf.~\cref{fig:mirror-planes}(d)].
Again, we choose coordinates $(k_x,k_z)$ in the mirror plane under consideration (the three mirror planes are equivalent due to rotational symmetry), then the Hamiltonian for all possible combinations of ICRs takes the form (see \cref{App:Classif:no_PT_mirror:kdotp:C3v} for the derivation)
\begin{equation}
    \Ham(k_x,k_z) = \mqty(ak_x & 0 & 0\\0 & -ak_x & Ak_x\\0 & \cconj{A}k_x & bk_z+ck_x),
    \label{eq:kdotp_C3v}
\end{equation}
where $a,b,c$ are real-valued functions and $A$ is a complex-valued function of $k_z$.

Degeneracies in the spectrum are obtained (1) if
\begin{equation}
    \left[(a+c)k_x+bk_z\right]^2 + 4\abs{A}kx^2 = 0,
\end{equation}
with the only solution $k_x=k_z=0$ (the TP), and (2) if
\begin{align}
    0 &= \det\left[h_{23}(k_x,k_z) - ak_x\id\right]\\
    &= k_x\left[-2abk_z + 2a(a-c)k_x - \abs{A}^2k_x\right].
\end{align}
The latter equation has two solutions.
First, $k_x=0$ (the central NL), and second, after approximating $a,b,c,A$ as constants in the vicinity of the TP,
\begin{equation}
    k_z = \left(\frac{a_0-c_0}{b_0}-\frac{\abs{A_0}^2}{2a_0b_0}\right)k_x
\end{equation}
(the NL arc).
Since $\lim_{k_x\to 0}k_z^\text{arc}(k_x)=0$, the NL arc attaches to the TP and it does so linearly as a function of $(k_x^2+k_y^2)^{1/2}$, such that the TP is always type $\mathsf{B}_l$.
We again find that $n_a^\mathrm{nexus}=6$ (due to three symmetry-related mirror planes with two NL arcs in each).

\section{Derivation in the presence of \texorpdfstring{$\mcP\mcT$}{PT} symmetry}\label{Sec:Classif:PT}

There are six relevant MPGs with $\mcP\mcT$ symmetry listed in \cref{tab:MPGs} that can protect a TP along a HSL: $\bar{3}'$, $\bar{3}'m$, $4/m'$, $4/m'mm$, $6/m'$ and $6/m'mm$.
The discussion of these MPGs is rather involved, because NLs can be stabilized by $\mcP\mcT$ symmetry \emph{anywhere} in the momentum space~\cite{Fang:2015,Bzdusek:2017}, i.e., they are not constrained to symmetric planes.
Although we ultimately find the resulting classification to be the same (up to a reduction of the codimension where applicable) as for the corresponding MPGs \emph{without} $\mcP\mcT$ symmetry (specifically $3m$, $4mm$ and $6mm$ analyzed in \cref{Sec:Classif:no_PT_mirror}), the minimal \kdotp{} models and their analysis are considerably more complicated, and much of the explicit algebra is carried out in a supplementary \Mathematica{} notebook~\cite{Lenggenhager:2022:TPClassif:SDC}.

Initially, we proceed similarly to the previous section: we construct minimal \kdotp{} models for the Hamiltonian near the TP in the various MPGs (see \cref{App:Classif:PT:kdotp}).
However, due to the presence of $\mcP\mcT$ symmetry, we cannot restrict to mirror planes and need to study the full 3D \kdotp{} models.
In contrast to \cref{Sec:Classif:no_PT}, we here find it more convenient to perform the expansion in all three momentum components of $\vec{k}$.
In the present section, we focus on introducing the relevant methods; in particular, in \cref{Sec:Classif:TP:NLarcs}, we describe the techniques we developed to determine the leading-order terms of the \kdotp{} expansion, while in \cref{Sec:Classif:TP:NLarcs-part2} we show how to deduce the NL structure near the TP from the obtained leading-order expansions.
While we defer concrete calculations and proofs to a supplementary \Mathematica{} notebook~\cite{Lenggenhager:2022:TPClassif:SDC}, several representative calculations are included in \cref{App:Classif:PT:nls}.

In addition, the presence of $\mcP\mcT$ symmetry implies further \emph{topological aspects} of the triple points:
the central NLs containing the triple points are characterized~\cite{Lenggenhager:2021:MBNLs} by the non-Abelian~\cite{Wu:2019,Bouhon:2020} (also called \enquote{generalized-quaternion}) invariant, and pairs of TPs are characterized~\cite{Lenggenhager:2021:TPHOT} by Euler and Stiefel-Whitney monopole charges~\cite{Fang:2015,Bzdusek:2017,Zhao:2017,Ahn:2018,Ahn:2019,Bouhon:2020b,Unal:2020}.
While an extensive discussion of the latter appears in a separate publication~\cite{Lenggenhager:2021:TPHOT}, we show in \cref{Sec:Classif:TP:winding} that the quaternion invariant computed on a closed contour surrounding only the central nodal line (indicated by the winding number of the 2D ICR computed on the same contour) is also determined by symmetry.
The full results of the classification in the presence of $\mcP\mcT$ symmetry are listed in \cref{tab:Classification_PT}.

\begin{table}
    \centering
    \begin{threeparttable}
    \caption{
        Classification of triple points (TPs) on high-symmetry lines with little co-group $\LcG{\vec{k}}$ containing $\mcP\mcT$ symmetry.
        The TP is characterized by whether nodal-line (NL) arcs attach to it (type $\mathsf{A}$) or not (type $\mathsf{B}$)~\cite{Zhu:2016}.
        We further distinguish type-$\mathsf{B}$ TPs according to the scaling of the attached NL arcs $k_z\propto(k_x^2+k_y^2)^{\mu/2}$: type-$\mathsf{B}_q$ TPs have $\mu=2$ and type-$\mathsf{B}_l$ TPs have $\mu=1$.
        For all TP types we further define $n_a^\mathrm{nexus}$ as the number$^\ddagger$ of NL arcs attached to a generic \emph{nexus} point (coinciding with the TP for type $\mathsf{B}$, in the vicinity of the TP for type $\mathsf{A}$).
        These properties depend on $\LcG{\vec{k}}$ and, in general, on the irreducible corepresentations (ICRs) of the bands involved in the TP formation.
        The notation for the ICRs follows Ref.~\onlinecite{Bradley:1972}, with $i=1,2$.
        For most little co-groups, all possible pairs of ICRs are equivalent (cf.~text); only for the $C_6$-symmetric groups do the pairs fall into \emph{two} equivalence classes (denoted by I and II in the second column).
        In the last column we list the winding number $w_\mathrm{2D}$ of the 2D ICR computed on a closed contour around the central NL.
    }
    \begin{ruledtabular}
    \begin{tabular}{CcCCCCC}
    	      \LcG{\vec{k}}       & Class &                                \text{ICRs}                                & \text{TP type} & \;n_a^{\mathrm{nexus}\,\ddagger}\; & \mu & \abs{w_{\mathrm{2D}}} \\ \hline\addlinespace
    	        \bar{3}'          &  \td  &                   (\irrep{E}{}{2}\!\irrep{E}{}{1};A_1)                    &  \mathsf{B}_l  &         6          &  1  &           1           \\
    	        \bar{3}'m         &  \td  &                                  (E;A_i)                                  &  \mathsf{B}_l  &         6          &  1  &           1           \\ \addlinespace
    	          4/m'            &  \td  &   (\irrep{E}{}{2}\!\irrep{E}{}{1};A),(\irrep{E}{}{2}\!\irrep{E}{}{1};B)   &   \mathsf{A}   &         4          &  2  &           2           \\
    	         4/m'mm           &  \td  &                              (E;A_i),(E;B_i)                              &   \mathsf{A}   &         4          &  2  &           2           \\ \addlinespace
    	 \multirow{2}{*}{$6/m'$}  &   I   & (\irrep{E}{2}{2}\!\irrep{E}{2}{1};A),(\irrep{E}{1}{2}\!\irrep{E}{1}{1};B) &   \mathsf{A}   &         12         &  2  &           2           \\
    	                          &  II   & (\irrep{E}{1}{2}\!\irrep{E}{1}{1};A),(\irrep{E}{2}{2}\!\irrep{E}{2}{1};B) &  \mathsf{B}_q  &         12         &  2  &           2           \\ \addlinespace[0.25\defaultaddspace]
    	\multirow{2}{*}{$6/m'mm$} &   I   &                            (E_1;A_i),(E_2;B_i)                            &   \mathsf{A}   &         12         &  2  &           2           \\
    	                          &  II   &                            (E_1;B_i),(E_2;A_i)                            &  \mathsf{B}_q  &         12         &  2  &           2
    \end{tabular}
    \end{ruledtabular}
    \begin{tablenotes}[flushleft]
        \item[$\ddagger$] {\small This quantity has to be distinguished from $N_a$ introduced in Ref.~\onlinecite{Lenggenhager:2021:MBNLs} which is the number of NL arcs attached to a type-$\mathsf{B}$ TP \emph{per gap}; for type-$\mathsf{B}$ TPs $N_a=\tfrac{1}{2}n_a^\mathrm{nexus}$.}
    \end{tablenotes}
    \label{tab:Classification_PT}
    \end{threeparttable}
\end{table}

\subsection{Condition for the occurrence of NL arcs near TPs}\label{Sec:Classif:TP:NLarcs}

The first step in deriving the classification is once more the construction of minimal \kdotp{} models describing the Hamiltonian near the TP (this time we use the full 3D \kdotp{} models).
We keep only traceless terms of leading order in $\vec{k}$ (leading in each momentum component separately) and without loss of generality we set the energy of the 2D ICR for $\vec{k}=(0,0,k_z)$ to zero and the TP position to the origin $\vec{k}=0$.
It is always possible to find a basis in which $\mcP\mcT$ is represented by complex conjugation $\mcK$~\cite{Bouhon:2020}; then, the Bloch Hamiltonian is a real symmetric matrix.
For a given MPG, the leading-order \kdotp{} expansions for various combinations of 1D and 2D ICRs result in \emph{Hamiltonian families} $\Ham_\vec{a}(\vec{k})$ parametrized by real parameters $a_i$, collected in the vector $\vec{a}$, which we provide in the supplementary data and code~\cite{Lenggenhager:2022:TPClassif:SDC}.
Further details on the derivation of the leading-order \kdotp{} models, including an explicit discussion for ICRs $(E_1;A_1)$ of $6/m'mm$, are presented in \cref{App:Classif:PT:kdotp}.

For a given MPG, most combinations of ICRs lead to \emph{equivalent} Hamiltonian families in the following sense.
Let $\chi[\Ham]$ be the characteristic polynomial of some Hamiltonian $\Ham$; then, we call two families $\Ham^{(1,2)}_\vec{a}(\vec{k})$ equivalent if
\begin{equation}
    \chi[\Ham^{(1)}_{\tilde{\vec{a}}}(\tilde{\vec{k}})] = \chi[\Ham^{(2)}_{\vec{a}}(\vec{k})]
    \label{eq:model_equivalence}
\end{equation}
for $\tilde{\vec{k}}$ and $\tilde{\vec{a}}$ related to $\vec{k}$ and $\vec{a}$ by linear transformations (in \cref{App:Classif:PT:nls:equiv_classes} we discuss one example of such an equivalence).
In particular, this implies that the Hamiltonian spectra are also equal up to the same linear transformations of $\vec{k}$ and $\vec{a}$.
Defining equivalence according to \cref{eq:model_equivalence}, we find~\cite{Lenggenhager:2021:MBNLs,Lenggenhager:2022:TPClassif:SDC} that the \kdotp{} models for $6/m'$ and $6/m'mm$ fall into two equivalence classes, while all the other MPGs only have a single equivalence class (\cref{tab:Classification_PT}).
Since NLs are properties of the spectrum alone, we restrict the analysis of the TP characteristics to one representative for each equivalence class.

We next determine the number of NL arcs attached to the TP by analyzing the discriminant $\Delta[\Ham]$ of the characteristic polynomial $\chi[\Ham]$, similar to our analysis in the previous section.
However, while in \cref{Sec:Classif:no_PT} the restriction to mirror planes led us to analyze the characteristic polynomial of a $2\times 2$ Hamiltonian block, for the presently studied MPGs we are led to consider the characteristic polynomial of the full $3\times 3$ \kdotp{} Hamiltonian.
Because zeroes of $\Delta[\Ham]$ correspond to NLs, we need to solve the multivariate polynomial equation
\begin{equation}
    \Delta[\Ham_{\vec{a}}(\vec{k})] = 0\label{eq:NLs-as-roots-of-Delta}
\end{equation}
with parameters $\vec{a}$
over $\vec{k}\in\mathbb{R}^3$ to find the NLs.
By construction, the line $k_x=k_y=0$ is always a solution with some multiplicity $r$. 
Determining the NL arcs (if there are any) corresponds to finding additional real roots of \cref{eq:NLs-as-roots-of-Delta}. 
However, since the characteristic polynomial of the $3\times 3$ \kdotp{} model is of high-order in the momentum components, this is a non-trivial task that requires methods beyond those described in \cref{Sec:Classif:no_PT}.

Because we are primarily interested in NLs attached to the TP at $\vec{k}=0$, we focus on the leading terms of $\Delta[\Ham_\vec{a}(\vec{k})]$.
In cylindrical coordinates $(k,\theta,k_z)$ we can consider $\theta\in[0,2\pi)$ as an additional parameter, such that the discriminant is a bivariate polynomial in $k,k_z$
\begin{equation}
    \begin{split}
        \Delta_{\vec{a},\theta}(k,k_z) &= \Delta[\Ham_\vec{a}(k\cos\theta, k\sin\theta, k_z)]\\
        &= k^r\sum_{\alpha,\beta}c_{\alpha\beta}(\theta,\vec{a})k^\alpha k_z^\beta
    \end{split}
    \label{eq:disc_polynomial}
\end{equation}
with real coefficients $c_{\alpha\beta}(\theta,\vec{a})$.
The particular exponents $(\alpha,\beta)$ of the bivariate monomials that appear in \cref{eq:disc_polynomial} depend on the choice of the MPG and of its ICRs.
To determine the leading-order monomials of $\Delta_{\vec{a},\theta}(k,k_z)$, note that NL arcs attached to the TP have an anticipated functional dependence
\begin{equation}
    k_z^\text{arc}(k)\propto k^\mu\quad\textrm{for some $\mu\in\mathbb{R}^+$}
    \label{eq:arc-scaling}
\end{equation} 
in the vicinity of the TP. Such a root of $ \Delta_{\vec{a},\theta}(k,k_z)$ is only attainable if the exponents $(\alpha',\beta')$ of the leading-order monomials $k^{\alpha'} k_z^{\beta'}$ obey $\alpha'+\mu\beta'=\textrm{const}$.
To determine \emph{all} leading terms in \cref{eq:arc-scaling}, we need to account for arbitrary prospective scalings $0<\mu<\infty$, i.e., we need to keep all the bivariate monomials $k^\alpha k_z^\beta$ such that $k^{\alpha + \mu \beta}$ is of leading order for \emph{at least one} value of $\mu$.

We now describe a systematic procedure for identifying the leading terms.
Let
\begin{equation}
    M = \left\{\left.(\alpha,\beta)\in\mathbb{R}^+\right|c_{\alpha\beta}(\theta,\vec{a})\not\equiv 0\right\}
    \label{eq:disc_terms}
\end{equation}
be the set of monomials that appear in $\Delta_{a,\theta}(k,k_z)$, where \enquote{$f_1\not\equiv f_2$} indicates that functions $f_1$ and $f_2$ are not identical; then, for each fixed scaling $0<\mu<\infty$, the set of leading monomials is
\begin{equation}
    L_\mu(M) = \argmin_{(\alpha,\,\beta)\in M}(\alpha+\mu\beta).
\end{equation}
Geometrically, $L_\mu(M)$ is the set of points in $M$ that lie on a line of slope $-1/\mu$, such that the origin $(\alpha,\beta)=(0,0)$ is on one side and all the other points of $M$ are on the other side of this line (cf.~\cref{App:fig:disc_C6vI}).
Naturally, the union of such sets gives the set of all leading monomials:
\begin{equation}
    L(M) = \bigcup_{0<\mu<\infty} L_\mu(M).
    \label{eq:disc_leading_terms}
\end{equation}
We note that this is equivalent to the part of the convex hull of $M$ that faces the origin, which is useful for explicitly computing $L(M)$.
For more details see the example discussed in \cref{App:Classif:PT:nls:typeA} and \cref{App:fig:disc_C6vI} therein.

\subsection{TP characterization from the leading-order expansion}\label{Sec:Classif:TP:NLarcs-part2}

Knowing the general principles that determine the leading-order terms of the discriminant in \cref{eq:disc_polynomial}, we next discuss how to derive the studied characteristics of the TPs.
The discussion is divided into several parts, corresponding to distinct collections of $\mcP\mcT$-symmetric MPGs.
First, in \cref{Sec:TP-type-A}, we consider the MPGs where the restriction to leading-order terms reveals the absence of non-trivial roots of the discriminant, resulting in type-$\mathsf{A}$ TPs.
In the remaining \cref{Sec:TP-type-B-with-mirror} (with mirror symmetry) and \cref{Sec:TP-type-B-wo-mirror} (without mirror symmetry), the discriminant in \cref{eq:disc_polynomial} turns out to be \emph{quasi-homogeneous}, i.e., there is a scaling factor $\mu$ for which all the monomials are of the same order.
This implies that \emph{all} the terms in the discriminant that are obtained from the leading-order \kdotp{} Hamiltonian are themselves leading. 
Here, we are led to develop additional arguments, which unambiguously reveal that the TPs in these MPGs are always of type $\mathsf{B}$.
Due to the extensiveness of the underlying algebraic manipulations, the detailed analysis for all the MPGs and all combinations of ICRs is made available as part of the supplementary data and code~\cite{Lenggenhager:2022:TPClassif:SDC}, with only a few representative calculations presented explicitly in \cref{App:Classif:PT}.

\subsubsection{MPGs \texorpdfstring{$4/m'\!(mm)$}{4/m'(mm)} and class I of MPGs \texorpdfstring{$6/m'\!(mm)$}{6/m'(mm)}}\label{Sec:TP-type-A}

We begin with MPGs $4/m'$, $4/m'mm$, and with class-I Hamiltonians of MPGs $6/m'$ and $6/m'mm$, when the restriction to leading terms results in a significant simplification.
Namely, $\Delta_{\vec{a},\theta}(k,k_z)/k^r$ is a quadratic polynomial in $k$ and $k_z$ with non-negative coefficients [as exemplified by \cref{App:eq:leading-C6vI} for class~I of $6/m'mm$].
We are then able to prove that for generic $\vec{a}$ and all $\theta$ there are no real roots other than the one at $k=0$ the central NL).
This implies that there are no NL arcs attached to the TP, such that the TP is classified as type $\mathsf{A}$ [\cref{fig:TPTypes}(a)].
The relevant point groups and corepresentations for which this situation arises are indicated in \cref{tab:Classification_PT}.

Note that while there are no NL arcs \emph{attached} to the TP, there is a possibility (analogous to the corresponding cases without $\mcP\mcT$ symmetry) of nexus points occurring \emph{near} the TP, i.e., additional NL arcs coalescing at the central NL away from the TP [\cref{fig:TPTypes}(b)].
To obtain those NL arcs, we depart from the same leading-order \kdotp{} Hamiltonian, but keep all terms in \cref{eq:disc_polynomial} [i.e., $M$ and not only $L(M)$].
We find~\cite{Lenggenhager:2022:TPClassif:SDC} that in the four-fold symmetric case there are two nexus points with four NL arcs each, $n_a^\mathrm{nexus}=4$, and in the six-fold symmetric case, there is one nexus point with 12 NL arcs, $n_a^\mathrm{nexus}=12$.
In \cref{App:Classif:PT:nls:nexus_NLarcs}, we illustrate this using the example $4/m'mm$. 
It is further manifest from the analytic solutions~\cite{Lenggenhager:2022:TPClassif:SDC} that a single real-valued parameter needs to be tuned to collide the nexus point with the TP, leading to codimension $1$ (in contrast to codimension $2$ in the analogous MPGs without $\mcP\mcT$ symmetry).
In analogy with the MPG $4mm$ (discussed in \cref{Sec:Classif:mirror_no_PT:C4vC6v-part2}), such fine-tuning collides the TP simultaneously with \emph{two} nexus points [cf.~\cref{fig:NLStructure:4mm}(c)]; these two nexus points could be either of the same or of opposite color.
In this respect, we illustrate in \cref{Sec:TPMaterials:examples} [\cref{fig:TPmat:ZrO_2}] a particular example of a material (\ce{ZrO}~\cite{Zhang:2017b}, the relevant HSL is $\Delta=\Gamma X$ with little co-group $4/m'mm$) for which the nexus point appears to closely coincide with a type-$\mathsf{A}$ TP due to such an accidental fine-tuning of the relevant model parameter, see \cref{Sec:TPMaterials:examples}.

\subsubsection{MPGs with mirror symmetries: \texorpdfstring{$\bar{3}'m$}{-3'm} and class II of \texorpdfstring{$6/m'mm$}{6/m'mm}}\label{Sec:TP-type-B-with-mirror}

For $\bar{3}'m$ and class-II $6/m'mm$ the discriminant $\Delta_{\vec{a},\theta}(k,k_z)/k^r$ of the leading-order Hamiltonian turns out to be quasi-homogeneous with $\mu=1$ ($\mu=2$) for $\bar{3'}m$ (class-II $6/m'mm$).
We find $\Delta_{a,\theta}(k,k_z)/k^r$ to be a fourth-order polynomial in $k_z$, such that the nature of the roots~\cite{Rees:1922} can be determined analytically.
More precisely, we determine~\cite{Lenggenhager:2022:TPClassif:SDC} the conditions on the parameters $\vec{a},\theta$ for the presence of a certain number of real roots using \Mathematica{} (see \cref{App:Classif:PT:nls:typeB} for a detailed discussion of how we formulate these conditions).

\begin{figure}[t]
    \centering
    \includegraphics{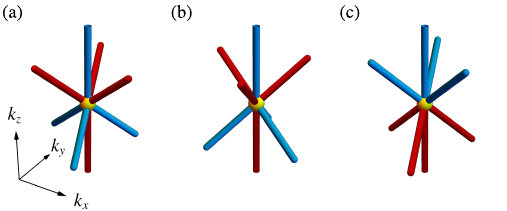}
    \caption{
        Nodal-line (NL) structure near a type-$\mathsf{B}_l$ triple point on a high-symmetry line with little co-group $\bar{3}'m$ (or $3m$).
        NLs are shown in red (blue) if they are in the lower (upper) band gap of the three-band model and the TP is indicated in yellow.
        The NL arc always changes gap (and therefore its color) when passing through the TP, but several different arrangements are possible as illustrated.
    }
    \label{fig:NLStructure:-3'm}
\end{figure}

The described analysis of the quartic polynomial reveals that there is either a single real root (indicating a NL) or no real root (indicating the absence of NLs away from the rotation axis), depending on the values of the parameters $\vec{a}$ and $\theta$.
In particular, the requirement of a real root restricts $\theta$ to discrete values, $\tfrac{\pi}{4},\tfrac{7\pi}{12}\,(\mathrm{mod}\,\tfrac{2\pi}{3})$ for $\bar{3}'m$ [$\tfrac{\pi}{12},\tfrac{\pi}{4}\,(\mathrm{mod}\,\tfrac{\pi}{3})$ for class-II $6/m'mm$], which correspond exactly to the mirror planes [cf.~\cref{fig:mirror-planes} (d) and (c), respectively], in which case generic values of $\vec{a}$ yield a root of the discriminant that continuously connects to $\vec{k}=\vec{0}$, meaning that the TPs are of type $\mathsf{B}$.
Additionally, the value of the scaling factor $\mu=1$ ($\mu=2$) fixes the attachment of the NL arcs to be linear (quadratic).
We therefore conclude that $\bar{3}'m$ gives type-$\mathsf{B}_l$ TPs with $n_a^\mathrm{nexus}=6$ [\cref{fig:TPTypes}(c)], while class-II $6/m'mm$ gives type-$\mathsf{B}_q$ TPs with $n_a^\mathrm{nexus}=12$ [\cref{fig:TPTypes}(d)].
Although the presence, number and scaling $\mu$ of the NL arcs do not depend on the parameters $\vec{a}$ (up to fine-tuning) of the model, the precise NL structure does.
We illustrate three examples of the possible variations for the MPG $\bar{3}'m$ in \cref{fig:NLStructure:-3'm}.

\subsubsection{MPGs without mirror symmetries: \texorpdfstring{$\bar{3}'$}{-3'} and class II of \texorpdfstring{$6/m'$}{6/m'}}\label{Sec:TP-type-B-wo-mirror}

We finally discuss the MPG $\bar{3}'$ and class-II Hamiltonians of the MPG $6/m'$.
In these cases, $\Delta_{\vec{a},\theta}(k,k_z)/k^r$ is again quasi-homogeneous with $\mu=1$ ($\mu=2$ for $6/m'$), and the discriminant is a fourth-order polynomial in $k_z$; however, in contrast to \cref{Sec:TP-type-B-with-mirror}, the presently considered MPGs have no mirror symmetries, leading to an increased number of terms in the leading-order \kdotp{} Hamiltonian.
While the procedure described above is in principle still applicable, we find that the conditions for real roots of $\Delta_{\vec{a},\theta}(k,k_z)/k^r$ become too complicated for \Mathematica{} to handle and simplify.
However, we argue below that the (qualitative) NL structure near the TP is the same as in the corresponding cases with mirror symmetry, up to a rotation of the coordinates that is determined by the model parameters.

More specifically, we find that the leading-order \kdotp{} model for the MPGs without mirror symmetry has two additional terms compared to the corresponding \kdotp{} models with mirror symmetry (discussed in \cref{Sec:TP-type-B-with-mirror}).
The particular structure of these additional terms is rather fortunate: they can both be generated from terms already present in the mirror-symmetric \kdotp{} expansion via an $\mathsf{SO}(2)$ rotation of $(k_x,k_y)$-coordinates.
In particular, one can always find a suitable rotation of the coordinates that removes one of these additional terms (the rotations needed to remove either of the two additional terms are generically \emph{different}, making it impossible to rotate away both of these new terms simultaneously).
Therefore, a suitable rotation of the momentum coordinates leaves us with only one extra term compared to the \kdotp{} models with mirror symmetry.
Additionally, while the one remaining new term cannot be removed on the level of the leading-order \kdotp{} Hamiltonian, it can be removed on the level of the characteristic polynomial by another momentum space rotation and reparametrization.
In \cref{App:Classif:PT:nls:mirror_absence} we discuss this reduction using a concrete example.

After performing both of these coordinate transformations and the reparametrization, we reduce the characteristic polynomials of $\bar{3}'$ and for class-II $6/m'$ to those of $\bar{3}'m$ and of class-II $6/m'mm$, respectively.
Therefore, the TP characterization derived in \cref{Sec:TP-type-B-with-mirror} translates directly to the MPG without mirror symmetry. 
The results are summarized in \cref{tab:Classification_PT}.

\subsection{Winding number of the 2D ICRs}\label{Sec:Classif:TP:winding}

For two-band spinless $\mcP\mcT$-symmetric systems, the winding number of a 2D ICR, i.e., computed on a contour around the corresponding NL, is an integer topological invariant~\cite{Bzdusek:2017}. 
This integer invariant is \emph{delicate}~\cite{Nelson:2021}, in the sense that it ceases to be defined in models with three or more bands.
Nevertheless, we are interested in the value of this integer invariant for the 2D ICRs involved in the TPs listed in \cref{tab:Classification_PT}, because it determines both the $\ztwo$-quantized Berry phase~\cite{Fang:2015a} as well as the non-Abelian topological invariant~\cite{Wu:2019} carried by the central nodal line in models with three or more bands~\cite{Tiwari:2020}, including the presently studied \kdotp{} models of TPs~\cite{Lenggenhager:2021:MBNLs}.

To determine the winding number of the 2D ICRs, we construct additional minimal two-band \kdotp{} models.
Because the $k_z$-dependence is not directly relevant to the winding number and is not restricted by symmetry, it can be absorbed in the coefficients of the \kdotp{} expansion.
This results in traceless models of the form
\begin{equation}
    \Ham_\vec{a}(\vec{k}) = \vec{h}_\vec{a}(k_x,k_y)\cdot\vb*{\sigma}
    \label{eq:kdotp2D}
\end{equation}
with $\vb*{\sigma}=(\sigma_x,\sigma_z)$ the real Pauli matrices, $\vec{h}_\vec{a}(k_x,k_y)$ a real two-component vector and $\vec{a}$ a (minimal) list of real parameters. 
By construction $\vec{h}_\vec{a}(0,0)=0$ for all $\vec{a}$, which corresponds to the central NL along the HSL.
The explicit models we used are provided in the supplementary data and code~\cite{Lenggenhager:2022:TPClassif:SDC}.

The winding number $w_\mathrm{2D}$ of $\vec{h}_\vec{a}(\vec{k})$ along a tight contour around that NL (suppressing the dependence on the parameter~$\vec{a}$) is given by
\begin{equation}
    w_\mathrm{2D} = \frac{1}{2\pi}\oint_C\dd{\vec{k}}\cdot\left(\frac{h_x(\vec{k})\grad_\vec{k}h_z(\vec{k})-h_z(\vec{k})\grad_\vec{k}h_x(\vec{k})}{h(\vec{k})^2}\right).
    \label{eq:w2D}
\end{equation}
This calculation can easily be completed analytically by simply plugging $\vec{h}_\vec{a}(k_x,k_y)$ in \cref{eq:w2D} for all cases except for $4/m'$; in the latter case, going to polar coordinates and deforming the contour appropriately is necessary to complete the integration. 
The integrations are performed in \Mathematica{} and we provide the corresponding notebook in the supplementary data and code~\cite{Lenggenhager:2022:TPClassif:SDC}; an illustrative calculation for ICR $E_1$ of $6/m'mm$ is shown in \cref{App:Classif:PT:nls:winding_number}.
The results are shown in \cref{tab:Classification_PT}.
Note that $\abs{w_\mathrm{2D}}$ depends only on the MPG and not on the particular choice of 2D ICR; furthermore, we observe that $\abs{w_\mathrm{2D}} = \mu$.

The fact that three-fold symmetric MPGs give rise to \mbox{type-$\mathsf{B}_l$} TPs with $\abs{w_\mathrm{2D}}=1$ can be intuitively understood (at least in the presence of mirror symmetry) based on Berry phases as follows.
Recall that the $\mcP\mcT$ symmetry quantizes the Berry phase $\phi_\textrm{B}$ on any closed contour (along which a specific energy gap of the spectrum is preserved) to $0$ vs.\ $\pi$.
In particular, this holds for contours around nodal lines, such that we can assign the quantized Berry phase to each NL (in analogy with assigning the integer winding number to NLs in two-band models).
Nodal lines that are protected by either $\mcP\mcT$ or mirror symmetry $m_v$ generally carry Berry phase $\phi_\textrm{B}=\pi$~\cite{Fang:2015}.
At a type-$\mathsf{B}$ TP, $\frac{1}{2}n_a^\mathrm{nexus}$ NL arcs, each carrying Berry phase $\pi$, annihilate together with the central NL (here, we count NLs in one energy gap), therefore the Berry phase of the central NL must be $\phi_\textrm{B}=\frac{1}{2}n_a^\mathrm{nexus}\pi\;(\textrm{mod}\,2\pi$), which results in $\phi_\textrm{B}=\pi$ for TPs in $C_3$-symmetric MPGs and $\phi_\textrm{B}=0$ in $C_6$-symmetric MPGs.

\begin{table*}
    \centering
    \caption{
        Examples of triple-point (TP) materials.
        For each material (rows), the following information is listed (in consecutive columns): the space group (SG), the number of its entry in the Inorganic Crystal Structure Database (ICSD) if applicable, the high-symmetry line (HSL) on which the TP of interest lies, the little co-group $\LcG{\vec{k}}$ of that HSL given in Hermann-Mauguin  notation~\cite{Bradley:1972}, the irreducible corepresentations (ICRs) of the bands involved in the TP formation, the energy of the TP $E_\mathrm{TP}$ relative to the Fermi energy, the type, the number $n_a^\mathrm{nexus}$ of nodal-line arcs attached to a \emph{nexus point} (either coinciding with the TP or not) if there is one, the scaling of those nodal-line arcs $\mu$ ($k_z\propto (k_x^2+k_y^2)^{\mu/2}$), and finally a reference to the figure containing the relevant data.
        The identified type, and values of $n_a^\mathrm{nexus}$, $\mu$ for all materials match our theoretical predictions in \cref{tab:Classification}.
    }
    \begin{ruledtabular}
\begin{tabular}{cCcCCCCCCCl}
	                  Material                   &       \text{SG}        &       ICSD number       &               \text{HSL}               &       \LcG{\vec{k}}       &             \text{ICRs}              &  E_\mathrm{TP}  & \text{Type}  & n_a^\mathrm{nexus} & \mu & Figure             \\ \hline\addlinespace
	                 \ce{SiO2}                   &           82           &          75647          &          \Lambda\, (\Gamma Z)          &         \bar{4}'          &                (BB;A)                & -0.62\nunit{eV} &  \mathsf{A}  &        \td         & \td & \matref{SiO2}      \\
	                 \ce{Li4HN}                  &           88           &         409633          &          \Lambda\, (\Gamma M)          &           4/m'            &  (\irrep{E}{}{2}\!\irrep{E}{}{1};B)  & -0.92\nunit{eV} &  \mathsf{A}  &         4          &  2  & \matref{Li4HN}     \\
	                 \ce{CaNaP}                  &          107           &           \td           &          \Lambda\, (\Gamma M)          &            4mm            &               (E;A_1)                &  4.1\nunit{eV}  &  \mathsf{A}  &         4          &  2  & \matref{CaNaP}     \\
	                \ce{Na2LiN}                  &          129           &          92309          &          \Lambda\, (\Gamma Z)          &          4/m'mm           &               (E;A_1)                & -1.0\nunit{eV}  &  \mathsf{A}  &        \td         & \td & \matref{Na2LiN}    \\
	                 \ce{B2CN}                   &          156           &         183791          &          \Delta\, (\Gamma A)           &            3m             &               (E;A_1)                &  9.3\nunit{eV}  & \mathsf{B}_l &         6          &  1  & \matref{B2CN}      \\
	                \ce{MgH2O2}                  &          164           &          34401          &               P\, (K H)                &         \bar{3}'          & (\irrep{E}{}{2}\!\irrep{E}{}{1};A_1) & -1.4\nunit{eV}  & \mathsf{B}_l &         6          &  1  & \matref{MgH2O2}    \\
	                   \ce{P}                    &          166           &          53301          &          \Lambda\, (\Gamma Z)          &         \bar{3}'m         &               (E;A_1)                &  1.5\nunit{eV}  & \mathsf{B}_l &         6          &  1  & \matref{P}         \\
	               \ce{Li2Co12P7}                &          174           &         656419          &          \Delta\, (\Gamma A)           &         \bar{6}'          & (\irrep{E}{}{2}\!\irrep{E}{}{1};A_1) & -0.17\nunit{eV} &  \mathsf{A}  &        \td         & \td & \matref{Li2Co12P7} \\ \addlinespace
	         \multirow{2}{*}{\ce{C3N4}}          & \multirow{2}{*}{$176$} & \multirow{2}{*}{246661} & \multirow{2}{*}{$\Delta\, (\Gamma A)$} &  \multirow{2}{*}{$6/m'$}  & (\irrep{E}{1}{2}\!\irrep{E}{1}{1};A) & -9.3\nunit{eV}  & \mathsf{B}_q &         12         &  2  & \matref{C3N4_6_1}  \\
	                                             &                        &                         &                                        &                           & (\irrep{E}{2}{2}\!\irrep{E}{2}{1};A) & -2.6\nunit{eV}  &  \mathsf{A}  &        \td         & \td & \matref{C3N4_6_2}  \\ \addlinespace
	         \multirow{2}{*}{\ce{AlN}}           & \multirow{2}{*}{$186$} & \multirow{2}{*}{31169}  & \multirow{2}{*}{$\Delta\, (\Gamma A)$} &  \multirow{2}{*}{$6mm$}   &              (E_1;A_1)               & -0.28\nunit{eV} &  \mathsf{A}  &         12         &  2  & \matref{AlN_1}     \\
	                                             &                        &                         &                                        &                           &              (E_2;A_1)               & -0.85\nunit{eV} & \mathsf{B}_q &         12         &  2  & \matref{AlN_2}     \\ \addlinespace
	                 \ce{Li4N}                   &          187           &         675123          &          \Delta\, (\Gamma A)           &        \bar{6}'m2'        &               (E;A_1)                & -0.68\nunit{eV} &  \mathsf{A}  &         6          &  2  & \matref{Li4N}      \\
	                 \ce{Na2O}                   &          189           &           \td           &          \Delta\, (\Gamma A)           &        \bar{6}'m2'        &               (E;A_1)                & -0.26\nunit{eV} &  \mathsf{A}  &         6          &  2  & \matref{Na2O}      \\
	        \ce{Li2NaN}~\cite{Jin:2019}          &          191           &          92308          &          \Delta\, (\Gamma A)           &          6/m'mm           &              (E_1;A_1)               & -48\nunit{meV}  &  \mathsf{A}  &        \td         & \td & \matref{Li2NaN}    \\ \addlinespace
	\multirow{2}{*}{\ce{TiB2}~\cite{Zhang:2017}} & \multirow{2}{*}{$191$} & \multirow{2}{*}{30330}  & \multirow{2}{*}{$\Delta\, (\Gamma A)$} & \multirow{2}{*}{$6/m'mm$} &              (E_1;A_1)               & 0.57\nunit{eV}  &  \mathsf{A}  &        \td         & \td & \matref{TiB2_1}    \\
	                                             &                        &                         &                                        &                           &              (E_2;A_1)               &  1.3\nunit{eV}  & \mathsf{B}_q &         12         &  2  & \matref{TiB2_2}    \\ \addlinespace
	 \multirow{2}{*}{\ce{Na3N}~\cite{Jin:2020}}  & \multirow{2}{*}{$194$} &  \multirow{2}{*}{\td}   & \multirow{2}{*}{$\Delta\, (\Gamma A)$} & \multirow{2}{*}{$6/m'mm$} &              (E_2;A_1)               & -93\nunit{meV}  & \mathsf{B}_q &         12         &  2  & \matref{Na3N_1}    \\
	                                             &                        &                         &                                        &                           &              (E_1;A_1)               & -38\nunit{meV}  &  \mathsf{A}  &        \td         & \td & \matref{Na3N_2}    \\ \addlinespace
	                 \ce{C3N4}                   &          215           &          83264          &          \Delta\, (\Gamma X)           &        \bar{4}'2'm        &             (B_2B_1;A_1)             & -6.5\nunit{eV}  &  \mathsf{A}  &        \td         & \td & \matref{C3N4_4}    \\ \addlinespace
	\multirow{2}{*}{\ce{ZrO}~\cite{Zhang:2017b}} & \multirow{2}{*}{$225$} & \multirow{2}{*}{76019}  & \multirow{2}{*}{$\Delta\, (\Gamma X)$} & \multirow{2}{*}{$4/m'mm$} &               (E;A_1)                & 0.12\nunit{eV}  &  \mathsf{A}  &         4          &  2  & \matref{ZrO_1}     \\
	                                             &                        &                         &                                        &                           &               (E;B_1)                &  2.0\nunit{eV}  &  \mathsf{A}  &         4          &  2  & \matref{ZrO_2}
\end{tabular}
    \end{ruledtabular}
    \label{tab:TPmaterials}
\end{table*}

Finally, we consider the implications of the winding number for the non-Abelian charge.
Note that the integer winding number is only defined for two-band blocks that are separated from the remaining bands by energy gaps; in particular, for the full three-band model exhibiting the TP, the winding number of the central NL \emph{cannot} be defined anymore.
In fact, the central NL is transferred from one gap to another at the TP, such that on the two sides of the TP the integer winding number would have to be computed with respect to different energy gaps.
In contrast, the non-Abelian band invariant~\cite{Wu:2019,Bouhon:2020} is sensitive to closing \emph{either} of the two energy gaps of the three-band model.
Crucially, the non-Abelian invariant  $\mathfrak{q}$ preserves partial information contained in the integer winding number when additional (trivial) bands are added; more concretely we have the reduction~\cite{Tiwari:2020} 
\begin{equation}
    \begin{split} 
    w_\mathrm{2D} &= 0\mod 4\quad \Rightarrow\quad \mathfrak{q} = 0 \qquad\textrm{and} \\ 
    w_\mathrm{2D} &= 2\mod 4\quad \Rightarrow\quad \mathfrak{q} = -1.
    \end{split} 
    \label{eq:quaternion-values}
\end{equation} 
We therefore conclude that TPs of type $\mathsf{A}$ and $\mathsf{B}_q$ are associated with a non-Abelian charge $\mathfrak{q}=-1$ computed on a contour encircling the central NL. 

The two values of the non-Abelian invariant in \cref{eq:quaternion-values} cannot be distinguished by the Berry phases on the individual bands~\cite{Wu:2019}.
However, the value $\mathfrak{q}=-1$ poses an obstruction to remove the enclosed band degeneracy as long as the $\mcP\mcT$ symmetry is present.
This aspect has been utilized in our recent work~\cite{Lenggenhager:2021:MBNLs} to uncover a relation between certain TPs and multi-band nodal links, and is considered in our parallel publication~\cite{Lenggenhager:2021:TPHOT} to reveal higher-order topology associated with pairs of TPs with a semimetallic band dispersion.

\section{Material Examples}\label{Sec:Materials}

The classification of TPs derived above allows us to predict, based on symmetry properties, the possibility of stable TPs (including their type) in real materials.
In \cref{tab:TPmaterials} we list several compounds as representative triple-point materials with weak SOC, which are subject to our derived classification. 
Some of these compounds have been previously described~\cite{Mikitik:2006,Zhang:2017b,Jin:2019,Zhang:2017,Jin:2020,Lenggenhager:2021:MBNLs}, while others have, to the best of our knowledge, not been reported as TP materials before.

For each listed material, we analyze selected TPs and verify their type against the predictions we made in the previous sections.
We provide access to all the first-principles data in the supplementary data and code~\cite{Lenggenhager:2022:TPClassif:SDC} and present relevant figures for all compounds in \cref{App:Materials}.
In \cref{Sec:TPMaterials:examples}, we discuss a few selected examples to illustrate our procedure and highlight some interesting aspects.
The results on the TP types are summarized in the last three columns of \cref{tab:TPmaterials} and agree with the classification in \cref{tab:Classification}.

\begin{figure*}
    \centering
    \includegraphics{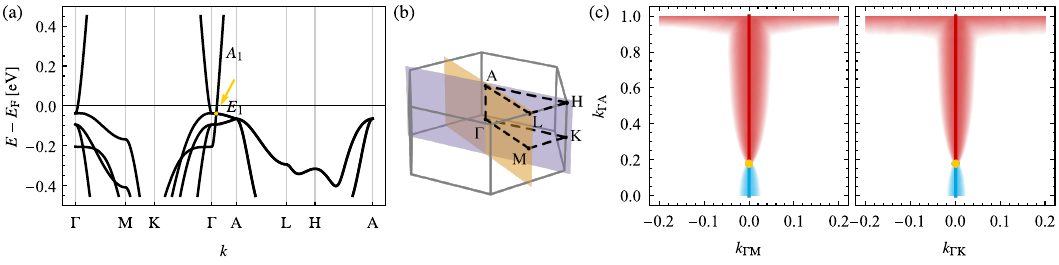}
    \caption{
        Nodal line structure of \ce{Na3N} near the triple point (TP) at $E_\mathrm{TP}=-38\nunit{meV}$ lying on the $\Delta$ line with little co-group $6/m'mm$ (cf.~\cref{tab:TPmaterials}).
        (a) Band structure along lines of symmetry. The TP is indicated by a yellow dot and arrow, and the bands forming the TP are labelled by their irreducible corepresentations.
        (b) Brillouin zone (boundary in gray) with points and lines of symmetry (black dashed lines) and the two inequivalent mirror planes (orange and purple planes). Note that there are additional symmetry related mirror planes for each of those two.
        (c) Size of the lower (red) and upper (blue) gap in the two mirror planes shown in panel (b) encoded by the intensity of the color (higher color saturation implies smaller energy gap between the corresponding pair of bands). The gap is only plotted up to a cutoff of $0.01\nunit{eV}$ such that white color indicates a gap larger than that.
        The TP (yellow) and the central nodal line are emphasized by appropriately colored overlays and the data shows that the TP is type $\mathsf{A}$.
        Furthermore, there are no nexus points present, even though they could be stabilized.
    }
    \label{fig:TPmat:Na3N_2}
\end{figure*}

\begin{figure*}
    \centering
    \includegraphics{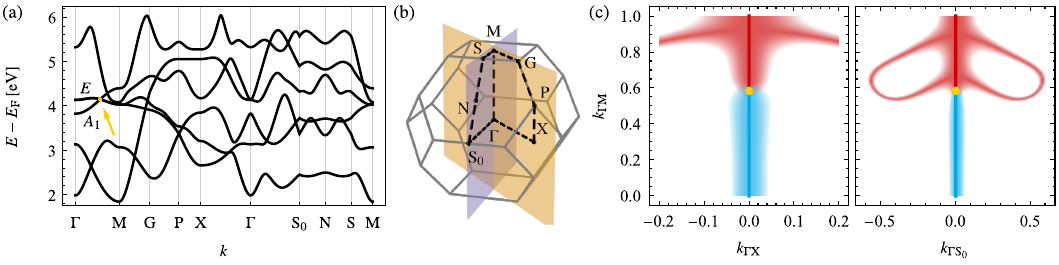}
    \caption{
        Nodal line structure of \ce{CaNaP} near the triple point (TP) on the $\Lambda$ line with little co-group $4mm$.
        The organization of the panels is in one-to-one correspondence with \cref{fig:TPmat:Na3N_2} and the cutoff on the gap size was chosen to be $0.05\nunit{eV}$.
        No nodal lines are attached to the TP, such that we conclude it to be type $\mathsf{A}$.
        Note the occurrence of a nexus of four (due to the four-fold rotational symmetry) red nodal lines near the TP, from which we can deduce that $n_a^\mathrm{nexus}=4$ and $\mu=2$.
        Our theoretical analysis predicts that two parameters need to be tuned to collide the nexus point with the TP.
    }
    \label{fig:TPmat:CaNaP}
\end{figure*}

\subsection{Methods}\label{Sec:TPMaterials:methods}

Based on the little groups that can stabilize TPs (listed in \cref{tab:MPGs}) and the program \textsc{Mkvec} on the Bilbao crystallographic server (BCS)~\cite{BCS, Aroyo:2006a, Aroyo:2006b, Elcoro:2017, Elcoro:2021, Xu:2020}, one can easily scan all magnetic space groups to identify those that support TPs (i.e., both 1D and 2D ICRs) on high-symmetry lines.
This search (for type-II magnetic space groups, i.e., those that exhibit no magnetic order) has been very recently performed independently by Feng et al. in Ref.~\onlinecite{Feng:2021} and their list of admissible little co-groups matches ours.
The relevant space groups and high-symmetry lines are listed in Table II in Ref.~\onlinecite{Feng:2021}.
Here, we also restrict to finding representative materials in type-II magnetic space groups.
Exhaustive lists of (magnetic) space groups of any type and high-symmetry lines supporting various quasiparticles, including TPs, have recently been published in Refs.~\onlinecite{Yu:2021,Liu:2022,Zhang:2021}.

For each of the type-II magnetic space groups listed in Table II in Ref.~\onlinecite{Feng:2021}, we performed a search of compounds with light elements (from the first three rows of the periodic system) on the Topological Materials Database~\cite{TopoMatDB, Vergniory:2019, Vergniory:2021}.
By looking at the irreducible representation on high-symmetry points (for the case without spin-orbit coupling) and using the program \textsc{Mcomprel}~\cite{BCS, Elcoro:2021, Xu:2020} on the BCS, we inferred the ICRs and their dimension along the relevant HSL and identified crossings of 2D with 1D ICRs, i.e., TPs.
This resulted in a list of several hundered candidate materials from which we selected the most promising ones [our arbitrary criteria adopted a tradeoff between TPs being close to the Fermi level (with the distance counted in number of bands rather than in energy), a small number of additional degeneracies, and well separated nodal lines] to illustrate and verify our classification.

We analyzed the selected materials as well as the materials from Refs.~\onlinecite{Mikitik:2006,Zhang:2017b,Jin:2019,Zhang:2017,Jin:2020,Lenggenhager:2021:MBNLs} in detail by performing first-principle calculations ourselves as detailed below.
For that, DFT calculations with the projected augmented wave (PAW) method are implemented in the Vienna ab initio simulation package (VASP)~\cite{Kresse:1996,Kresse:1999} with generalized gradient approximation (GGA) using PBE functional pseudopotentials~\cite{Perdew:1996}.
A $6\times6\times6$ uniform mesh for bulk $k$-space provided converged total energies.
For the 2D planes a $140\times140$ ($100\times100$ for \ce{Si2O} and \ce{Li2Co12P_7}) mesh is used to calculate the band gap.
Using plane-wave-based wavefunctions and space group operators generated by VASP, we calculate the traces of matrix representations to get the irreducible representations of the energy bands at high-symmetry points in the first Brillouin zone with the help of \textsc{IrRep}~\cite{Iraola:2022}.
Using compatibility relations~\cite{Elcoro:2017} we then deduce the irreducible \mbox{(co-)representations} of the lines of symmetry shown in the sixth row of \cref{tab:TPmaterials}.

\begin{figure*}
    \centering
    \includegraphics{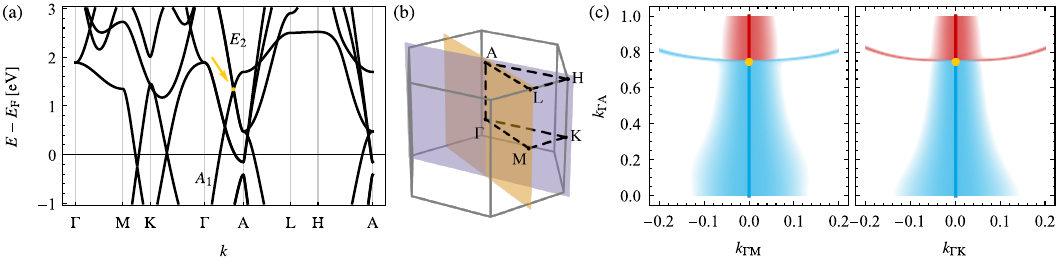}
    \caption{
        Nodal line structure of \ce{TiB2} near the triple point (TP) at $E_\mathrm{TP}=1.3\nunit{eV}$ lying on the $\Delta$ line with little co-group $6/m'mm$.
        The organization of the panels is in one-to-one correspondence with \cref{fig:TPmat:Na3N_2} with a cutoff of $0.05\nunit{eV}$ on the gap size that is shown in panel (c).
        Note the NL arcs attaching quadratically ($k_z\propto k_x^2+k_y^2$) to the TP, i.e., the two nexus points coincide with the TP, implying that the TP is type $\mathsf{B}_q$.
        Due to the rotational symmetry, there are six nodal-line arcs in each of the two gaps (shown in red and blue, respectively), implying $n_a^\mathrm{nexus}=12$.
    }
    \label{fig:TPmat:TiB2_2}
\end{figure*}

\begin{figure*}
    \centering
    \includegraphics{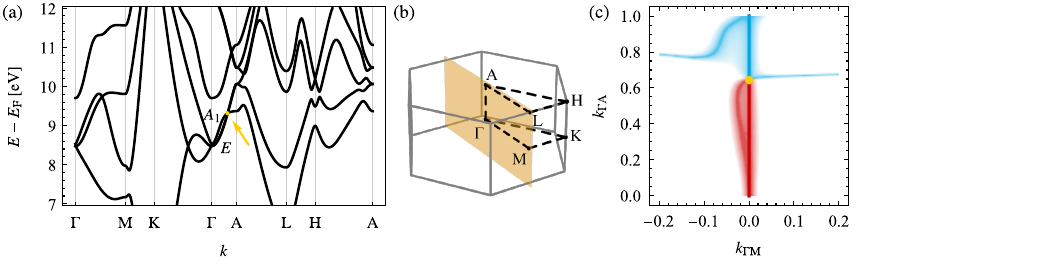}
    \caption{
        Nodal line structure of \ce{B2CN} near the triple point (TP) on the $\Delta$ line with little co-group $3m$.
        The organization of the panels is in one-to-one correspondence with \cref{fig:TPmat:Na3N_2} (except for the fact that only one inequivalent high-symmetry plane is now present) and the cutoff on the gap size was chosen to be $0.02\nunit{eV}$.
        Note the NL arcs attaching to the TP, i.e., the two nexus points coincide with the TP and that the NLs scale linearly, $k_z\propto \sqrt{k_x^2+k_y^2}$, implying a type-$\mathsf{B}_l$ TP. Due to the three-fold rotational symmetry, there are $n_a^\mathrm{nexus}=6$ nodal-line arcs in total.
    }
    \label{fig:TPmat:B2CN}
\end{figure*}

\subsection{Examples}\label{Sec:TPMaterials:examples}

Here we discuss some examples of TP materials in order to illustrate how we used first-principle calculations to verify the predictions of our classification for the NL structure near the TPs.
We start with the following four materials: \ce{Na3N} hosting a type-$\mathsf{A}$ TP without nexus points, \ce{CaNaP} hosting a type-$\mathsf{A}$ TP \emph{with} nexus points, \ce{TiB2} hosting a type-$\mathsf{B}_q$ TP and \ce{B2CN} hosting a type-$\mathsf{B}_l$ TP.
The first-principles data for these four compounds are shown in \cref{fig:TPmat:Na3N_2,fig:TPmat:CaNaP,fig:TPmat:TiB2_2,fig:TPmat:B2CN}, in the given order.
In each figure, the band structure on high-symmetry lines is shown in panel (a); therein, the TP is indicated by a yellow dot and arrow, and the bands involved in the TP formation are labelled by their irreducible corepresentations.
The corresponding Brillouin zone is illustrated in panel (b).

To detect NLs, we perform additional DFT calculations in the appropriate planes containing the nodal lines close to the TP.
For compounds with mirror symmetries, these are the mirror planes;
for the other compounds we first study slices of fixed $k_z$ to detect any NLs close to the central nodal line, and if there are any, we determine the plane in which they lie in the vicinity of the central NL.
In panel (c) we then plot the magnitude of the two gaps between the three bands involved in the TP formation (larger color saturation implies smaller energy gap), with the lower (upper) gap data shown in red (blue) color, as usual.
The TP is located where the central NL changes color (and sum of the two gaps is minimal).
Choosing a suitable cutoff for the gap size, we can also easily recognize the additional NLs and infer the number of NLs attached to the TP as well as their momentum space behaviour $\mu$, which directly determines the type of the TP.

\begin{figure*}[t]
    \centering
    \includegraphics{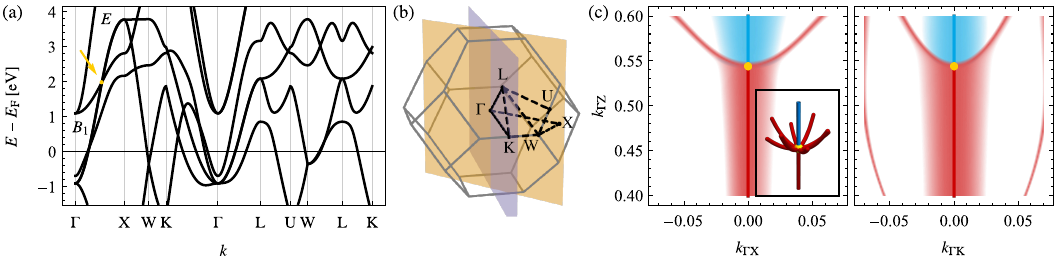}
    \caption{
        Nodal line structure of \ce{ZrO} near the triple point (TP) at $E_\mathrm{TP}=2.0\nunit{eV}$ lying on the $\Delta$ line with little co-group $4/m'mm$.
        The organization of the panels is in one-to-one correspondence with \cref{fig:TPmat:Na3N_2} and the cutoff on the gap size was chosen to be $0.02\nunit{eV}$.
        The inset of panel (c) shows the nodal-line (NL) structure of the minimal \kdotp{} model given in \cref{App:Classif:PT:nls:nexus_NLarcs} with parameters tuned to qualitatively reproduce the situation in \ce{ZrO}. 
        We observe that there are two red nexus points, each with four NL arcs, coinciding with the TP, such that $\sign(a_3 a_4)=+1$ in the \kdotp{} model in \cref{App:eq:4/m'mm-nexi}  (in contrast to another TP in this compound, illustrated in \cref{fig:TPmat:ZrO_1}).
        Therefore, there are in total eight red NLs attached to the TP.
        This is accidental (see text and \cref{fig:TPmat:ZrO_2_strain}) and we therefore classify the TP despite the NL arcs as type $\mathsf{A}$ and consider the two nexus points (separately) with $n_a^\mathrm{nexus}=4$ and $\mu=2$.
    }
    \label{fig:TPmat:ZrO_2}
\end{figure*}

We briefly discuss one peculiar TP found in \ce{ZrO}.
According to \cref{tab:TPmaterials,tab:Classification}, \ce{ZrO} supports type-$\mathsf{A}$ TPs on the HSL $\Delta$ (little co-group $4/m'mm$).
However, in \cref{fig:TPmat:ZrO_2}, we can clearly identify eight additional NLs attached to the TP: for each of the two inequivalent mirror planes shown in \cref{fig:TPmat:ZrO_2}(c) there are two NL arcs and another two in the symmetry related mirror plane.
Note, however, that all those NL arcs are red (located in the lower gap) meaning that there are two red nexus points at the TP.
This is incompatible with type-$\mathsf{B}$ TPs where the TP coincides with both a red and a blue nexus.
We therefore conclude that the TP in \ce{ZrO} is type $\mathsf{A}$ but fine-tuned such that the two red nexus points coincide with it.
Interestingly, such a configuration arises by fine-tuning a single parameter, as we discussed in \cref{Sec:TP-type-A} above (see also the explicit \kdotp{} model for the little group $4/m'mm$ in \cref{App:Classif:PT:nls:nexus_NLarcs}).
The fine-tuned NL structure of the minimal \kdotp{} model for a type-$\mathsf{A}$ TP on a HSL with little group $4/m'mm$ is shown in the inset in \cref{fig:TPmat:ZrO_2}(c).
To verify that the NL structure is the consequence of fine-tuning, we include a perturbation in the form of $5\%$ uniaxial tensile strain, which preserves the little co-group of the $\Delta$ line.
This splits the two nexus points and the TP demonstrating that the \emph{stable} TP is type $\mathsf{A}$ and that we indeed have two separate nexus points with $n_a^\mathrm{nexus}=4$, cf.~\cref{fig:TPmat:ZrO_2_strain} in the Appendix.

\section{Conclusion}\label{Sec:Conclusion}

In this work we have presented a classification of triple nodal points (TPs) on high-symmetry lines (HSLs) in spinless systems.
In contrast to earlier works, which focused on the classification of such TPs based on their dispersion in $k$ as linear vs.~quadratic~\cite{Park:2021,Feng:2021,Tian:2021}, our classification is based on the nodal-line (NL) structure in the vicinity of the TP. 
We specifically distinguish TPs without additional NL arcs  attached to it (besides the central NL along the HSL), which we call type-$\mathsf{A}$ (to maintain consistency with the classification of TPs in strongly spin-orbit coupled systems~\cite{Zhu:2016}), vs.\ TPs \emph{with} additional NL arcs which are either attached linearly (type $\mathsf{B}_l$) or quadratically (type $\mathsf{B}_q$). 
We have shown that the TP type is fully determined by the symmetry, i.e., the little co-group of the HSL and in some cases (specifically for HSLs with sixfold rotational symmetry) the particular choice of irreducible corepresentations of the bands involved in the TP.
To demonstrate the validity of our result, we have compiled a table of materials hosting TPs with different symmetry and of different types, and we used first-principles calculations to determine the NL structure near these TPs.
We observe agreement between our the theoretical characterization of the TPs and their realizations in our first-principles studies.

Curiously, we find (both in \kdotp{} models and in first-principles calculations) that sometimes a collection of NL arcs meet at a nexus point along the HSL in close vicinity of a type-$\mathsf{A}$ TP. 
By fine-tuning the model parameters it is possible to accidentally collide the two, i.e., to accidentally attach NL arcs to the otherwise type-$\mathsf{A}$ TP. 
To estimate the likelihood of this scenario, we have determined the codimension for realizing such fine-tuned models. 
We have shown that for certain symmetries, this is achieved by tuning a single parameter; indeed, we have identified a concrete material (\ce{ZrO}) in which such an accidental fine-tuning happens to be realized. 
Additionally, while type-$\mathsf{B}$ TPs are convergence points of an equal number of NL arcs in both adjacent energy gaps, the NL arcs meeting at such fine-tuned type-$\mathsf{A}$ TPs can (for a suitable choice of model parameters) all be realized in the same energy gap.

It is worth noting that TPs of a remarkable variety of types and codimensions are realizable in spinless band structures, especially if contrasted to the spinful case.
In this respect, note that a HSL can only exhibit TPs if its little group exhibits both 1D and 2D irreducible corepresentation (ICRs). 
While for spinful systems there are only two magnetic points groups (MPGs) that have both 1D and 2D ICRs and that arise as little co-groups along HSLs~\cite{Zhu:2016}, there are as many as $13$ such MPGs for spinless systems. 
This suggests that TPs might be potentially much more common among weakly spin-orbit coupled materials.
We anticipate that the revealed richness of TPs in spinless band structures will motivate research of their experimental realizations, associated transport features, and topological characteristics.
Regarding the latter, we note that our parallel publication~\cite{Lenggenhager:2021:TPHOT} reveals a universal higher-order bulk-boundary correspondence of TP pairs, which is associated with a filling anomaly~\cite{Benalcazar:2019} and fractional charges at nanowire hinges.

We further find that non-symmorphicity influences the derived classification of TPs in spinless band structures in a rather trivial way: it either renders TPs impossible [for HSLs on the Brillouin zone (BZ) boundary when the projective factor system is non-trivial] or keeps the characterization of TPs unchanged (for HSLs inside BZ and when the projective factor system belongs to the trivial equivalence class).

The presented characterization of TPs applies to spinless representations in \emph{all} space groups (magnetic or not, non-symmorphic or not), and as such provides that next essential piece of information towards the growing catalogue of symmetry-protected and symmetry-enforced band nodes~\cite{Chan:2019,Hirschmann:2021,Leonhardt:2021,Yu:2021,Yang:2021}.
Nevertheless, it may be hard to find examples of electronic material which are well described by spinless band structures.
Indeed, several recent works have argued that magnetic space groups might not be ideally suited to capture band structure features of certain magnetic materials with negligible spin-orbit coupling, dubbed \enquote{altermagnets} by Refs.~\onlinecite{Smejkal:2021,Smejkal:2021b}.
On the other hand, spinless representations of magnetic space groups could more readily be realized in classical metamaterials such as discrete spring systems, elastic, acoustic and phononic systems, as well as classical electric circuits.
For example, it is easily revealed that the displacement vector of coupled two-dimensional pendula transforms according to a 2D spinless representation.
Time-reversal symmetry can be broken by applying a magnetic field perpendicular to the plane of motion and charging the pendula electrically or by placing a gyroscope with certain angular momentum in each pendulum~\cite{Nash:2015,Cheng:2016,Suesstrunk:2016}.
The symmetry of the resulting system is then captured by a magnetic space group.

We conclude with one speculation that may constitute an interesting research avenue for the near future. 
For the \enquote{altermagnets} mentioned above, \enquote{spin symmetry groups}~\cite{Litvin:1974,Litvin:1977} (where \enquote{spin} describes a magnetic pattern, and should not be confused for the double cover of orthogonal groups) were proposed as the proper substitute of magnetic space groups.
The characterization of band degeneracies in spin symmetry groups was systematically studied only recently~\cite{Yang:2021b}, and it is likely that many of them support HSLs with both 1D and 2D ICRs, leading potentially to a further variety of TPs beyond the ones classified here. 
More generally, the very recent work~\cite{Yang:2021b} foreshadows that such altermagnets may admit novel types of band structure nodes and topologies that are impossible with magnetic space groups alone, and we anticipate exciting future discoveries in this direction.

\begin{acknowledgments}
P.~M.~L. and T.~B. were supported by the Ambizione grant No.~185806 by the Swiss National Science Foundation.
T.~N. acknowledges support from the European Research Council (ERC) under the European Union’s Horizon 2020 research and innovation programm (ERC-StG-Neupert-757867-PARATOP).
T.~B. and T.~N. were supported by the NCCR MARVEL funded by the Swiss National Science Foundation.
X.~L acknowledges the support by the China Scholarship Council (CSC). 
\end{acknowledgments}


\appendix
\renewcommand{\thefigure}{\thesection.\arabic{figure}}

\section{Antiunitary symmetries and corepresentations}\label{App:antiunitary_symmetries}

In this appendix we briefly review how to deal with groups that contain antiunitary symmetries, in particular we discuss the consequences on their representation theory.
More details can be found, e.g., in Ref.~\onlinecite{Bradley:1972}.

Given a unitary group $\uG$ we consider the addition of an arbitrary antiunitary symmetry element $a$, e.g., time reversal or time reversal combined with another symmetry operation, to the generators, such that we obtain the \emph{magnetic group}
\begin{equation}
	\mG = \uG \cup a\uG.
\end{equation}
All magnetic (point or space) groups can be written in this form.
Given that we know the irreducible representations $\rep$ of the unitary subgroup $\uG\subg \mG$ we can construct the irreducible \emph{corepresentations} of $\mG$, see below.

By construction $\mG$ contains two (left) cosets: $1\uG$ and $a\uG$, where $1$ is the identity, such that $\uG$ is a subgroup of index $2$.
Elements of the coset $1\uG$ ($a\uG$) are unitary (antiunitary) such that $\mG$ has an equal number of unitary and antiunitary elements.
It follows that $\uG$ is a \emph{normal subgroup} of $\mG$ and consequently the cosets form a group, the \emph{quotient group} $\mG/\uG\cong\{1,a\}\cong \ztwo$ that obeys $a^{-1}=a$ and $a^2=1$.
According to the natural group homomorphism, i.e., the canonical projection of $\mG$ onto $\mG/\uG$, this implies that $bc\in \uG$ for $b,c\in a\uG$, and that $a^{-1}\in a\uG$.

\subsection{Corepresentations}
For a non-unitary group we study \emph{corepresentations} $\corep$ instead of representations $\rep$.
These satisfy the relations
\begin{subequations}\label{App:eq:corep_def}
    \begin{align}
    	\corep(g)\corep(h) &= \corep(gh),\\
    	\corep(g)\corep(b) &= \corep(gb),\\
    	\corep(b)\cconj{\corep(g)} &= \corep(bg),\\
    	\corep(b)\cconj{\corep(c)} &= \corep(bc)
    \end{align}
\end{subequations}
for $g,h\in \uG$ and $b,c,\in a\uG$.
Similarly, a change of basis by some unitary $U$ acts as follows:
\begin{subequations}\label{App:eq:corep_similarity}
    \begin{align}
    	\corep(g) &\mapsto U\corep(g)U^{-1},\\
    	\corep(b) &\mapsto U\corep(b)\cconj{(U^{-1})}.
    \end{align}
\end{subequations}
We note that by (formally) considering the matrices $\corep(g)$ for $g\in \uG$ and $\corep(b)\mcK$, where $\mcK$ is complex conjugation, for $b\in a\uG$, they form a representation.

If the irreducible representations (IRs) $\rep$ of $\uG$ are known, the irreducible corepresentations (ICRs) $\corep$ of $\mG$ can be determined as follows (see Ref.~\onlinecite{Bradley:1972} for more details and proofs of the statements summarized here).
In general we need to distinguish three cases based on the reality of $\rep$: (a) real, (b) pseudoreal and (c) complex.
For our purposes only (a) and (c) are relevant.
\begin{enumerate}[(a)]
    \item[(a)] Reality of $\rep$ implies that there is a unitary matrix $N$ such that for all $g\in\uG$
    \begin{subequations}\label{App:eq:ICRN_real}
        \begin{align}
            \rep(g) &= N\cconj{\rep}(a^{-1}ga)N^{-1},\\
		    N\cconj{N} &= +\rep(a^2).
        \end{align}
    \end{subequations}
	Then, $\corep$, defined by
	\begin{equation}
		\corep(g) = \rep(g),\quad \corep(b) = \pm \rep(ba^{-1})N
		\label{App:eq:ICR_real}
	\end{equation}
	for all $g\in\uG$ and $b\in a\uG$, is an ICR of $\mG$. The corepresentation with the $+$ sign in \cref{App:eq:ICR_real} is equivalent to the one with the $-$ sign.
	
	\item[(c)] If $\rep$ is a complex representation, then $\corep$, defined by
	\begin{equation}
		\corep(g) = \mat{\rep(g)&0\\0&\corep(g)},\quad \corep(b) = \mat{0 & \rep(ba)\\\corep(ba^{-1}) & 0}
	\end{equation}
	for all $g\in\uG$ and $b\in a\uG$, is an ICR of $\mG$.
\end{enumerate}

The representation matrices $\rep(g_i)$ of a set of generators of a unitary point group $\uG$ can be obtained from the Bilbao crystallographic server (BCS) using the \textsc{Representations PG} application~\cite{Elcoro:2017}.
This application also gives the reality of each IR.
Applying the above procedure, we determine the relevant corepresentation matrices $\corep(g_i)$.
Note that recently, a new program \textsc{Corepresentations PG}~\cite{Xu:2020,Elcoro:2021} has been added to the BCS, which allows for direct extraction of the matrix corepresentations of magnetic point groups.

\subsection{Symmetry constraints on the Bloch Hamiltonian}

Recall that the Bloch Hamiltonian $\Ham(\vec{k})$ close to a high-symmetry point or line (e.g., obtained from a \kdotp{} expansion) is constrained by the symmetries that leave that point or line invariant.
These symmetries form a subgroup of the space group called the \emph{little group} $\LG{\vec{k}}$, which generally contains both unitary and antiunitary elements.
The $n$-band Hamiltonian transforms in a corepresentation $\corep\,:\, \LG{\vec{k}}\to \mathsf{U}(n)$, such that
\begin{subequations}\label{App:eq:kdotp_constraint}
    \begin{align}
        \forall g\in\LG{\vec{k}}_\textrm{u} &\,:\, & \corep(g)\Ham(g^{-1}\vec{k})\corep(g)^{-1} &= \Ham(\vec{k}),\label{App:eq:kdotp_constraint:unitary}\\\
        \forall g\in\LG{\vec{k}}_\textrm{au} &\,:\, & \corep(g)\cconj{\Ham(g^{-1}\vec{k})}\corep(g)^{-1} &= \Ham(\vec{k}),\label{App:eq:kdotp_constraint:antiunitary}
    \end{align}
\end{subequations}
where $\LG{\vec{k}}_\textrm{u}$ and $\LG{\vec{k}}_\textrm{au}$ are the unitary subgroup of $\LG{\vec{k}}$ and the anitunitary complement, respectively.
Due to the group structure of $\LG{\vec{k}}$, the only independent constraints are those due to the \emph{generators} of $\LG{\vec{k}}$.
Furthermore, elements of the translation subgroup $\TG\subg\LG{\vec{k}}$ lead to trivial constraints and can thus be neglected.
If the space group is symmorphic, then $\LG{\vec{k}}$ is a semidirect product of the little \emph{co-group} of $\LcG{\vec{k}}$ (the group formed by the point group elements in $\LG{\vec{k}}$) and $\TG$, such that it is sufficient to consider the constraints due to elements of $\LcG{\vec{k}}$.
For non-symmorphic space groups this is not as straightforward and is discussed in \cref{App:non-symm_SGs}.

Given such a set of constraints, a family of \kdotp{} Hamiltonians can be determined by expanding the full Bloch Hamiltonian $\Ham(\vec{q})$ around the momentum vector $\vec{k}$ under consideration up to some order $n$ in $\vec{k}$, and restricting to terms that satisfy \cref{App:eq:kdotp_constraint}.
The space of all \kdotp{} Hamiltonians then is the tensor product of the space of Hermitian matrices of the appropriate dimensions (given by the number of bands) and the space of polynomials in $k_x,k_y,k_z$ up to the order $n$.
\Cref{App:eq:kdotp_constraint} then constrains symmetry compatible \kdotp{} Hamiltonians $\Ham(\vec{k})$ to some subspace of it.
In simple cases this analysis can be performed by hand (see \cref{App:Classif:no_PT_mirror:kdotp:C4vC6v,App:Classif:no_PT_mirror:kdotp:C6PT,App:Classif:no_PT_mirror:kdotp:C3v}), however for groups with a large number of generators, we use the \Python{} package \textsc{kdotp-symmetry}~\cite{Gresch:PhD}, which implements an algorithm for finding this subspace as a span of a family of symmetry allowed \kdotp{} Hamiltonians.
For brevity we refer to such a parametrized family as a \emph{\kdotp{} model}.

\section{Effect of non-symmorphicity}\label{App:non-symm_SGs}

According to \cref{App:eq:kdotp_constraint}, the symmetry constraints of the Hamiltonian near a high-symmetry point or line $\vec{k}$ are determined by the corepresentations (CRs) of the little group $\LG{\vec{k}}$, but translations lead to trivial constraints.
For symmorphic space groups the irreducible corepresentations (ICRs) of $\LG{\vec{k}}$ are deduced from the ICRs of the little co-group $\LcG{\vec{k}}$, which is isomorphic to $\LG{\vec{k}}/\TG$ (and equals one of the magnetic point groups)~\cite{Bradley:1972}, allowing us to rewrite the constraints in terms of CRs of $\LcG{\vec{k}}$.
On the other hand, if the space group is non-symmorphic, it becomes necessary to study the \emph{projective} corepresentations (PCRs) of $\LcG{\vec{k}}$ and the symmetry constraints on \kdotp{} models are formulated in terms of those.

Despite this apparent additional level of difficulty, in this appendix we prove the following important result.
For the specific little groups that are relevant for triple points (TPs) the resulting symmetry constraints are equivalent to those given by an appropriate \emph{ordinary} corepresentation of $\LcG{\vec{k}}$ \emph{corresponding} to the relevant corepresentation of $\LG{\vec{k}}$.

We begin by briefly reviewing how to find all ICRs of a little group in \cref{App:non-symm_SGs:LGirreps} and by reviewing some properties of PCRs in \cref{App:non-symm_SGs:proj_coreps}.
The review follows Ref.~\onlinecite{Bradley:1972} but directly deals with magnetic groups, i.e., groups containing both unitary and antiunitary elements~\cite{Dimmock:1963,Janssen:1972}.
In \cref{App:non-symm_SGs:LGwith1Dirreps} we then prove that any little group with at least one 1D ICR has ICRs deduced from the PCRs of the corresponding little co-group with a factor system that lies in the trivial factor system class (these notions are reviewed in \cref{App:non-symm_SGs:proj_coreps}).
We discuss the consequences on the classification of TPs in non-symmorphic space groups in \cref{App:non-symm_SGs:TPclassif} and argue that the classification reduces to the one derived for symmorphic space groups with an appropriate identification of ICRs.

\subsection{Irreducible representations of the little group}\label{App:non-symm_SGs:LGirreps}
Given a little group $\LG{\vec{k}}$ (which generally is non-unitary) we are interested in finding all ICRs.
We start by factorizing $\LG{\vec{k}}$ into left cosets with respect to the translation group $\TG$:
\begin{equation}
	\LG{\vec{k}} = \sum_{\alpha}\{S_{\!\alpha}|\vec{w}_\alpha\}\circ\TG,
	\label{App:eq:LG_decomposition}
\end{equation}
where $\{R|\vec{v}\}$ denotes the symmetry operation that acts on a position vector as $\{R|\vec{v}\}\vec{r} = R\vec{r}+\vec{v}$ and $R$ is either unitary or antiunitary.
We make the unitarity and antiunitarity more explicit by using unprimed and primed indices for unitary elements and antiunitary elements, respectively:
\begin{equation}
	\LG{\vec{k}} = \sum_{i}\{S_{\!i}|\vec{w}_i\}\circ\TG + \sum_{i'}\{S_{\!i'}|\vec{w}_{i'}\}\circ\TG.
\end{equation}
This decomposition is unique only up to changes of each $\vec{w}_\alpha$ by arbitrary lattice vectors $\vec{t}_\alpha\in\TG$, where the Greek subscripts encompass both unitary and antiunitary elements.

The coset representatives $\{S_{\!\alpha}|\vec{w}_\alpha\}$ do not form a group, but satisfy
\begin{equation}
    \{S_{\!\alpha}|\vec{w}_\alpha\}\{S_{\!\beta}|\vec{w}_\beta\} = \{\id|\vec{t}_{\alpha\beta}\}\{S_{\!\gamma}|\vec{w}_\gamma\},
\end{equation}
where $\id$ is the identity rotation, $S_{\!\gamma} = S_{\!\alpha} S_{\!\beta}$ with $\vec{w}_\gamma$ the corresponding translation in the coset decomposition and
\begin{equation}
    \vec{t}_{\alpha\beta} = \vec{w}_\alpha + S_{\!\alpha}\vec{w}_\beta - \vec{w}_\gamma.
\end{equation}
Given a CR $\corep$ of $\LG{\vec{k}}$, then according to \cref{App:eq:corep_def} and using
\begin{equation}
    \corep(\{\id|\vec{t}_{\alpha\beta}\})=\e^{-\i\vec{k}\cdot\vec{t}_{\alpha\beta}}=:f(\vec{t}_{\alpha\beta})
\end{equation}
it holds that
\begin{subequations}
    \begin{align}
    	\corep(\{S_{\!i}|\vec{w}_{i}\})\corep(\{S_{\!j}|\vec{w}_{j}\}) &= f(\vec{t}_{ij})\corep(\{S_{\!l}|\vec{w}_{l}\}),\\
    	\corep(\{S_{\!i}|\vec{w}_{i}\})\corep(\{S_{\!j'}|\vec{w}_{j'}\}) &= f(\vec{t}_{ij'})\corep(\{S_{\!l'}|\vec{w}_{l'}\}),\\
    	\corep(\{S_{\!i'}|\vec{w}_{i'}\})\cconj{\corep(\{S_{\!j}|\vec{w}_{j}\})} &= f(\vec{t}_{i'j})\corep(\{S_{\!l'}|\vec{w}_{l'}\}),\\
    	\corep(\{S_{\!i'}|\vec{w}_{i'}\})\cconj{\corep(\{S_{\!j'}|\vec{w}_{j'}\})} &= f(\vec{t}_{i'j'})\corep(\{S_{\!l}|\vec{w}_{l}\}),
    \end{align}
\end{subequations}
where $S_{\!l}\!=\!S_{\!i}S_{\!j}$, $S_{\!l'}\!=\!S_{\!i}S_{\!j'}$, $S_{\!l'}\!=\!S_{\!i'}S_{\!j}$, and $S_{\!l}\!=\!S_{\!i'}S_{\!j'}$, respectively.
One can show that $\corep$ only depends on the coset in the decomposition shown in \cref{App:eq:LG_decomposition} and therefore is a matrix-valued function on the quotient group $\LG{\vec{k}}/\TG$, which is isomorphic to the little co-group $\LcG{\vec{k}}$.
This implies that we can write 
\begin{equation}
    \corep(\{S_{\!\alpha}|\vec{w}_\alpha\})=\projcorep(S_{\!\alpha})
\end{equation}
and $\projcorep$ forms a \emph{projective} CR of $\LcG{\vec{k}}$ with factor system $\mu(S_{\!\alpha},S_{\!\beta})=f(\vec{t}_{\alpha\beta})$ (the definition of PCRs and some of their properties are reviewed in \cref{App:non-symm_SGs:proj_coreps}).
Furthermore, all ICRs of $\LG{\vec{k}}$ can be found by finding the projective ICRs of $\LcG{\vec{k}}$ and only keeping those with the correct factor system
\begin{equation}
	\mu(S_{\!\alpha},S_{\!\beta})=\e^{-\i\vec{k}\cdot\left(\vec{w}_\alpha + S_{\!\alpha}\vec{w}_\beta - \vec{w}_\gamma\right)}.
	\label{App:eq:LG_factor_system}
\end{equation}

\subsection{Properties of projective corepresentations}\label{App:non-symm_SGs:proj_coreps}
Projective corepresentations of a finite-order group $G$ with unitary subgroup $H$ and antiunitary complement $G-H$ generalize the notion of (ordinary) representations.
A PCR $\projcorep$ is a map from $G$ to the group of invertible matrices that satisfies
\begin{subequations}\label{App:eq:projcorep-def}
    \begin{align}
        \projcorep(h)\projcorep(g) &= \mu(h,g)\projcorep(hg),\\
        \projcorep(a)\cconj{\projcorep(g)} &= \mu(a,g)\projcorep(ag),
    \end{align}
\end{subequations}
for any $g\in G$, $h\in H$ and $a\in G-H$ [in contrast, for \emph{ordinary} CRs one demands, $\mu(g,g')=1$ for all $g,g'\in G$].
The map $\mu:G\times G\mapsto\mathbb{C}$ forms a so-called \emph{factor system} and satisfies
\begin{subequations}\label{App:eq:factor_system_condition}
    \begin{align}
        \mu(h,g)\mu(hg,g') &= \mu(h,gg')\mu(g,g'),\\
        \mu(a,g)\mu(ag,g') &= \mu(a,gg')\cconj{\mu(g,g')},
    \end{align}
\end{subequations}
for all $g,g'\in G$, $h\in H$ and $a\in G-H$.

As for ordinary CRs, similarity transformations [cf.~\cref{App:eq:corep_similarity}] map a PCR to an equivalent PCR.
If the factor system satisfies $\abs{\mu(g,g')}=1$ for all $g,g'\in G$ [which it does in our use case, cf.~\cref{App:eq:LG_factor_system}], we can always find a transformation such that $\projcorep(g)$ is unitary for all $g\in G$.
Additionally, given a PCR $\projcorep$
\begin{equation}
	\projcorep'(g) = C(g)\projcorep(g)
	\label{App:eq:proj_equiv_transf}
\end{equation}
for all $g\in G$ and $C:G\to \cmplx\setminus\{0\}$ a function into non-zero complex numbers, $\projcorep'$ forms another projective representation with factor system
\begin{subequations}\label{App:eq:proj_equiv_transf:factor_system}
    \begin{align}
        \nu(h,g) &= \frac{C(h)C(g)}{C(hg)}\mu(h,g),\\
        \nu(a,g) &= \frac{C(a)\cconj{C(g)}}{C(ag)}\mu(a,g),
    \end{align}
\end{subequations}
for all $g\in G$, $h\in H$ and $a\in G-H$.
The transformation in \cref{App:eq:proj_equiv_transf} defines \emph{equivalence classes of factor systems}: two factor systems $\mu$ and $\nu$ are equivalent if and only if there exists a complex-valued function $C$ such that \cref{App:eq:proj_equiv_transf:factor_system} is satisfied.

\subsection{Little groups with 1D irreducible representations}\label{App:non-symm_SGs:LGwith1Dirreps}
We now prove the following theorem.
Although we expect the theorem to be known among the experts, we failed to find it explicitly stated within the literature on representation theory of (magnetic) space groups.
\smallskip

\noindent\emph{Theorem.}
Let $\LG{\vec{k}}$ be a little group that has at least one 1D ICR $\corep_1$.
Then, the ICRs of $\LG{\vec{k}}$ are deduced from projective ICRs of the corresponding little co-group $\LcG{\vec{k}}$ with factor system belonging to the trivial equivalence class.
\smallskip

\noindent\emph{Proof.}
The little group $\LG{\vec{k}}$ has CRs deduced from PCRs of $\LcG{\vec{k}}$ with factor system $\mu$ given by \cref{App:eq:LG_factor_system}.
Because $\LG{\vec{k}}$ has a 1D ICR $\corep_1$, there must be a corresponding 1D projective ICR $\projcorep_1$ of $\LcG{\vec{k}}$ with the same factor system $\mu$.
Consider the equivalence transformation of factor systems induced by
\begin{equation}
    C(g) = \projcorep_1(g)^{-1}.
    \label{App:eq:factor-system-transfo}
\end{equation}
Then, according to \cref{App:eq:projcorep-def,App:eq:proj_equiv_transf:factor_system}, one easily verifies that the transformed PCRs will have factor system
\begin{equation}
    \nu(g,g') = \mu(g,g')^{-1}\mu(g,g') = 1
\end{equation}
for all $g,g'\in G$.
This immediately implies that $\mu$ is a factor system in the trivial \emph{equivalence class} (but not necessarily trivial itself).

\subsection{Consequences for triple point classification}\label{App:non-symm_SGs:TPclassif}
Now that we know the relationship between CRs of $\LG{\vec{k}}$ and PCRs of $\LcG{\vec{k}}$ we can rewrite \cref{App:eq:kdotp_constraint} in terms of elements of $\LcG{\vec{k}}$ and the corresponding PCR $\projcorep$:
\begin{subequations}
    \begin{align}
        \projcorep(S_{\!i})\Ham(S_{\!i}^{-1}\vec{k})\projcorep(S_{\!i})^{-1} &= \Ham(\vec{k}),\\
        \projcorep(S_{\!i'})\cconj{\Ham(S_{\!i'}^{-1}\vec{k})}\projcorep(S_{\!i'})^{-1} &= \Ham(\vec{k}),
    \end{align}
\end{subequations}
for $S_{\!i},S_{\!i'}\in\LcG{\vec{k}}$, $S_{\!i}$ unitary and $S_{\!i'}$ antiunitary.

The little groups we study support stable TPs.
Therefore, they must have at least one 1D ICR, such that, according to the theorem in \cref{App:non-symm_SGs:LGwith1Dirreps}, $\projcorep$ has factor system in the trivial equivalence class.
We therefore consider
\begin{equation}
	\projcorep'(g) = \e^{\i\phi(g)}\projcorep(g),
	\label{App:eq:ord_reps_from_proj}
\end{equation}
where $\e^{\i\phi(g)}=\Delta_1^{-1}(g)$ plays the role of the equivalence transformation $C(g)$ by the 1D unitary projective ICR [cf.~\cref{App:eq:factor-system-transfo}], and $\projcorep'$ is an \emph{ordinary} CR of $\LcG{\vec{k}}$ (but not a CR of $\LG{\vec{k}})$.
Then,
\begin{subequations}
    \begin{align}
        \projcorep'(S_{\!i})\Ham(S_{\!i}^{-1}\vec{k})\projcorep'(S_{\!i})^{-1} &= \Ham(\vec{k}),\\
        \projcorep'(S_{\!i'})\cconj{\Ham(S_{\!i'}^{-1}\vec{k})}\projcorep'(S_{\!i'})^{-1} &= \Ham(\vec{k}),
    \end{align}
\end{subequations}
because the factors $\e^{\i\phi(S_{\!\alpha})}$ and $\e^{-\i\phi(S_{\!\alpha})}$ cancel. 
The key observation here is that the symmetry constraints on the \kdotp{} expansion are unaffected by the equivalence transformation in \cref{App:eq:ord_reps_from_proj}.

By matching the ICRs of $\LG{\vec{k}}$ to \emph{ordinary} ICRs of $\LcG{\vec{k}}$ via \cref{App:eq:ord_reps_from_proj}, the set of constraints for any given CR $\corep$ of $\LG{\vec{k}}$ is therefore reduced to a set of constraints for the corresponding ordinary CRs $\corep'$ of $\LcG{\vec{k}}$.
The classification of triple points in non-symmorphic space groups is therefore reduced to their classification in symmorphic space groups by properly identifying ICRs of the little group with ICRs of the little co-group.
That identification of ICRs can be easily performed by looking up ICRs of both groups, matching generators of $\LG{\vec{k}}$ to those of $\LcG{\vec{k}}=\LG{\vec{k}}/\TG$ and identifying the necessary phases $\phi(S_{\!\alpha})$.

\section{Classification in the absence of \texorpdfstring{$\mcP\mcT$}{PT} symmetry and presence of mirror symmetry}\label{App:Classif:no_PT_mirror}

In this appendix we discuss some details of the derivation of the classification of triple points (TPs) in the absence of $\mcP\mcT$ symmetry and presence of mirror symmetry that have been left out in \cref{Sec:Classif:no_PT} in the main text.
In \cref{App:Classif:no_PT_mirror:kdotp} we derive the minimal \kdotp{} models given in \cref{eq:kdotp_C4vC6v,eq:kdotp_C6PT,eq:kdotp_C3v} based on the symmetry constraints due to the corresponding magnetic points groups (MPGs).
In \cref{App:Classif:no_PT_mirror:mirror_planes} we comment on the relation between the \kdotp{} expansions in symmetry inequivalent sets of planes (cf.~\cref{fig:mirror-planes}), deriving specifically \cref{eq:inequiv-m-relation-1,eq:inequiv-m-relation-2,eq:inequiv-m-relation-3}.
The derived models and constraints are used in \cref{Sec:Classif:no_PT_mirror} to deduce the type of the TP, the number of nodal-line arcs attached to a nexus point and the codimension of the nexus point.

\subsection{Derivation of \texorpdfstring{\kdotp{}}{k.p} models}\label{App:Classif:no_PT_mirror:kdotp}

\subsubsection{\texorpdfstring{$4mm$, $\bar{4}'2'm$ and $6mm$}{4mm, -4'2'm and 6mm}}\label{App:Classif:no_PT_mirror:kdotp:C4vC6v}

As discussed in \cref{Sec:Classif:mirror_no_PT:C4vC6v}, the MPGs $4mm$, $\bar{4}'2'm$ and $6mm$, have mirror symmetries that appear in orthogonal pairs $(m,m_\perp)$.
With $\rep$ the representation capturing the transformation of the three bands involved in the triple point, $\rep=\rho^\mathrm{2D}\oplus\rho^\mathrm{1D}$, a basis can be chosen such that the two mirror symmetries have the following matrix representations
\begin{equation}
    \!\rep(m) = \mqty(\dmat{\pm_p1,\mp_p1,\mp_p1}),\;\, \rep(m_\perp) = \mqty(\dmat{\pm_q1,\mp_q1,\pm_t1}).
    \label{App:eq:rep_m}
\end{equation}
Note that in the main text \cref{tab:mirror_eigvals} we have introduced a fourth free sign `$\pm_r$', which in the present discussion is already fixed by a proper choice of basis as in \cref{eq:signs-a-c}.

We choose orthogonal coordinates $(k_x,k_z)$ in the mirror plane of $m$, with $k_z$ along the rotation axis, i.e., the HSL containing the TP.
Then, $m$ constrains the Hamiltonian as follows:
\begin{equation}
    \Ham(k_x,k_z) = \rep(m)\Ham(k_x,k_z)\rep(m)^{-1},
    \label{App:eq:the-first-constraint}
\end{equation}
and implies that
\begin{equation}
    \Ham(k_x,k_z) = \mqty(\epsilon^1&0&0\\0&\epsilon^2&f\\0&\cconj{f}&\epsilon^3)
    \label{App:eq:Ham_C4vC6v_form}
\end{equation}
with $\epsilon^i$ being real-valued functions and $f$ a complex-valued function of $k_x$ and $k_z$. 

Note that, as emphasized in the main text, for groups without $\mcP\mcT$ symmetry we derive the low-order expansion only in $k_x$, while keeping the full dependence on $k_z$.
To that end, first note that $m_\perp$ constrains the Hamiltonian by
\begin{equation}
    \Ham(k_x,k_z) = \rep(m_\perp)\Ham(-k_x,k_z)\rep(m_\perp)^{-1}.
    \label{App:eq:the-second-constraint}
\end{equation}
This implies that $\epsilon^i$ are even in $k_x$, and that
\begin{equation}
    f(k_x,k_z) = (\mp_q1)(\pm_t1)f(-k_x,k_z),
\end{equation}
i.e., $f$ is even (odd) in $k_x$ if the product of the signs of the $C_2$-characters of $\rho^\mathrm{2D}$ and $\rho^\mathrm{1D}$ is positive (negative), cf.~\cref{tab:mirror_eigvals}.
Let us denote the sign of that number by
\begin{equation}
    s=-(\pm_q1)(\pm_t1).
\end{equation}
Finally, the (antiunitary) rotational symmetry ($C_4$, $C_4\mcP\mcT$, and $C_6$ for the three discussed MPGs, respectively) leads to the following additional constraints: $\epsilon^1(0,k_z)=\epsilon^2(0,k_z)$ and $f(0,k_z)=0$.

Assuming there is a TP at $k_z=0$ at energy $E=0$, we have $\epsilon^i(0,0)=0$ for $i=1,2,3$ and
\begin{subequations}
    \begin{align}
        \tilde{\epsilon}^1(k_x,k_z) &= \tilde{a}k_z + \tilde{b}k_x^2,\\
        \tilde{\epsilon}^2(k_x,k_z) &= \tilde{a}k_z + \tilde{c}k_x^2,\\
        \tilde{\epsilon}^3(k_x,k_z) &= \tilde{d}k_z + \tilde{e}k_x^2,\\
        f(k_x,k_z) &= Ak_x^{\frac{3+s}{2}},
    \end{align}
\end{subequations}
with  $\tilde{a},\tilde{b},\tilde{c},\tilde{d},\tilde{e}$ real-valued functions and $A$ a complex-valued function of $k_z$.
The degeneracies of the spectrum are clearly independent of constant energy shifts (i.e., terms in the Hamiltonian that are proportional to the identity matrix), such that we can subtract $\tfrac{1}{2}(\tilde{\epsilon}^1+\tilde{\epsilon}^2)\id$ and arrive at the following Hamiltonian
\begin{equation}
    \Ham(k_x,k_z) = \mqty(ak_x^2 & 0 & 0\\0 & -ak_x^2 & Ak_x^{\frac{3+s}{2}}\\0 & \cconj{A}k_x^{\frac{3+s}{2}} & bk_z+ck_x^2)
    \label{App:eq:kdotp_Ham_C4vC6v}
\end{equation}
with $a,b,c$ real-valued functions and $A$ a complex-valued functions of $k_z$.

\subsubsection{\texorpdfstring{$\bar{6}'m2'$}{-6'm2'}}\label{App:Classif:no_PT_mirror:kdotp:C6PT}

The MPG $\bar{6}'m2'$ has three vertical mirror planes $m$, which are related by $C_3$ symmetry, and another set of symmetry-related pseudo-mirror symmetry $m_\perp\mcP\mcT$, where $m_\perp$ is the mirror perpendicular to $m$.
Therefore, we proceed analogously to \cref{App:Classif:no_PT_mirror:kdotp:C4vC6v}, with the key difference that the analog to \cref{App:eq:the-second-constraint} involves the antiunitary symmetry $m_\perp\mcP\mcT$ instead of $m_\perp$.
Therefore, in the mirror planes of $m$, the Hamiltonian takes the form given in \cref{App:eq:Ham_C4vC6v_form} and is subjected to the constraint [cf.~\cref{App:eq:kdotp_constraint:antiunitary}]:
\begin{equation}
    \Ham(k_x,k_z) = \corep(m_\perp\mcP\mcT)\cconj{\Ham(-k_x,k_z)}\corep(m_\perp\mcP\mcT)^{-1}.
\end{equation}
We find that in an appropriate choice of basis~\cite{Xu:2020,Elcoro:2021}
\begin{equation}
    \corep(m_\perp\mcP\mcT) = \rho^\mathrm{2D}\oplus\rho^\mathrm{1D} = \mqty(\dmat{-\i,-\i,\pm 1})
\end{equation}
where the upper (lower) sign corresponds to choosing representation $\rho^\mathrm{1D}=A_1$ ($\rho^\mathrm{1D}=A_2$).

The above constraint implies that $\epsilon^i$ are even functions of $k_x$, and that
\begin{equation}
    f(k_x,k_z) = \mp\i\cconj{f(-k_x,k_z)},
\end{equation}
further implying that
\begin{equation}
    f(k_x,k_z) = d(1\mp\i)k_x
\end{equation}
for $d\in\mathbb{R}$.
By repeating the final steps of \cref{App:Classif:no_PT_mirror:kdotp:C4vC6v}, we arrive at
\begin{equation}
    \Ham(k_x,k_z) = \mqty(ak_x^2 & 0 & 0\\0 & -ak_x^2 & d(1\mp\i)k_x\\0 & d(1\pm\i)k_x & bk_z+ck_x^2)
    \label{App:eq:kdotp-C4PT}
\end{equation}
with $a,b,c,d$ being real-valued functions of $k_z$.
Note that the final result in \cref{App:eq:kdotp-C4PT} corresponds exactly to \cref{App:eq:kdotp_Ham_C4vC6v} with $A=d(1\mp\i)$ and $s=-1$.

\subsubsection{\texorpdfstring{$3m$}{3m}}\label{App:Classif:no_PT_mirror:kdotp:C3v}

Finally, we discuss the remaining MPG $3m$, which has three-fold rotational symmetry with three vertical mirrors that do \emph{not} come in pairs with orthogonal mirror planes.
Again, we choose coordinates $(k_x,k_z)$ in the mirror plane under consideration (the three mirror planes are equivalent due to rotational symmetry), then the Hamiltonian takes again the form given in \cref{App:eq:Ham_C4vC6v_form}.
In this case, however, there are no constraints on $\epsilon^i$ and $f$ to be even or odd functions of $k_x$, such that to leading order in $k_x$:
\begin{equation}
    \Ham(k_x,k_z) = \mqty(ak_x & 0 & 0\\0 & -ak_x & Ak_x\\0 & \cconj{A}k_x & bk_z+ck_x),
\end{equation}
with $a,b,c$ real-valued functions and $A$ a complex-valued function of $k_z$.

\subsection{Relationship between independent sets of mirror planes}\label{App:Classif:no_PT_mirror:mirror_planes}

The functional form of the \kdotp{} models in \cref{App:eq:kdotp_Ham_C4vC6v} has been derived with rather general assumptions, such that it applies to \emph{arbitrary} mirror planes containing the rotation axis.
In this respect, note that point groups $4mm$ and $6mm$ each have two sets of symmetry-related mirror planes (cf.\ differently colored planes in \cref{fig:mirror-planes}).
Within each set, the mirror planes are related to each other by rotational symmetry, such that the parameters $a,b,c$ and $A$ are identical.
In contrast, momenta in the two sets are \emph{not} related by symmetry; therefore, the functions $a,b,c,A$ and $a',b',c',A'$ that encode the Hamiltonian in the two planes, respectively, are \emph{a priori} not related to each other.

Nevertheless, a relationship between the two sets of functions can be established. 
This is achieved by considering the full 3D \kdotp{} model, i.e., one not restricted to mirror planes, but which reduces to the functional form in \cref{App:eq:kdotp_Ham_C4vC6v} (with suitable choice of functions $a,b,c,A$) for both sets of mirror planes.
In this section, we clarify how to relate the parameters of the \kdotp{} model in the independent sets of mirror planes of point groups $4mm$ and $6mm$ without determining the full 3D \kdotp{} model.

We consider a point group $G$ with two generators: an $n$-fold rotation $C_n$ (for simplicity, $n$ is assumed even; in our application $n=4,6$) and a vertical mirror $m_0$ (such that the mirror plane contains the axis of rotation).
The elements of the group correspond to powers of the $n$-fold rotation, $C_n^l$ for $l\in\{0,1,\dotsc,n-1$\}, and two types of mirror symmetries
\begin{subequations}
    \label{App:eq:two-sets-of-mirrors}
    \begin{align}
        m_l  &= C_n^lm_0C_n^{-l},\\
        m_l' &= C_n^lm_0'C_n^{-l}
    \end{align}
\end{subequations}
where $m_0'=C_n m_0$ and $l\in\{0,1,\dotsc,n/2-1\}$.
Note that, additionally, all mirrors $m$ satisfy
\begin{equation}
    m = C_n^l m C_n^l
    \label{App:eq:mirror-rotation-rel}
\end{equation}
for any $l$.
The partitioning in \cref{App:eq:two-sets-of-mirrors} corresponds to two conjugacy classes of mirrors:
\begin{subequations}
    \begin{align}
        \mathsf{c}(m_0) &= \left\{m_0, m_1, \dotsc, m_{n/2-1}\right\},\\
        \mathsf{c}(m_0') &= \left\{m_0', m_1', \dotsc, m_{n/2-1}'\right\}.
    \end{align}
\end{subequations}
In contrast, rotations with $l\notin\{0,n/2\}$ appear in two-element conjugacy classes
\begin{equation}
    \left\{C_n^l,C_n^{-l}\right\}
\end{equation}
(with $mC_n^lm^{-1}=C_n^{-l}$) whereas rotations with $l\in\{0,n/2\}$ constitute single-element conjugacy classes.
Furthermore, each mirror $m$, is accompanied by a perpendicular mirror $m_\perp$ (for $4mm$ both are in the same conjugacy class, while for $6mm$ they are in different conjugacy classes).

The Hamiltonian satisfies the constraint in \cref{eq:sym_constraint} for both generators of the point group, i.e.,
\begin{subequations}\label{App:eq:constraint_full}
    \begin{align}
        \Ham(\vec{k}) &= D(C_n)\Ham\left(C_n^{-1}\vec{k}\right)D(C_n)^{-1},\label{App:eq:constraint_full-line1}\\
        \Ham(\vec{k}) &= D(m)\Ham\left(m^{-1}\vec{k}\right)D(m)^{-1}.\label{App:eq:constraint_full-line2}
    \end{align}
\end{subequations}
Restricting to the plane $M$ in momentum space, which we define\footnote{Note that the set $M$ is called $m$-plane in the main text; however, the mathematically involved discussion that follows requires a more compact notation (only adopted in the present \cref{App:Classif:no_PT_mirror:mirror_planes}).} as the set of $\vec{k}$ points invariant under $m$, \cref{App:eq:constraint_full} leads to the two constraints
\begin{subequations}\label{App:eq:constraint_mirrorplane}
    \begin{align}
        \Ham(\vec{k}) &= D(m)\Ham\left(\vec{k}\right)D(m)^{-1},\label{App:eq:constraint_mirrorplane:m}\\
        \Ham(\vec{k}) &= D(m_\perp)\Ham\left(m_\perp^{-1}\vec{k}\right)D(m_\perp)^{-1},
    \end{align}
\end{subequations}
for $\vec{k}\in M$.
We recognize \cref{App:eq:constraint_mirrorplane} as an incarnation of \cref{App:eq:the-first-constraint,App:eq:the-second-constraint}, which we employed in the derivation of the \kdotp{} Hamiltonians in $M$ (cf.~\cref{App:Classif:no_PT_mirror:kdotp:C4vC6v}).

We now formally perform the \kdotp{} analysis based on \cref{App:eq:constraint_mirrorplane}.
To do that, we decompose in-plane momentum vectors into orthogonal coordinates $\vec{k}=(k_M,k_z)$ (with $k_z$, as usual, along the rotation axis), and sometimes utilize the unit vectors $\vec{e}_M$ and $\vec{e}_z$ in the two directions, respectively.
Next, note that any \kdotp{} Hamiltonian is encoded by a collection of parameters $a_i$ (which in our case are taken to be functions of $k_z$, since the $k_z$-dependence is not constrained by elements of $G$; for brevity we call them \emph{coefficients}) that multiply matrix-valued polynomial functions $h_i(k_M)$ subject to \cref{App:eq:constraint_mirrorplane} (for brevity we call $h_i(k_M)$ \emph{terms in the \kdotp{} model}).
Finally, we truncate the expansion such that the polynomials have order of at most $N$ in $k_M$.
Then, for $\vec{k}\in M$, and collecting the parameters $a_i$ in the vector $\vec{a}$, we write
\begin{equation}
    \Ham_{M}^{(N)}[\vec{a}](\vec{k}) = \sum_i a_i(k_z) h_i\left(k_M\right).\label{App:eq:H_M^N[a](k)}
\end{equation}
To summarize, the multiple decorations of the Hamiltonian appearing on the left-hand side of \cref{App:eq:H_M^N[a](k)} remind us that $\Ham_{M}^{(N)}[\vec{a}]$ is a parameterized (parameters $\vec{a}$) and truncated (order $N$) expansion of the Hamiltonian $\Ham(\vec{k})$ restricted to the plane $M$.
Note that \cref{App:eq:H_M^N[a](k)} is incarnated, for example, in \cref{eq:kdotp_C4vC6v}, when setting $\vec{a}=(a,b,c,A)$.

It is important to keep in mind that an equation with the same functional form as \cref{App:eq:H_M^N[a](k)} applies to \emph{any} of the $n$ mirror planes [because for each such a plane one can choose a pair of orthogonal mirror symmetries characterized by constraints analogous to \cref{App:eq:constraint_mirrorplane}]; in particular, the functional form of $h_i$ can be considered to be independent of $M$ (while both its argument $k_M$ as well as the basis in which the resulting $\Ham_{M}^{(N)}$ is given depend on the mirror plane under consideration).
To simplify our subsequent analysis of the symmetry constraints, we find it convenient to further adopt the notation
\begin{equation}
    h^i_M(\vec{k}) := h_i\left(k_M\right) = h_i\left(\vec{k}\cdot\vec{e}_M\right)
\end{equation}
(one should keep in mind here that $h^i_M$ depends non-trivially only on the $k_M$-component of $\vec{k}$).
Generically, the parameters $\vec{a}$ are different in different mirror planes.
However, due to rotational symmetry, they are constrained to be identical in all mirror planes in the same conjugacy class.
We are therefore left with two sets of parameters $\vec{a}$ and $\vec{a}'$ for the mirror planes $M_0$ and $M_0'$, respectively.
The goal of this appendix is to relate $\vec{a}$ and~$\vec{a}'$.

To motivate the subsequent technical analysis, let us summarize here the steps that have to be taken to arrive at the sought relation (working out these steps fills the remainder of this Appendix).
\begin{enumerate}
    \item[1.] Given any term $h_M^i(\vec{k})$ that appears in the \kdotp{} expansion of the Hamiltonian restricted to a mirror plane, i.e., satisfies \cref{App:eq:constraint_mirrorplane}, we show in \cref{App:eq:h-hat-definiton,App:eq:k-tranfso-rotation,App:eq:h_hat_condition1,App:eq:k-transfo-mirror,App:eq:h_hat_condition2} that by symmetrizing $h_M^i(\vec{k})$ we obtain a term $\hat{h}_{\mathsf{c}(m)}^i(\vec{k})$ that appears in the \kdotp{} expansion of the full Hamiltonian, i.e., satisfies \cref{App:eq:constraint_full}.
\end{enumerate}
As indicated by the notation, the terms in the full Hamiltonian produced by such a symmetrization, only cover one conjugacy class of mirror planes; some terms that would appear in a \kdotp{} expansion of the full Hamiltonian and are relevant in its restriction to mirror planes from the other conjugacy class are missing.
\begin{enumerate}
    \item[2.] We next turn to relating the two conjugacy classes.
    We \emph{observe} that the point groups and representations under consideration satisfy a useful property that relates the representations of symmetry-unrelated mirror symmetries to each other, cf.~\cref{App:eq:mirror_relation}, giving us the previously unknown transformation $U$ between the bases in which $h_{M_0}^i(\vec{k})$ and $h_{M_0'}^i(\vec{k})$ are expressed in terms of the same functions $h_i$, cf.~\cref{App:eq:h-at-M0-vs-M0'}.
    \item[3.] Based on the symmetrized terms $\hat{h}_{\mathsf{c}(m_0)}^i(\vec{k})$ and $\hat{h}_{\mathsf{c}(m_0')}^i(\vec{k})$, we next define the \emph{reconstructed \kdotp{} model} $\hat{\Ham}^{(N)}[\vec{B}](\vec{k})$ as the \kdotp{} model obtained from all those symmetrized terms, i.e., from both $\mathsf{c}(m_0)$ and $\mathsf{c}(m_0')$, cf.~\cref{App:eq:Ham_reconstructed_def}.
    The basis transformation $U$ then allows us to write all those terms using the matrix-valued polynomial functions $h_i$, cf.~\cref{App:eq:Ham_reconstructed}.
    \item[4.] Finally, we evaluate $\hat{\Ham}^{(N)}[\vec{B}](\vec{k})$ in the planes $M_0$ and $M_0'$ and compare the result to $\Ham_{M_0}^{(N)}[\vec{a}](\vec{k})$ and $\Ham_{M_0'}^{(N)}[\vec{a}'](\vec{k})$, respectively.
    This allows use to express $\vec{a}$ and $\vec{a}'$ in terms of $\vec{B}$ and therefore determine if and how $\vec{a}$ and $\vec{a}'$ are related.
\end{enumerate}

To start the analysis of step 1 in the above list, note that each term $h_M^i(\vec{k})$ in $\Ham_{M}^{(N)}[\vec{a}](\vec{k})$ satisfies \cref{App:eq:constraint_mirrorplane} and originates from a term in the full \kdotp{} Hamiltonian $\Ham^{(N)}(\vec{k})$ that satisfies \cref{App:eq:constraint_full}.
Let further $\vec{n}_M=C_n^{n/2}\vec{e}_M$ be the normal of the mirror plane $M$ corresponding to the mirror symmetry $m$ (here $m$ is chosen to be one of the representatives of the conjugacy classes, i.e., either $m_0$ or $m_0'$).
In order to find symmetry-compatible terms appearing in the 3D expansion $\Ham^{(N)}(\vec{k})$, we define for any $\vec{k}\in\mathbb{R}^3$ a \emph{symmetrized} matrix-valued polynomial function
\begin{equation}
    \hat{h}_{\mathsf{c}(m)}^i(\vec{k}) = \sum_{l=0}^{n-1}D(C_n)^lh_M^i\left(C_n^{-l}\left(\left.\vec{k}\right|_{M_l}\right)\right)D(C_n)^{-l},
    \label{App:eq:h-hat-definiton}
\end{equation}
where $M_l=C_n^lM$ and
\begin{equation}
    \left.\vec{k}\right|_M=\vec{k}-\left(\vec{n}_M\cdot\vec{k}\right)\vec{n}_M.
\end{equation}
Note that
\begin{equation}
    \begin{split}
    C_n^{-l}\left(\left.\vec{k}\right|_{M_l}\right) &= C_n^{-l}\left[\vec{k}-\left(\vec{n}_{M_l}\cdot \vec{k}\right)\vec{n}_{M_l}\right]\\
    &= C_n^{-l}\vec{k} - \left[\left(C_n^{-l}\vec{n}_{M_l}\right)\cdot\left(C_n^{-l}\vec{k}\right)\right]\left(C_n^{-l} \vec{n}_{M_l}\right)\\
    &= C_n^{-l}\vec{k} - \left[\vec{n}_{M}\cdot\left(C_n^{-l}\vec{k}\right)\right]\vec{n}_{M}\\
    &= \left.\left(C_n^{-l}\vec{k}\right)\right|_M,
    \end{split}
    \label{App:eq:k-tranfso-rotation}
\end{equation}
where we utilized that $\vec{n}\cdot\vec{k}=(S\vec{n})\cdot(S\vec{k})$ for any orthogonal symmetry $S$.
The final line of \cref{App:eq:k-tranfso-rotation} reveals that the argument of $h_M^i$ in \cref{App:eq:h-hat-definiton} lies in $M$ as it is supposed to.

We proceed to show that $\hat{h}_{\mathsf{c}(m)}^i(\vec{k})$ defined by \cref{App:eq:h-hat-definiton} satisfies \cref{App:eq:constraint_full}.
First, by substituting $l=p+1$ and using $C_n^0=C_n^n$ (both corresponding to the identity element), we easily find that
\begin{equation}
    \begin{split}
        \hat{h}_{\mathsf{c}(m)}^i(\vec{k}) &= \sum_{l=1}^{n}D(C_n)^lh_M^i\left(\left.\left(C_n^{-l}\vec{k}\right)\right|_M\right)D(C_n)^{-l}\\
        &= \sum_{p=0}^{n-1}D(C_n)^{p+1}h_M^i\left(\left.\left(C_n^{-(p+1)}\vec{k}\right)\right|_M\right)D(C_n)^{-(p+1)}\\
        &= D(C_n)\hat{h}_{\mathsf{c}(m)}^i(C_n^{-1}\vec{k})D(C_n)^{-1},
    \end{split}
    \label{App:eq:h_hat_condition1}
\end{equation}
which reveals that $\hat{h}_{\mathsf{c}(m)}^i(\vec{k})$ obeys \cref{App:eq:constraint_full-line1}.
Furthermore, for any mirror symmetry $\tilde{m}$, it holds that
\begin{equation}
    \begin{split}
        \tilde{m}^{-1}\left(\left.\vec{k}\right|_{M}\right) &=
        \tilde{m}^{-1}\left[\vec{k}-\left(\vec{n}_M\cdot\vec{k}\right)\vec{n}_M\right] \\
        &= \tilde{m}^{-1}\vec{k}-\left[(\tilde{m}^{-1}\vec{n}_M)\cdot\left(\tilde{m}^{-1}\vec{k}\right)\right]\left(\tilde{m}^{-1}\vec{n}_M\right)  \\
        &= \left.\left(\tilde{m}^{-1}\vec{k}\right)\right|_{\tilde{m}M},
    \end{split}
    \label{App:eq:k-transfo-mirror}
\end{equation}
where we again used that $\vec{n}\cdot\vec{k}=(S\vec{n})\cdot(S\vec{k})$, and additionally that mirrors are their own inverses, $\tilde{m} = \tilde{m}^{-1}$.
Using \cref{App:eq:h-hat-definiton,App:eq:k-tranfso-rotation,App:eq:constraint_mirrorplane:m,App:eq:mirror-rotation-rel} and $C_n^{-l}=C_n^{n-l}$, we find
\begin{equation}
    \begin{split}
        \hat{h}_{\mathsf{c}(m)}^i(\vec{k}) &\stackrel{\textrm{(\ref{App:eq:k-tranfso-rotation})}}{=}
        \sum_{l=0}^{n-1}D(C_n)^lh_M^i\left(\left.\left(C_n^{-l}\vec{k}\right)\right|_M\right)D(C_n)^{-l} \\
        & \stackrel{\textrm{(\ref{App:eq:constraint_mirrorplane:m})}}{=}\!\sum_{l=1}^{n}\!D\left(C_n^lm\right)h_M^i\!\left(m^{-1}\!\!\left.\left(C_n^{-l}\vec{k}\right)\right|_M\right)D\left(m^{-1}C_n^{-l}\right)\\
        &\stackrel{\textrm{(\ref{App:eq:mirror-rotation-rel})}}{=} \sum_{l=1}^{n}D\left(mC_n^{-l}\right)h_M^i\left(\left.\left(C_n^{l}m^{-1}\vec{k}\right)\right|_M\right)D\left(C_n^{l}m^{-1}\right)\\
        &= \sum_{p=1}^{n}D\left(mC_n^p\right)h_M^i\left(\left.\left(C_n^{-p}m^{-1}\vec{k}\right)\right|_M\right)D\left(C_n^{-p}m^{-1}\right)\\
        &= D(m)\hat{h}_{\mathsf{c}(m)}^i\left(m^{-1}\vec{k}\right)D(m)^{-1},
    \end{split}
    \label{App:eq:h_hat_condition2}
\end{equation}
where in the second line we further used hat $m$ acts trivially inside $M$ (and therefore its addition to the argument of $h^i_M$ does not alter the expression), and in the fourth line we substituted $p=n-l$.
The final line above confirms that $\hat{h}_{\mathsf{c}(m)}^i(\vec{k})$ also respects \cref{App:eq:constraint_full-line2}.
Therefore, we have shown that $\hat{h}_{\mathsf{c}(m)}^i(\vec{k})$ is a term that appears in the 3D expansion $\Ham^{(N)}(\vec{k})$ (even though it does not generally reduce to $h_M^i(\vec{k})$ in the mirror plane $M$).

Up to here, we have treated each of the two independent sets of symmetry-related mirror planes independently, reflected in the use of the generic mirror $m$.
We now turn to step 2 of the above list.
Looking up~\cite{BCS} representations under consideration (representations that stabilize TPs) for the two point groups $4mm$ and $6mm$, we observe that the following property is always satisfied:
\begin{subequations}
    \begin{align}\label{App:eq:mirror_relation}
        D(m_0') &= \e^{\i\phi}UD(m_0)\adjo{U},\\
        D((m_0')_\perp) &= \e^{\i\phi_\perp}UD((m_0)_\perp)\adjo{U},
    \end{align}
\end{subequations}
for some $\phi,\phi_\perp\in\mathbb{R}$ and $U\in\U(3)$, with the particular values of $\phi$, $\phi_\perp$ and $U$ depending on the choice of ICRs $\rep=\rho^\mathrm{2D}\oplus\rho^\mathrm{1D}$~\cite{Lenggenhager:2022:TPClassif:SDC}.
If we rotate the band basis by $\adjo{U}$ (denoted by a tilde `$\tilde{\;\,}$' over the representation symbol), then
\begin{subequations}
    \begin{align}
        \tilde{D}(m_0') &= \e^{\i\phi}D(m_0),\\
        \tilde{D}((m_0')_\perp) &= \e^{\i\phi_\perp}D((m_0)_\perp).
    \end{align}
    \label{App:eq:mirror_reps_tilde}
\end{subequations}

\Cref{App:eq:mirror_reps_tilde} imply that the constraints for the corresponding mirror plane $M_0'$ in the rotated basis are identical to \cref{App:eq:constraint_mirrorplane} (which captures constraints for plane $M_0$ in the original basis).
Thus, any term $\tilde{h}_{M_0'}^i(\vec{k})$ in the \kdotp{} model for $M_0'$ in that basis is given by precisely the same matrix-valued polynomial function $h_i$ (with the argument $k_{M_0}$ replaced by $k_{M_0'}$) as the corresponding term $h_{M_0}^i(\vec{k})$ in \kdotp{} model for $M_0$ in the original basis:
\begin{equation}
    \begin{split}
        \tilde{h}_{M_0'}^i(\vec{k}) &= \adjo{U}h_{M_0'}^i(\vec{k})U\\
        &= h_i(k_{M_0'}) = h_{M_0}^i(C_{2n}\vec{k}),
    \end{split}
    \label{App:eq:h-at-M0-vs-M0'}
\end{equation}
where in indicating the momentum arguments we used that the (symmetry inequivalent) planes $M_0$ and $M_0'$ (and the corresponding in-plane components $k_{M_0}$ and $k_{M_0'}$) are related by a $C_{2n}$ rotation (not an element of $G$).
Note that in the resulting \kdotp{} Hamiltonian these terms appear with different coefficient functions $\vec{a}'$ compared to the ones that appear in the expansion for $M_0$:
\begin{equation}
    \tilde{\Ham}_{M_0'}^{(N)}[\vec{a}'] = \sum_i a_i'(k_z) h_i\left(k_{M_0'}\right).
    \label{App:eq:Ham-at-M0-vs-M0'}
\end{equation}

Next, in step 3, we define the \emph{reconstructed \kdotp{} model} as the family of Hamiltonians with symmetrized terms $\hat{h}_{\mathsf{c}(m_0)}^i(\vec{k})$ and $\hat{h}_{\mathsf{c}(m_0')}^i(\vec{k})$ parametrized by the $k_z$-dependent components of some vector $\vec{B}$.
Since there are twice as many terms, $\vec{B}$ has twice as many components as $\vec{a}$ making it convenient to introduce vectors $\vec{b}$ and $\vec{b}'$, $\vec{B}=(\vec{b},\vec{b}')$, such that the terms originating from $\mathsf{c}(m_0)$ and $\mathsf{c}(m_0')$ appear with components of $\vec{b}$ and $\vec{b}'$, respectively, as coefficients:
\begin{equation}
    \hat{\Ham}^{(N)}[\vec{B}](\vec{k}) = \sum_i\left(b_i(k_z)\hat{h}_{\mathsf{c}(m_0)}^i(\vec{k})+b_i'(k_z)\hat{h}_{\mathsf{c}(m_0')}^i(\vec{k})\right).
    \label{App:eq:Ham_reconstructed_def}
\end{equation}
Substituting \cref{App:eq:h-hat-definiton}, and comparing to \cref{App:eq:kdotpmodel} we can rewrite the terms as symmetrizations of $\Ham_{M_0}^{(N)}[\vec{b}]$ and $\Ham_{M_0'}^{(N)}[\vec{b}']$
\begin{equation}
    \begin{split}
        \hat{\Ham}^{(N)}[\vec{B}](\vec{k}) &= \sum_{l=0}^{n-1}D(C_n)^l\left\{\Ham_{M_0}^{(N)}[\vec{b}]\left(\left.\left(C_n^{-l}\vec{k}\right)\right|_{M_0}\right)\vphantom{\left(\left.\left(C_n^{-l}\vec{k}\right)\right|_{M_0'}\right)}\right.\\
        &\qquad\left.+\,\Ham_{M_0'}^{(N)}[\vec{b}']\left(\left.\left(C_n^{-l}\vec{k}\right)\right|_{M_0'}\right)\right\}D(C_n)^{-l}\\
        &= \sum_{l=0}^{n-1}D(C_n)^l\left\{\Ham_{M_0}^{(N)}[\vec{b}]\left(\left.\left(C_n^{-l}\vec{k}\right)\right|_{M_0}\right)\vphantom{\left(\left.\left(C_n^{-l}\vec{k}\right)\right|_{M_0'}\right)}\right.\\
        &\qquad\left.+\,U\Ham_{M_0}^{(N)}[\vec{b}']\left(\left.\left(C_{2n}C_n^{-l}\vec{k}\right)\right|_{M_0}\right)\adjo{U}\right\}D(C_n)^{-l},
    \end{split}
    \label{App:eq:Ham_reconstructed}
\end{equation}
where for the last equality we applied \cref{App:eq:h-at-M0-vs-M0'} to the terms in $\Ham_{M_0'}^{(N)}[\vec{b}']$.

The family of Hamiltonians $\hat{\Ham}^{(N)}[\vec{B}]$ given by \cref{App:eq:Ham_reconstructed} is generally both incomplete (as a \kdotp{} in the full 3D momentum space), because potential terms that vanish in all mirror planes are missed out, and overparametrized, because some terms originating from $\mathsf{c}(m_0)$ and $\mathsf{c}(m_0')$ are linearly dependent (e.g., the constant terms).
In step 4, we evaluate $\hat{\Ham}^{(N)}[\vec{B}]$ in the mirror planes $M_0$ and $M_0'$, where the incompleteness does not affect the result, and compare that result to $\Ham_{M_0}^{(N)}[\vec{a}](\vec{k})$ and $\Ham_{M_0'}^{(N)}[\vec{a}'](\vec{k})$, respectively.
Let $\vec{k}^{M}=k\vec{e}_{M}+k_z\vec{e}_z\in M$ for $M\in\{M_0,M_0'\}$, then
\begin{subequations}
    \begin{align}
        \begin{split}
            \hat{\Ham}^{(N)}[\vec{B}]\left(\vec{k}_{M_0}\right) &= \Ham_{M_0}^{(N)}[\vec{a}]\left(\vec{k}_{M_0}\right)\\
            &= \sum_i a_i(k_z)h_i(k),
        \end{split}\\
        \begin{split}
            \adjo{U}\hat{\Ham}^{(N)}[\vec{B}]\left(\vec{k}_{M_0'}\right)U &= \tilde{\Ham}_{M_0'}^{(N)}[\vec{a}']\left(\vec{k}_{M_0'}\right)\\
            &= \sum_i a_i'(k_z)h_i(k).
        \end{split}
    \end{align}
\end{subequations}
Recognizing the terms $h_i(k)$ on the both sides of the above two equations and comparing their respective coefficients, allows us to express $\vec{a}$ and $\vec{a}'$ in terms of $\vec{B}$.
Subsequently, we can simply read off the relationship between $\vec{a}$ and $\vec{a}'$ (cf.~\cref{eq:inequiv-m-relation-1,eq:inequiv-m-relation-2,eq:inequiv-m-relation-3} in the main text) from their expressions in terms of $\vec{B}$.
This comparison of coefficients automatically takes care of the overparametrization problem mentioned in the beginning of the present paragraph.

Performing this analysis for $4mm$ we find that the coefficients in \cref{App:eq:kdotp_Ham_C4vC6v} for $(k,0)$ and $k\left(\tfrac{1}{\sqrt{2}},\tfrac{1}{\sqrt{2}}\right)$ (dashed parameters) are related as follows: $a$ and $a'$ are independent, $b'=b$, $c'=c$ and $A'=\mp_p A$ [cf.~\cref{eq:inequiv-m-relation-2}] with the sign defined in \cref{App:eq:rep_m}.
For $6mm$ we compare the coefficients for the planes defined by $(k,0)$ and $k\left(\tfrac{\sqrt{3}}{2},\tfrac{1}{2}\right)$ (dashed parameters) and find, for type-$\mathsf{A}$ TPs that $a'=a$, $b'=b$, $c'=c$ and $A'=(\mp_p)(\pm_q)A$ [cf.~\cref{eq:inequiv-m-relation-3}] with the signs again defined in \cref{App:eq:rep_m}.
Note that for type-$\mathsf{A}$ TPs $\mp_t=\mp_q$.
For type-$\mathsf{B}$ TPs on the other hand, we find that $a'=-a$, $b'=b$, $c'=c$ and $A'=-A$ [cf.~\cref{eq:inequiv-m-relation-1}].
Detailed derivations of the above three results are provided in a supplementary \Mathematica{} notebook~\cite{Lenggenhager:2022:TPClassif:SDC}.

\section{Classification in the presence of \texorpdfstring{$\mcP\mcT$}{PT} symmetry}\label{App:Classif:PT}

In this Appendix we provide several examples illustrating the rather abstract discussion of the derivation of the classification of triple points in the presence of $\mcP\mcT$ symmetry in \cref{Sec:Classif:PT}.
We first work through one example for a derivation of the minimal \kdotp{} model in \cref{App:Classif:PT:kdotp}.
Then, in \cref{App:Classif:PT:nls}, we present a number of examples illustrating different aspects of the derivation of the nodal-line (NL) structure, in particular the type of the triple point and the equations for the NL arcs.
Full derivations are given in the supplementary data and code~\cite{Lenggenhager:2022:TPClassif:SDC}.

\subsection{Minimal \texorpdfstring{\kdotp{}}{k.p} models}\label{App:Classif:PT:kdotp}

In \cref{App:antiunitary_symmetries} we have described how to obtain \kdotp{} models for magnetic point groups (MPGs) based on the knowledge of only the irreducible representations of the unitary subgroup.
Below, we work through an example and in the supplementary data and code~\cite{Lenggenhager:2021:MBNLs:SDC} we provide a small database and \Python{} scripts that output corepresentations and \kdotp{} models for the MPGs with $\mcP\mcT$ symmetry relevant to our studies.

As an example we consider the MPG $6/m'mm$ and representation $(E_1;A_1)$.
The representation matrices of the two generators of the unitary subgroup $6mm$ of $6/m'mm$ can be easily obtained from the Bilbao crystallographic server (BCS) [first two rows of \cref{App:tab:kdotp_sym}].
We determine the corresponding corepresentation (CR) of the non-unitary group $6/m'mm=6mm\cup(\mcP\mcT)6mm$.
Both irreducible representations (IRs) of the unitary subgroup $6mm$ are real, which we look up on the BCS; therefore, according to \cref{App:eq:ICRN_real} one can find for each IR a unitary matrix $N$ such that
\begin{equation}
	\Gamma(g) = N\cconj{\Gamma}(g)N^{-1},\quad N\cconj{N} = \id,
	\label{App:eq:what-N-must-obey}
\end{equation}
where we used that $\mcP\mcT$ commutes with all $g\in 6mm$ and $(\mcP\mcT)^2=\id$.
For the 1D IR $A_1$ a solution is $N_{A_1}=1$ and for $E_1$ $N_{E_1}=\sigma_x$, [this can be verified using the lower (upper) diagonal $1\times 1$ ($2\times 2$) block in the corepresentation matrices of the generators given in \cref{App:tab:kdotp_sym}, which correspond to the representation matrices of the irreducible representation $A_1$ ($E_1$).]
The resulting CR according to \cref{App:eq:ICR_real} is given in \cref{App:tab:kdotp_sym}.

\begin{table}[t]
    \centering
    \caption{
        Generators of the magnetic point group $6/m'mm$, including their action on position as well as the matrix corepresentation $(E_1;A_1)$.
    }
    \begin{ruledtabular}
    \begin{tabular}{ccc}
		Generator         	& Action on position		        & Matrix corepresentation		\\
		\hline\addextralinespace{2}
		$C_6^+$				& $\mat{\tfrac{1}{2}&-\tfrac{\sqrt{3}}{2}&0\\\tfrac{\sqrt{3}}{2}&\tfrac{1}{2}&0\\0&0&1}$ & $\mat{\e^{\i\pi/3}&0&0\\0&\e^{-\i\pi/3}&0\\0&0&1}$ \\
		\hline\addextralinespace{2}
		$\sigma_{d3}$		& $\mat{0&1&0\\1&0&0\\0&0&-1}$ 		& $\mat{0&1&0\\1&0&0\\0&0&1}$ \\
		\hline\addextralinespace{2}
		$\mcP\mcT$				& $\mat{-1&0&0\\0&-1&0\\0&0&-1}$ 	& $\mat{0&1&0\\1&0&0\\0&0&1}$ \\
    \end{tabular}
    \end{ruledtabular}
    \label{App:tab:kdotp_sym}
\end{table}

Since $(\mcP\mcT)^2=\id$, we can find a change of basis such that $\corep(\mcP\mcT)=\id$~\cite{Bouhon:2020}, i.e., for the 2D irreducible corepresentation (ICR) (corresponding to the upper $2\times 2$ block of the matrices given in \cref{App:tab:kdotp_sym})
\begin{equation}
	U\mat{0&1\\1&0}U^\top = \id,\qquad U = \frac{1}{\sqrt{2}}\mat{1&0\\-\i&\i}.
\end{equation}
The resulting CR matrices $U\corep(g)\adjo{U}$ for $g=C_6^+,\sigma_{d3}$ and $U\corep(\mcP\mcT)U^\top$ are provided as input to the \Python{} package \textsc{kdotp-symmetry}.
That returns a set of eleven Bloch matrices $H_i(\vec{k})$ with entries that are polynomials in $k_x,k_y,k_z$.

Not all of those terms are actually relevant for our purposes.
Diagonal matrices that are $\vec{k}$-independent and those that are proportional to the identity encode only constant shifts in energy and can thus be dropped.
Therefore, we make all matrices traceless and then keep only an independent set of  those.
Finally, we are interested in the behavior near the expansion point, such that leading order matrices are sufficient.

After this procedure we end up with four matrices
\begin{subequations}\label{App:eq:kdotpmodel}
    \begin{equation}
        \begin{split}
            	H_1(\vec{k}) &= \left(
            	\begin{array}{ccc}
            	k_x k_y & \frac{1}{2} \left(k_y^2-k_x^2\right) & 0 \\
            	\frac{1}{2} \left(k_y^2-k_x^2\right) & -k_x k_y & 0 \\
            	0 & 0 & 0 \\
            	\end{array}
            	\right),\\
            	H_2(\vec{k}) &= \left(
            	\begin{array}{ccc}
            	k_z & 0 & 0 \\
            	0 & k_z & 0 \\
            	0 & 0 & -2 k_z \\
            	\end{array}
            	\right),\\
            	H_3(\vec{k}) &= \left(
            	\begin{array}{ccc}
            	k_x^2+k_y^2 & 0 & 0 \\
            	0 & k_x^2+k_y^2 & 0 \\
            	0 & 0 & -2 \left(k_x^2+k_y^2\right) \\
            	\end{array}
            	\right),\\
            	H_4(\vec{k}) &= \left(
            	\begin{array}{ccc}
            	0 & 0 & -k_x+k_y \\
            	0 & 0 & -k_x-k_y \\
            	-k_x+k_y & -k_x-k_y & 0 \\
            	\end{array}
            	\right)
            \end{split}
    \end{equation}
    and the full \kdotp{} model can be written as
    \begin{equation}
    	\Ham_\vec{a}(\vec{k}) = \sum_{i=1}^4 a_i H_i(\vec{k}),
    \end{equation}
\end{subequations}
where $\vec{a}$ is the vector with components $a_i$.
The terms $H_{2,3}$ contribute directly to the three eigenenergies, while $H_1$ couples the two bands transforming in the same 2D ICR and $H_4$ couples the 2D ICR to the 1D ICR.

\subsection{Nodal-line structure}\label{App:Classif:PT:nls}

Here, we illustrate the derivation of the classification of triple points (TPs) for magnetic point groups with $\mcP\mcT$ symmetry, i.e., the MPGs $\bar{3}'$, $4/m'$, $6/m'$, $\bar{3}'m$, $4/m'mm$ and $6/m'mm$, by presenting a number of representative example calculations.
The full calculations for all MPGs are provided in a \Mathematica{} notebook~\cite{Lenggenhager:2022:TPClassif:SDC}

We first utilize the MPG $6/m'mm$ as an example.
The MPG $6/m'mm$ has two 2D ICRs $E_1$, $E_2$ and four 1D ICRs $A_1$, $A_2$, $B_1$, $B_2$.
We can build eight different combinations of one 2D and one 1D ICR, which fall into two equivalence classes [cf.~\cref{eq:model_equivalence,tab:Classification_PT}].
We demonstrate this equivalence for the ICR combinations $(E_2;A_1)$ and $(E_2;A_2)$ in \cref{App:Classif:PT:nls:equiv_classes}, before making in \cref{App:Classif:PT:nls:winding_number} a short detour to discuss the winding number around the central NL formed by the 2D ICR $E_1$.
Then, in \cref{App:Classif:PT:nls:typeA}, we apply the method we developed to determine the leading-order terms in the discriminant to the ICR combination $(E_1;A_1)$ of $6/m'mm$.
This method only leads to a simplification if the TP turns out to be type $\mathsf{A}$.
Therefore, we discuss the more complicated analysis of the discriminant of models hosting type-$\mathsf{B}$ TPs in \cref{App:Classif:PT:nls:typeB} using $(E_1;B_1)$ of $6/m'mm$ as an example.

In the subsequent subsections we show some representative calculations for MPGs without sixfold rotational symmetry.
First, in \cref{App:Classif:PT:nls:nexus_NLarcs}, we discuss the MPG $4/m'mm$ giving rise to a type-$\mathsf{A}$ TP, while highlighting that although there are no additional nodal lines attached directly to the TP, there might be a nexus of NL arcs a short distance away from the TP.
Finally, we consider the MPG $\bar{3}'$ to illustrate the reduction of the characteristic polynomial of models \emph{without} mirror symmetry to the characteristic polynomial of the corresponding models \emph{with} mirror symmetry in \cref{App:Classif:PT:nls:mirror_absence}.

\subsubsection{Equivalence classes of Hamiltonians}\label{App:Classif:PT:nls:equiv_classes}

Let us briefly illustrate the equivalence of certain ICR combinations as discussed in \cref{Sec:Classif:PT} using $(E_2;A_1)$ and $(E_2;A_2)$ of $6/m'mm$ as an example.
The characteristic polynomials $\chi[\Ham^{(i)}_{\vec{a}}(\vec{k})](E)$ of the \kdotp{} Hamiltonians for the ICR combinations $(E_2;A_i)$ are (in polar coordinates $(k_x,k_y)=k(\cos(\theta),\sin(\theta))$ and with energy $E$ the polynomial variable)
\begin{widetext}
\begin{align}
    \begin{split}\label{App:eq:equiv-first}
    	\chi\left[\Ham^{(1)}_{\vec{a}}(k,\theta,k_z)\right](E) &= \frac{1}{8} \Bigg[-8 \left(a_3 k^2+a_2 k_z\right)^2 \left(2 a_3 k^2+2 a_2 k_z-3 E \right)-a_4^2 k^4 \left(a_1 k^2 \sin (6 \theta )+2 a_3 k^2+2 a_2 k_z-2 E \right)\\
    	&\qquad\quad+\,2 a_1^2 k^4 \left(2 a_3 k^2+2 a_2 k_z+E \right)-8 E ^3\Bigg],
    \end{split}\\
    \begin{split}
        \chi\left[\Ham^{(2)}_{\vec{a}}(k,\theta,k_z)\right](E) &= \frac{1}{8} \Bigg[-8 \left(a_3 k^2+a_2 k_z\right)^2 \left(2 a_3 k^2+2 a_2 k_z-3 E \right)-4 a_4^2 k^4 \left(-a_1 k^2 \sin (6 \theta )+2 a_3 k^2+2 a_2 k_z-2 E \right)\\
    	&\qquad\quad+\,2 a_1^2 k^4 \left(2 a_3 k^2+2 a_2 k_z+E \right)-8 E ^3\Bigg].
    \end{split}
\end{align}
\end{widetext}
We observe that
\begin{equation}
	\chi\left[\Ham^{(1)}_{\tilde{\vec{a}}}\left(k,\theta+\frac{\pi}{6},k_z\right)\right](E) = \chi\left[\Ham^{(2)}_{\vec{a}}\left(k,\theta,k_z\right)\right](E)
\end{equation}
with $\tilde{a}_4 = 2a_4$ and $\tilde{a}_i=a_i$ for $i\neq 4$. 
An analysis along these lines of reasoning reveals that all combinations of 2D+1D ICR fall into two distinct equivalence classes in the presence of sixfold rotational symmetry, and to a single equivalence class for all other MPGs.

\subsubsection{Winding number (\texorpdfstring{ICR $E_1$ of $6/m'mm$}{ICR E1 of 6/m'mm})}\label{App:Classif:PT:nls:winding_number}

We first construct a \kdotp{} model for the 2D ICR $E_1$ of $6/m'mm$ as described in \cref{App:Classif:PT:kdotp}:
\begin{equation}
    \Ham(\vec{k}) = \left[a_0+a_2\left(k_x^2+k_y^2\right)\right]\id + a_1\mat{-2k_xk_y & k_x^2-k_y^2\\k_x^2-k_y^2 & 2k_xk_y}.\label{App:eq:winding-E1-of-6/m'mm}
\end{equation}
The resulting model can be written in terms of Pauli matrices as indicated in~\cref{eq:kdotp2D}, while dropping the topologically unimportant term proportional to the identity; then
\begin{equation}
    \vec{h}_\vec{a}(k_x,k_y) = a_1\mat{k_x^2-k_y^2\\-2k_xk_y}.
\end{equation}
The winding number defined in \cref{eq:w2D} then becomes
\begin{equation}
    w_\mathrm{2D} = \frac{1}{2\pi}\oint_C\dd{\vec{k}}\cdot\left(\frac{1}{k_x^2+k_y^2}\mat{2k_y\\-2k_x}\right) = -2.
\end{equation}

\begin{figure}[h]
    \centering
    \includegraphics{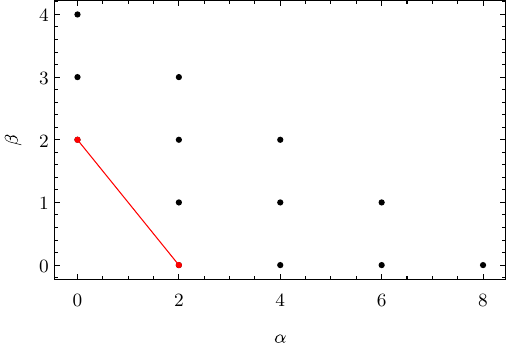}
    \caption{
        Construction of the leading terms contributing to the discriminant of $6/m'mm$ class~I given the Hamiltonian in \cref{App:eq:disc_C6vI} (red line).
        The black dots denote elements of the set $M$ as defined in \cref{eq:disc_terms}, and the red dots are elements of $L(M)$ as defined in \cref{eq:disc_leading_terms}.
    }
    \label{App:fig:disc_C6vI}
\end{figure}

\subsubsection{Analysis for type-\texorpdfstring{$\mathsf{A}$}{A} triple points (\texorpdfstring{$6/m'mm$}{6/m'mm} class I)}\label{App:Classif:PT:nls:typeA}

To study the NL structure near a TP, we need three-band models.
For equivalence class I of $6/m'mm$, we use the ICR combination $(E_1;A_1)$, for which the \kdotp{} model is given in \cref{App:eq:kdotpmodel}.
The corresponding characteristic polynomial in energy $E$ is (in cylindrical coordinates; $\theta$ drops out)
\begin{widetext}
\begin{equation}
    \chi\left[\Ham_{\vec{a}}(k,\theta,k_z)\right](E) = -\frac{1}{4} \left[2 \left(E -a_2 k_z\right)-\left(a_1+2 a_3\right) k^2\right] \left\{\left(2 a_3 k^2+2 a_2 k_z+E \right) \left[\left(a_1-2 a_3\right) k^2+2 \left(E -a_2 k_z\right)\right]-4 a_4^2 k^2\right\}.
    \label{App:eq:char-poly_C6vI}
\end{equation}
Noting that the discriminant of a third-order polynomial is defined as
\begin{equation}
    \Delta\left[\sum_{n=0}^3b_nx^n\right] = b_1^2b_2^2 - 4 b_1^3b_3 - 4 b_0b_2^3 - 27b_0^2b_3^2 + 18 b_0b_1b_2b_3,
\end{equation}
we find (by fully utilizing \Mathematica{}) the discriminant of the characteristic polynomial to be   
\begin{equation}
    \Delta_{\vec{a},\theta}(k,k_z) = \frac{1}{16} k^4 \left[a_1\left(a_1+6a_3\right) k^2+6 a_1a_2 k_z-4 a_4^2\right]^2 \left\{\left[\left(a_1-6a_3\right) k^2-6a_2 k_z\right]^2+32 a_4^2 k^2\right\}.
    \label{App:eq:disc_C6vI}
\end{equation}
\end{widetext}

We next determine the leading-order terms in the discriminant following the method outlined in \cref{eq:disc_polynomial,eq:disc_leading_terms} in the main text.
The set $M$ (black dots) of all monomials appearing in $\Delta_{\vec{a},\theta}$ with non-vanishing coefficients together with the leading terms $L(M)$ (red dots) is shown in \cref{App:fig:disc_C6vI}.
There are only two leading terms, such that we can approximate
\begin{equation}
    \Delta_{\vec{a},\theta}(k,k_z) \approx 4 a_4^4 k^4 \left(8 a_4^2 k^2+9 a_2^2 k_z^2\right)
    \label{App:eq:leading-C6vI}
\end{equation}
near the TP.
From the obtained leading-order expansion it is apparent that there are no real roots apart from $k=0$ (the central NL), and we conclude that  the TP is type $\mathsf{A}$.

\subsubsection{Analysis for type-\texorpdfstring{$\mathsf{B}$}{B} triple points (\texorpdfstring{$6/m'mm$}{6/m'mm} class II)}\label{App:Classif:PT:nls:typeB}

For class II the algebra is much more involved, mainly because none of the terms in the discriminant of the characteristic polynomial is subleading.
Again, we first determine the relevant \kdotp{} model for the corepresentation $(E_1;B_1)$.
The only difference to \cref{App:eq:kdotpmodel} is the replacement of $H_4(\vec{k})$ by
\begin{equation}
    H_4^\text{(II)}(\vec{k}) = \left(
        \begin{array}{ccc}
         0 & 0 & k_x k_y \\
         0 & 0 & \frac{1}{2} \left(k_x^2-k_y^2\right)\\
         k_x k_y & \frac{1}{2} \left(k_x^2-k_y^2\right) &
           0 \\
        \end{array}
        \right).
\end{equation}
We omit the explicit expressions for the characteristic polynomial and for the discriminant $\Delta_{\vec{a},\theta}(k,k_z)$ here; they can be found in the supplementary data and code~\cite{Lenggenhager:2022:TPClassif:SDC}.

Similar to \cref{App:eq:disc_C6vI}, the discriminant of the characteristic polynomial is a fourth-order polynomial in $k_z$ and takes the form
\begin{equation}
	\Delta_{\vec{a},\theta}(k,k_z) = \sum_{\beta=0}^4b_\beta(\vec{a},k,\theta)k_z^\beta.
\end{equation}
The nature of its real roots, i.e., the number of distinct real roots and their multiplicities, is determined~\cite{Rees:1922} by the following five quantities (suppressing the dependence on $\vec{a}$, $k$ and $\theta$): the \emph{discriminant $\bar{\Delta}$ of the discriminant $\Delta_{\vec{a},\theta}(k,k_z)$ seen as a polynomial in $k_z$},
\begin{subequations}
    \begin{align}
    	P &= 8 b_2 b_4-3 b_3^2,\\
    	R &= b_3^3-4 b_2 b_4 b_3+8 b_1 b_4^2,\\
    	\bar{\Delta}_0 &= b_2^2-3 b_1 b_3+12 b_0 b_4,\\
    	D &= -3 b_3^4+16 b_2 b_4 b_3^2+64 b_0 b_4^3-16 \left(b_2^2+b_1 b_3\right) b_4^2.
    \end{align}
\end{subequations}

The following cases are possible:
\begin{enumerate}[(1),leftmargin=*]
	\item a single real root of multiplicity $2$ if $\bar{\Delta }{=}0\land \{D{>}0\lor [P{>}0\land (D{\neq} 0\lor R{\neq} 0)]\}$,
	\item a single real root of multiplicity $4$ if $\bar{\Delta }{=}0\land D{=}0\land \bar{\Delta }_0{=}0$,
	\item one real root of multiplicity $3$ and one of multiplicity $1$ if $\bar{\Delta }{=}0\land \bar{\Delta }_0{=}0\land D{\neq} 0$,
	\item two real roots of multiplicity $2$ each if $\bar{\Delta }{=}0\land D{=}0\land P{<}0$,
	\item two real roots of multiplicity $1$ each if $\bar{\Delta }{<}0$,
	\item one real root of multiplicity $2$ and two of multiplicity $1$ each if $\bar{\Delta }{=}0\land P{<}0\land D{<}0\land \bar{\Delta }_0{\neq} 0$,
	\item four real roots of multiplicity $1$ if $\bar{\Delta }{>}0\land P{<}0\land D{<}0$, and
	\item no real roots if none of the above is satisfied.
\end{enumerate}

Evaluating all these conditions with \Mathematica{}, we find that only cases (1) and (8) arise for valid choices of the parameters $\vec{a}$, $k$ and $\theta$.
For generic $\vec{a}$, i.e., excluding fine-tuned models, case (1) is satisfied if
\begin{equation}
	\theta = \frac{\pi}{12} + \frac{2\pi}{6}\mathbb{Z}\quad\vee\quad\theta = \frac{\pi}{4} + \frac{2\pi}{6}\mathbb{Z}
	\label{App:eq:C6vII:NL_theta}
\end{equation}
and case (8) for any other value of $\theta$ and for the fine-tuned models.
Thus, we conclude that generically, there are NL arcs in the two sets of six symmetry-related mirror planes and no other NLs near the TP, resulting in $n_a^\mathrm{nexus}=12$.
In the fine-tuned case some or all of those NL arcs are not present.

Because for $6/m'mm$ the NL arcs lie in the mirror planes, we can find explicit expressions $k_z^\text{arc}(k)$ for them by substituting the conditions on $\theta$ given in \cref{App:eq:C6vII:NL_theta} into $\Delta_{\vec{a},\theta}(k,k_z)=0$ and solving for $k_z$:
\begin{subequations}\label{App:eq:NLarcs_C6vII}
    \begin{alignat}{3}
        \theta &= \frac{\pi}{12}+\frac{\pi}{3}\mathbb{Z} && \, : \quad & k_z^\text{arc}(k) &= -\frac{2a_1 (a_1 + 6a_3) - a_4^2}{12a_1a_2}k^2,\\
        \theta &= \frac{\pi}{4}+\frac{\pi}{3}\mathbb{Z} && \, : \quad & k_z^\text{arc}(k) &= \frac{2a_1 (a_1 - 6a_3) - a_4^2}{12a_1a_2}k^2.
    \end{alignat}
\end{subequations}
We observe that $\lim_{k\to 0}k_z^\text{arc}(k)=0$, such that the NL arcs attach to the TP and from \cref{App:eq:NLarcs_C6vII} we can read off that $\mu=2$, since $k_z(k)\propto k^2$; thus, the TP is type $\mathsf{B}_q$.

\subsubsection{Nodal-line arcs attached to a nexus point (\texorpdfstring{$4/m'mm$}{4/m'mm})}\label{App:Classif:PT:nls:nexus_NLarcs}

While there are no NL arcs attached to the TP in the cases $4/m'mm$ and for class-I Hamiltonians of $6/m'mm$, the corresponding \kdotp{} models predict the possibility of NL arcs attached to the central NL \emph{near} the TP, forming a nexus point of NL arcs [cf.~\cref{fig:TPTypes}(b)]. 
To see this, we consider the corepresentation $(E;A_1)$ of $4/m'mm$.
The \kdotp{} model is
\begin{subequations}
    \begin{equation}
    	\Ham_{\vec{a}}(\vec{k}) = \sum_{i=1}^5 a_i H_i(\vec{k}).
    \end{equation}
    with
    \begin{equation}
        \begin{split}
        	H_1(\vec{k}) &= \left(
            \begin{array}{ccc}
             k_z & 0 & 0 \\
             0 & k_z & 0 \\
             0 & 0 & -2 k_z \\
            \end{array}
            \right),\\
        	H_2(\vec{k}) &= \left(
            \begin{array}{ccc}
             k_x^2+k_y^2 & 0 & 0 \\
             0 & k_x^2+k_y^2 & 0 \\
             0 & 0 & -2 \left(k_x^2+k_y^2\right) \\
            \end{array}
            \right),\\
        	H_3(\vec{k}) &= \left(
            \begin{array}{ccc}
             k_x k_y & 0 & 0 \\
             0 & -k_x k_y & 0 \\
             0 & 0 & 0 \\
            \end{array}
            \right),\\
        	H_4(\vec{k}) &= \left(
            \begin{array}{ccc}
             0 & k_y^2-k_x^2 & 0 \\
             k_y^2-k_x^2 & 0 & 0 \\
             0 & 0 & 0 \\
            \end{array}
            \right),\\
        	H_5(\vec{k}) &= \left(
            \begin{array}{ccc}
             0 & 0 & k_y-k_x \\
             0 & 0 & -k_x-k_y \\
             k_y-k_x & -k_x-k_y & 0 \\
            \end{array}
            \right).
        \end{split}
    \end{equation}
\end{subequations}

We consider the discriminant of the characteristic polynomial of $\Ham_{\vec{a},\theta}(k,k_z)$ \emph{before} restricting to the leading terms, since then any additional NLs not attached to the TP would be lost.
Since one anticipates such NLs to appear in one of the mirror planes, we restrict the subsequent analysis to $\theta\in\{0,\pi/4\}\,\mathrm{mod}\,\pi/2$.
Restricting to those values of $\theta$, equation $\Delta_{\vec{a},\theta}(k,k_z)=0$ has near the rotation axis analytic solutions
\begin{subequations}\label{App:eq:4/m'mm-nexi}
    \begin{alignat}{3}
        \theta &= \frac{\pi}{2}\mathbb{Z} && \, : \quad & k_z^\text{arc}(k) &= \frac{a_5^2}{3 a_1 a_4} - \frac{3a_2+a_4}{3a_1}k^2,\\
        \theta &= \frac{\pi}{4}+\frac{\pi}{2}\mathbb{Z} && \, : \quad & k_z^\text{arc}(k) &= \frac{2a_5^2}{3 a_1 a_3} - \frac{6a_2+a_3}{6a_1}k^2.
    \end{alignat}
\end{subequations}
As long as $\abs{a_5}\ll \abs{a_{1,3,4}}$, i.e., as long as the coupling of the 1D to the 2D ICR is small, these NLs attach to the central NL close to the TP, such that the \kdotp{} expansion is still reliable enough.
In both sets of mirror planes ($\theta=0$ and $\theta=\pi/4$), the position of the nexus is proportional to the same parameter $a_5\in\mathbb{R}$, which implies codimension $1$.
The two sets of NL arcs that lie within the two inequivalent sets of mirror planes generically lie at different distance from the TP, implying $n_a^\mathrm{nexus}=4$.
They are of the same (different) color if $\sign(a_4a_5)=+1$ ($-1$).
Fine-tuning to $a_5=0$ collides both nexus points with the type-$\mathsf{A}$ \emph{simultaneously} [cf.~inset to \cref{fig:TPmat:ZrO_2}(c)].

Analogously, we find for class-I $6/m'mm$
\begin{subequations}\label{App:eq:6/m'mm-nexi}
    \begin{alignat}{3}
        \theta &= \frac{\pi}{12}+\frac{\pi}{3}\mathbb{Z} && \, : \quad & k_z^\text{arc}(k) &= \frac{2a_4^2}{3 a_1 a_2} - \frac{a_1+6a_3}{6a_2}k^2,\\
        \theta &= \frac{\pi}{4}+\frac{\pi}{3}\mathbb{Z} && \, : \quad & k_z^\text{arc}(k) &= \frac{2a_4^2}{3 a_1 a_2} - \frac{a_1+6a_3}{6a_1}k^2.
    \end{alignat}
\end{subequations}
Note that the expressions for $k_z^\text{arc}(k)$ in the two sets of planes are identical, such that there is only a single nexus point with $n_a^\mathrm{nexus}=12$ NL arcs in the same gap and with the same functional dependence on $k$ for small $k$ (including higher-order terms reveals that the NL arcs generically do behave differently as a function of $k$ in the two sets of planes).
Fine-tuning to $a_4=0$ collides the nexus point with the type-$\mathsf{A}$ TP, such that the codimension is $1$.

\subsubsection{Analysis in the absence of mirror symmetry (\texorpdfstring{$\bar{3}'$}{-3'})}\label{App:Classif:PT:nls:mirror_absence}

Finally, we discuss the reduction of the characteristic polynomial of the \kdotp{} Hamiltonian for the MPG $\bar{3}'$ to the one resulting from $\bar{3}'m$ as a concrete example of the abstract discussion presented in \cref{Sec:TP-type-B-wo-mirror}.
After a reparametrization of the \kdotp{} model for the ICR combination $(E;A_1)$ of $\bar{3}'$, the Hamiltonian takes the form
\begin{subequations}\label{App:eq:Ham_-3'}
    \begin{equation}
    	\Ham_{\vec{a}}^{\bar{3}'}(\vec{k}) = \sum_{i=1}^5 a_i H_i(\vec{k}).
    \end{equation}
    with
    \begingroup
    \allowdisplaybreaks
    \begin{align}
    	H_1(\vec{k}) &= \left(
        \begin{array}{ccc}
         k_x+k_y & k_x-k_y & 0 \\
         k_x-k_y & -k_x-k_y & 0 \\
         0 & 0 & 0 \\
        \end{array}
        \right),\nonumber \\
        H_2(\vec{k}) &= \left(
        \begin{array}{ccc}
         k_z & 0 & 0 \\
         0 & k_z & 0 \\
         0 & 0 & -2 k_z \\
        \end{array}
        \right),\nonumber \\
    	H_3(\vec{k}) &= \left(
        \begin{array}{ccc}
         0 & 0 & k_x+k_y \\
         0 & 0 & -k_x+k_y \\
         k_x+k_y & -k_x+k_y & 0 \\
        \end{array}
        \right),\\
    	H_4(\vec{k}) &= \left(
        \begin{array}{ccc}
         k_y-k_x & k_x+k_y & 0 \\
         k_x+k_y & k_x-k_y & 0 \\
         0 & 0 & 0 \\
        \end{array}
        \right),\nonumber \\
    	H_5(\vec{k}) &= \left(
        \begin{array}{ccc}
         0 & 0 & k_y-k_x \\
         0 & 0 & -k_x-k_y \\
         k_y-k_x & -k_x-k_y & 0 \\
        \end{array}
        \right),\nonumber
    \end{align}
    \endgroup
\end{subequations}
while the \kdotp{} Hamiltonian for the same ICR combination $(E;A_1)$ of the MPG $\bar{3}'m$ is
\begin{equation}
    \Ham_{\vec{a}}^{\bar{3}'m}(\vec{k}) = \sum_{i=1}^3 a_i H_i(\vec{k}),
\end{equation}
i.e., formed by only the first three terms in \cref{App:eq:Ham_-3'}.
In the following we show that by appropriate rotations of $\vec{k}$ and reparametrizations, i.e., transformations of $\vec{a}$, the characteristic polynomial $\chi[\Ham_{\vec{a}}^{\bar{3}'}(\vec{k})](E)$ can be reduced to $\chi[\Ham_{\vec{a}}^{\bar{3}'m}(\vec{k})](E)$, implying that the nodal structures close to the respective TPs are qualitatively the same.

We first observe that the momentum-space coordinate transformation $\vec{k}'=R_z(-\theta_1)\vec{k}$, where $R_z(\theta)$ denotes a rotation around the $k_z$-axis about the angle $\theta$, with
\begin{equation}
    \theta_1 = \arctan\left(\frac{a_5}{a_3}\right)
\end{equation}
(\emph{i}) $a_3H_3(\vec{k})+a_5H_5(\vec{k})$ to
\begin{equation}
    a_3H_3(R_z(\theta_1)\vec{k}')+a_5H_5(R_z(\theta_1)\vec{k}') = a_3'H_3(\vec{k}')
\end{equation}
with $a_3'=\sqrt{a_3^2+a_5^2}$, (\emph{ii}) leaves $H_2(\vec{k})$ invariant, and (\emph{iii}) transforms $a_1H_1(\vec{k})+a_4H_4(\vec{k})$ to
\begin{equation}
    a_1'H_1(\vec{k}')+a_4'H_4(\vec{k}')
\end{equation}
with $(a_1',a_4')=R(-\theta_1)(a_1,a_4)$, where $R(\theta)\in\SO(2)$ is the $2\times 2$ rotation matrix about the angle $\theta$.
Therefore, we find that
\begin{equation}
    \Ham_{\vec{a}}^{\bar{3}'}(R_z(\theta_1)\vec{k}') = \Ham_{\vec{a}'}^{\bar{3}'}(\vec{k}')
\end{equation}
with $\vec{a}'=(a_1',a_2,a_3',a_4',0)$.

Next, we compute the characteristic polynomial of $\Ham_{\vec{a}'}^{\bar{3}'}(\vec{k}')$:
\begin{widetext}
    \begin{equation}\label{App:eq:chi_-3'_transformed1}
        \begin{split}
            \chi\left[\Ham_{\vec{a}'}^{\bar{3}'}(\vec{k}')\right](E) &= 2(a_3')^2 (k')^3\left[\left(a_1'-a_4'\right)\sin(3 \theta' )-\left(a_1'+a_4'\right)\cos(3 \theta')\right]
            +2E\left[(a_1')^2+(a_3')^2+(a_4')^2\right] (k')^2\\
            &\qquad+\,2 a_2'\left[2(a_1')^2-(a_3')^2+2(a_4')^2\right] (k')^2 k_z'+3E(a_2')^2(k_z')^2-2(a_2')^3 (k_z')^3-E^3.
        \end{split}
    \end{equation}
\end{widetext}
Another coordinate transformation $\vec{k}''=R_z(-\theta_2)\vec{k}'$ with
\begin{equation}
    \theta_2 = \frac{1}{3}\arctan\left(\frac{a_1'}{a_4'}\right)
\end{equation}
transforms the first term in \cref{App:eq:chi_-3'_transformed1} to
\begin{equation}
    2(a_3')^2(k'')^3a_1'\sqrt{1+\left(\frac{a_4'}{a_1'}\right)^2}\left[\sin(3\theta'')-\cos(3\theta'')\right],
\end{equation}
while leaving all the other terms invariant.
Setting $\vec{a}''=(a_1'',a_2,a_3',0,0)$ with
\begin{equation}
    a_1'' = \sign(a_1')\sqrt{(a_1')^2+(a_4')^2} = \sign(a_1a_3+a_4a_5)\sqrt{a_1^2+a_4^2},
\end{equation}
we arrive at
\begin{equation}
    \chi\left[\Ham_{\vec{a}'}^{\bar{3}'}(R_z(\theta_2)\vec{k}'')\right](E) = \chi\left[\Ham_{\vec{a}''}^{\bar{3}'}(\vec{k}'')\right](E),
\end{equation}
which is the characteristic polynomial obtained from $\Ham_{\vec{b}}^{\bar{3}'m}(\vec{k}'')$ with
\begin{equation}
    \vec{b}=\left(\sign(a_1a_3+a_4a_5)\sqrt{a_1^2+a_4^2}, a_2, \sqrt{a_3^2+a_5^2}\right).
\end{equation}

We conclude that in the rotated coordinates $\vec{k}''$ the characteristic polynomial of the leading-order \kdotp{} model for $\bar{3}'$ is, up to reparametrization, identical to the characteristic polynomial obtained from the leading-order \kdotp{} model for $\bar{3}'m$.
Any properties that only depend on the characteristic polynomial, which include nodal structures such as TPs and NL arcs attached to TPs, are therefore the same.
Therefore, we find that TPs on HSLs with little co-group $\bar{3}'$ are always type $\mathsf{B}_l$ with $n_a^\mathrm{nexus}=6$ and $\mu=1$.

\section{Data on Material Examples}\label{App:Materials}
In this appendix we provide figures supporting the results in \cref{tab:TPmaterials} that are not already included in the main text.
The presented data clarify the types and values for $n_a^\textrm{nexus}$ and $\mu$ listed in the table.
All the first-principles data and the code used to analyze them is published in a data and code collection~\cite{Lenggenhager:2022:TPClassif:SDC}.

\makeatletter\onecolumngrid@push\makeatother

\begin{figure*}
    \centering
    \includegraphics{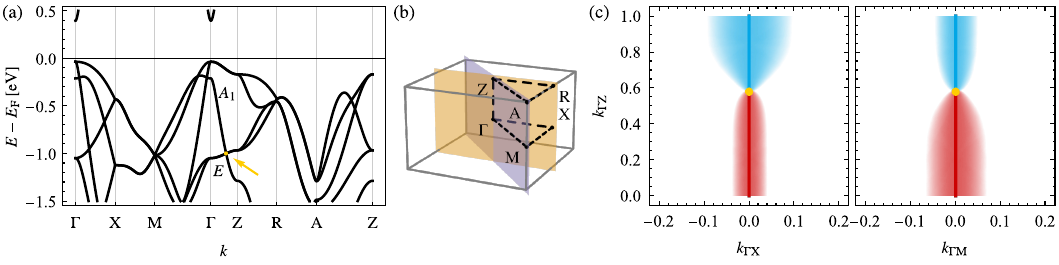}
    \caption{
    Nodal line structure of \ce{Na2LiN} near the type-$\mathsf{A}$ triple point on the $\Lambda$ line with little co-group $4/m'mm$.
    (a) Band structure along lines of symmetry. The triple point is indicated by a yellow dot and arrow and the bands forming the triple point are labelled by their irreducible corepresentations.
    (b) Brillouin zone (boundary in gray) with points and lines of symmetry (black dashed lines) and the two inequivalent mirror planes (orange and purple planes).
    (c) Size of the lower (red) and upper (blue) gap in the two mirror planes shown in panel (b) encoded by the intensity of the color (with a cutoff at a gap size of $0.01\nunit{eV}$, i.e., gaps larger than that are shown in white). The triple point (yellow) and the central nodal line are emphasized by appropriately colored overlays.
    }
    \label{fig:TPmat:Na2LiN}
\end{figure*}

\begin{figure*}
    \centering
    \includegraphics{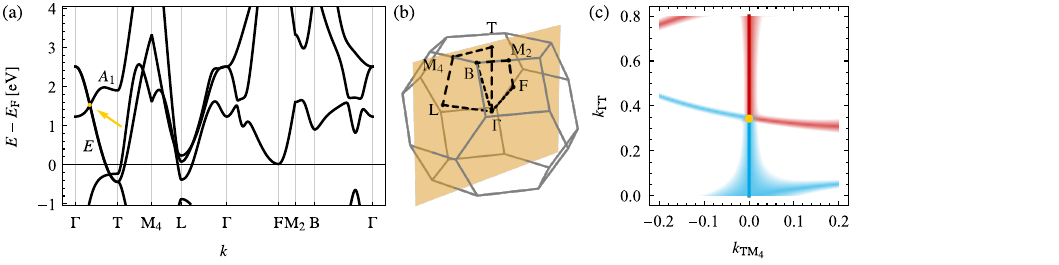}
    \caption{
    Nodal line structure of \ce{P} near the type-$\mathsf{B}_l$ triple point on the $\Lambda$ line with little co-group $\bar{3}'m$.
    The organization of the panels is in one-to-one correspondence with \cref{fig:TPmat:Na2LiN} with the difference that there is only one mirror plane. The cutoff on the gap size is $0.1\nunit{eV}$.
    Due to the three-fold rotational symmetry, we conclude that $n_a^\mathrm{nexus}=6$.
    }
    \label{fig:TPmat:P}
\end{figure*}

\begin{figure*}
    \centering
    \includegraphics{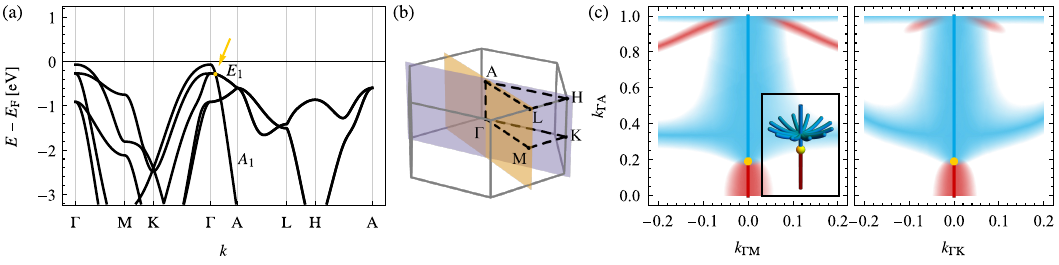}
    \caption{
    Nodal line structure of \ce{AlN} near the type-$\mathsf{A}$ triple point at $E_\mathrm{TP}=-0.28\nunit{eV}$ on the $\Delta$ line with little co-group $6mm$.
    The organization of the panels is in one-to-one correspondence with \cref{fig:TPmat:Na2LiN} and the cutoff on the gap size is $0.05\nunit{eV}$.
    Note the occurrence of the nearby nexus of $n_a^\mathrm{nexus}=12$ (due to the six-fold rotational symmetry) blue nodal-line (NL) arcs with $\mu=2$ in panel (c).
    The inset of panel (c) shows the NL structure of the minimal \kdotp{} model given in \cref{App:Classif:PT:nls:nexus_NLarcs} with parameters tuned to qualitatively reproduce the situation in \ce{AlN}.
    As expected, the \kdotp{} model does not reproduce the different curvature of the NL arcs in the two inequivalent sets of mirror planes which is clearly visible in the data. As remarked upon in \cref{Sec:Classif:mirror_no_PT:C4vC6v-part2} this would be reflected in terms of higher order in $k_x,k_y$ in the \kdotp{} expansion.
    Our theoretical arguments suggest that two parameters need be tuned to collide the nexus with the type-A TP.
    }
    \label{fig:TPmat:AlN_1}
\end{figure*}

\begin{figure*}
    \centering
    \includegraphics{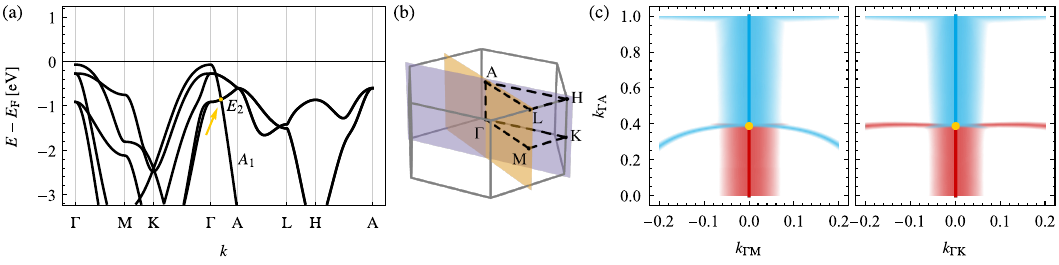}
    \caption{
    Nodal line structure of \ce{AlN} near the type-$\mathsf{B}_q$ triple point at $E_\mathrm{TP}=-0.85\nunit{eV}$ on the $\Delta$ line with little co-group $6mm$.
    The organization of the panels is in one-to-one correspondence with \cref{fig:TPmat:Na2LiN} and the cutoff on the gap size is $0.05\nunit{eV}$.
    Due to the six-fold rotational symmetry, we conclude that $n_a^\mathrm{nexus}=12$.
    }
    \label{fig:TPmat:AlN_2}
\end{figure*}

\begin{figure*}
    \centering
    \includegraphics{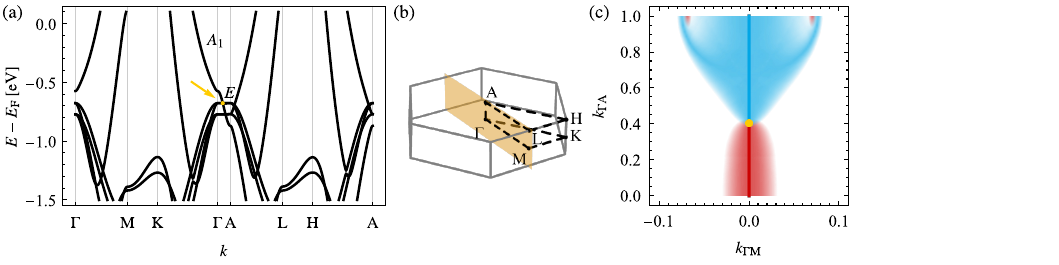}
    \caption{
    Nodal line structure of \ce{Li4N} near the type-$\mathsf{A}$ triple point on the $\Delta$ line with little co-group $\bar{6}'m2'$.
    The organization of the panels is in one-to-one correspondence with \cref{fig:TPmat:Na2LiN} with the difference that there is only one mirror plane. The cutoff on the gap size is $0.01\nunit{eV}$.
    Note the occurrence of a nearby nexus of $n_a^\mathrm{nexus}=6$ (due to the three-fold rotational symmetry) blue nodal-line arcs with $\mu=2$ in panel (c). Our theoretical arguments suggest that only a single parameter needs be tuned to collide the nexus with the type-$\mathsf{A}$ TP.
    }
    \label{fig:TPmat:Li4N}
\end{figure*}

\begin{figure*}
    \centering
    \includegraphics{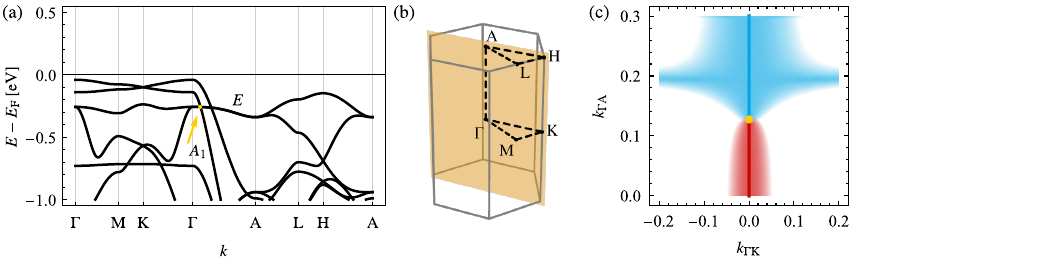}
    \caption{
    Nodal line structure \ce{Na2O} near the type-$\mathsf{A}$ triple point on the $\Delta$ line with little co-group $\bar{6}'m2'$.
    The organization of the panels is in one-to-one correspondence with \cref{fig:TPmat:Na2LiN} with the difference that there is only one mirror plane. The cutoff on the gap size is $0.01\nunit{eV}$.
    Note again the occurrence of a nearby nexus of $n_a^\mathrm{nexus}=6$ (due to the three-fold rotational symmetry) blue nodal-line arcs with $\mu=2$ in panel (c). Our theoretical arguments suggest that only a single parameter needs be tuned to collide the nexus with the type-$\mathsf{A}$ TP.
    }
    \label{fig:TPmat:Na2O}
\end{figure*}

\begin{figure*}
    \centering
    \includegraphics{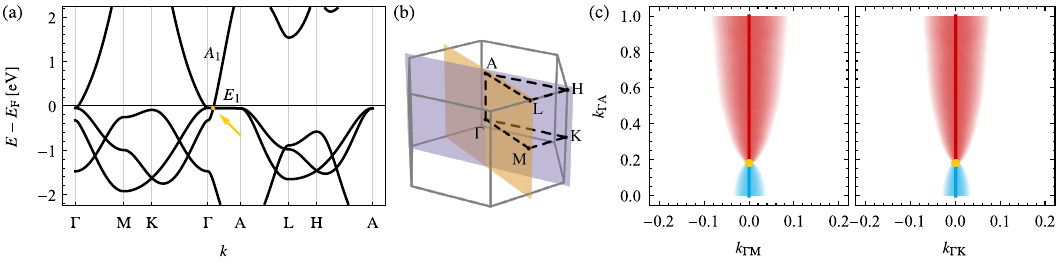}
    \caption{
    Nodal line structure of \ce{Li2NaN} near the type-$\mathsf{A}$ triple point on the $\Delta$ line with little co-group $6/m'mm$.
    The organization of the panels is in one-to-one correspondence with \cref{fig:TPmat:Na2LiN} and the cutoff on the gap size is $0.02\nunit{eV}$.
    Our recent work~\cite{Lenggenhager:2021:MBNLs} used this material to illustrate a relation between type-$\mathsf{A}$ TPs of spinless $\mcP\mcT$-symmetric crystals to multi-band nodal links.
    }
    \label{fig:TPmat:Li2NaN}
\end{figure*}

\begin{figure*}
    \centering
    \includegraphics{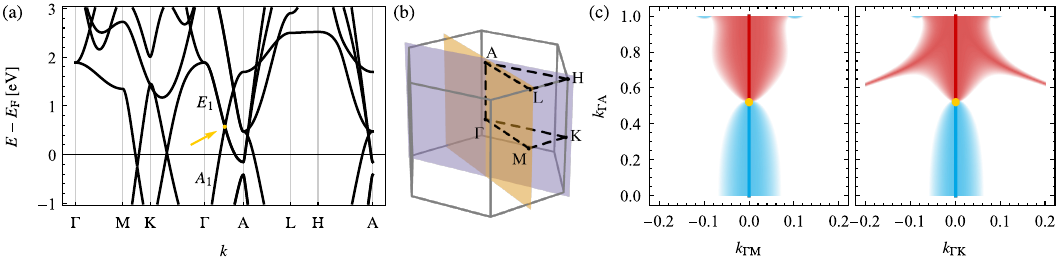}
    \caption{
    Nodal line structure of \ce{TiB2} near the type-$\mathsf{A}$ triple point at $E_\mathrm{TP}=0.57\nunit{eV}$ on the $\Delta$ line with little co-group $6/m'mm$.
    The organization of the panels is in one-to-one correspondence with \cref{fig:TPmat:Na2LiN} and the cutoff on the gap size is $0.05\nunit{eV}$.
    We do not identify a nexus point in this case; the nodal lines discernible in the right part of panel (c) seem to connect to the high-symmetry point $K$ such that they cannot be considered to be near the triple point.
    }
    \label{fig:TPmat:TiB2_1}
\end{figure*}

\begin{figure*}
    \centering
    \includegraphics{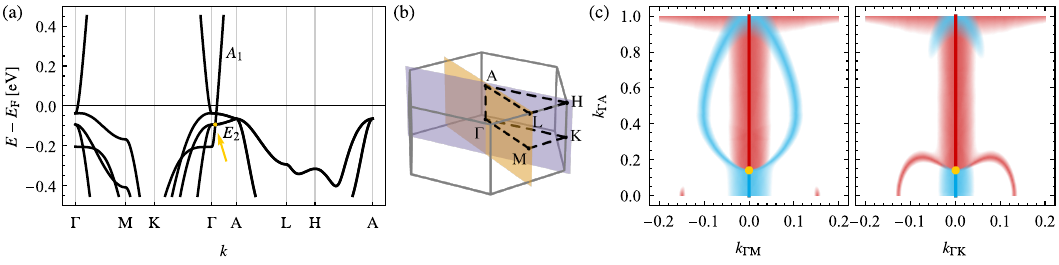}
    \caption{
    Nodal line structure of \ce{Na3N} near the type-$\mathsf{B}_q$ triple point at $E_\mathrm{TP}=-93\nunit{meV}$ on the $\Delta$ line with little co-group $6/m'mm$.
    The organization of the panels is in one-to-one correspondence with \cref{fig:TPmat:Na2LiN} and the cutoff on the gap size is $0.01\nunit{eV}$.
    Due to the six-fold rotational symmetry, we conclude that $n_a^\mathrm{nexus}=12$.
    }
    \label{fig:TPmat:Na3N_1}
\end{figure*}

\begin{figure*}
    \centering
    \includegraphics{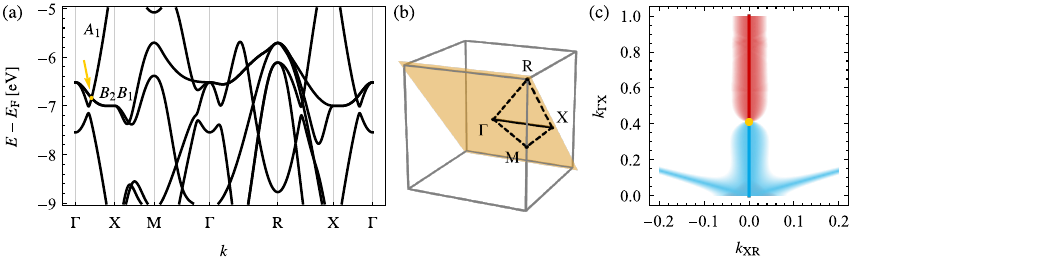}
    \caption{
    Nodal line structure of \ce{C3N4} (SG 215) near the type-$\mathsf{A}$ triple point on the $\Delta$ line with little co-group $\bar{4}'2'm$.
    The organization of the panels is in one-to-one correspondence with \cref{fig:TPmat:Na2LiN} with the difference that there is only one mirror plane. The cutoff on the gap size is $0.02\nunit{eV}$.
    We do not identify a nexus point in this case; the nodal lines discernible in panel (c) connect to the high-symmetry point $\Gamma$ such that they cannot be considered to be near the triple point.
    }
    \label{fig:TPmat:C3N4_4}
\end{figure*}

\begin{figure*}
    \centering
    \includegraphics{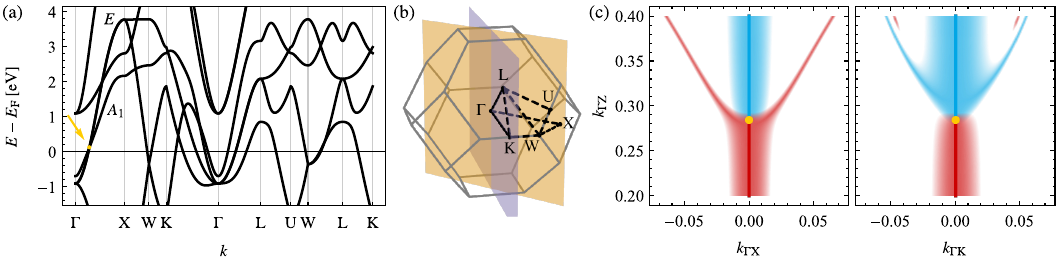}
    \caption{
    Nodal line structure of \ce{ZrO} near the type-$\mathsf{A}$ triple point at $E_\mathrm{TP}=0.12\nunit{eV}$ on the $\Delta$ line with little co-group $4/m'mm$.
    The organization of the panels is in one-to-one correspondence with \cref{fig:TPmat:Na2LiN} and the cutoff on the gap size is $0.02\nunit{eV}$.
    Note the extremely close adjacency of the triple-point to two nexus points of $n_a^\mathrm{nexus}=4$ (due to the four-fold rotational symmetry) nodal-line arcs each.
    According to our theoretical arguments, this is achieved by tuning a single model parameter. Furthermore, observe that the two nexus points are of \emph{different} colors (in contrast to \cref{fig:TPmat:ZrO_2}), suggesting $\sign(a_3 a_4) = -1$ in the \kdotp{} model in \cref{App:eq:4/m'mm-nexi}.
    }
    \label{fig:TPmat:ZrO_1}
\end{figure*}

\begin{figure*}
    \centering
    \includegraphics{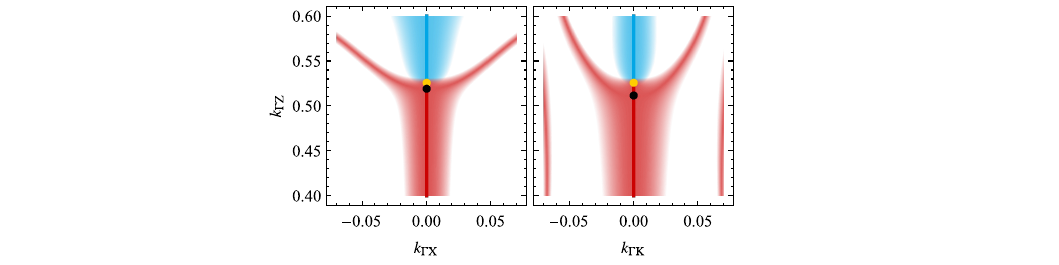}
    \caption{
    Nodal line structure of \ce{ZrO} with $5\%$ uniaxial tensile strain in $z$-direction near the type-$\mathsf{A}$ triple point at $E_\mathrm{TP}=2.0\nunit{eV}$ on the $\Delta$ line with little co-group $4/m'mm$, cf.~\cref{fig:TPmat:ZrO_2} for the same data in \ce{ZrO} without strain.
    Size of the lower (red) and upper (blue) gap in the two mirror planes shown in \cref{fig:TPmat:ZrO_2}(b) encoded by the intensity of the color (higher color saturation implies smaller energy gap between the corresponding pair of bands). The gap is only plotted up to a cutoff of $0.02\nunit{eV}$ such that white color indicates a gap larger than that.
    The triple point (yellow) and the central nodal line are emphasized by appropriately colored overlays.
    In contrast to \cref{fig:TPmat:ZrO_2}, we recognize two separate nexus points (indicated by black disks) not coinciding with the triple point.
    Thus, we conclude that the triple point is indeed type $\mathsf{A}$ and (due to the four-fold rotational symmetry) that $n_a^\mathrm{nexus}=4$.
    }
    \label{fig:TPmat:ZrO_2_strain}
\end{figure*}

\begin{figure*}
    \centering
    \includegraphics{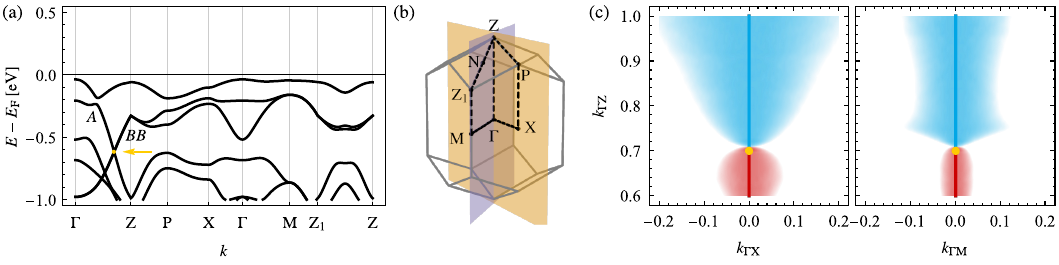}
    \caption{
    Nodal line structure of \ce{SiO2} near the type-$\mathsf{A}$ triple point on the $\Lambda$ line with little co-group $\bar{4}'$.
    The organization of the panels is in one-to-one correspondence with \cref{fig:TPmat:Na2LiN} with the difference that the shown planes are \emph{not} mirror planes. The cutoff on the gap size is $0.005\nunit{eV}$.
    The little co-group $\bar{4}'$ contains neither a vertical mirror symmetry nor $\mcP\mcT$ symmetry, such that nodal lines away from the rotation axis cannot be stabilized.
    We have performed DFT calculations in planes perpendicular to the central line (see supplementary data and code~\cite{Lenggenhager:2022:TPClassif:SDC}) and verified that.
    }
    \label{fig:TPmat:SiO2}
\end{figure*}

\begin{figure*}
    \centering
    \includegraphics{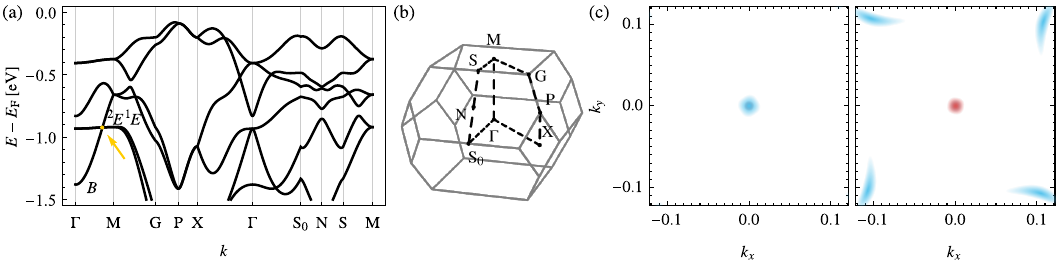}
    \caption{
    Nodal line structure of \ce{Li4HN} near the type-$\mathsf{A}$ triple point on the $\Lambda$ line with little co-group $4/m'$.
    (a) Band structure along lines of symmetry. The triple point is indicated by a yellow dot and arrow and the bands forming the triple point are labelled by their irreducible corepresentations.
    (b) Brillouin zone (boundary in gray) with points and lines of symmetry (black dashed lines). There are no mirror planes in this space group.
    (c) Size of the lower (red) and upper (blue) gap, encoded by the intensity of the color (with a cutoff in gap size at $0.02\nunit{eV}$), in two horizontal planes (at fixed $k_z$) one slightly below (left) and one slightly above (right) the triple point.
    We find that no nodal-line arcs attach to the triple point, but we show the attachment of nodal-line arcs to a nexus point in \cref{fig:TPmat:Li4HN:NP}.
    }
    \label{fig:TPmat:Li4HN}
\end{figure*}

\begin{figure*}
    \centering
    \includegraphics{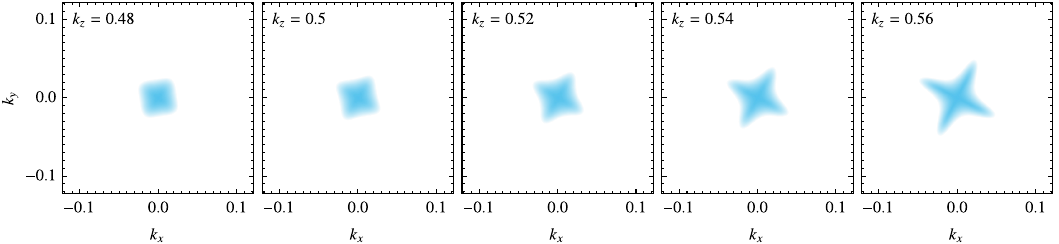}
    \caption{
    Nodal line structure of \ce{Li4HN} near the nexus point on the $\Lambda$ line with little co-group $4/m'$.
    Size of the lower (red) and upper (blue) gap, encoded by the intensity of the color (with a cutoff in gap size at $0.02\nunit{eV}$), in horizontal planes at $k_z\in\{0.48,0.5,0.52,0.54,0.56\}$.
    We observe that four red nodal-line arcs attach to the nexus point lying on the red central nodal line, such that $n_a^\mathrm{nexus}=4$.
    Furthermore, the nodal lines' symmetric arrangement indicates a quadratic attachment to the nexus point and therefore $\mu=2$.
    }
    \label{fig:TPmat:Li4HN:NP}
\end{figure*}

\begin{figure*}
    \centering
    \includegraphics{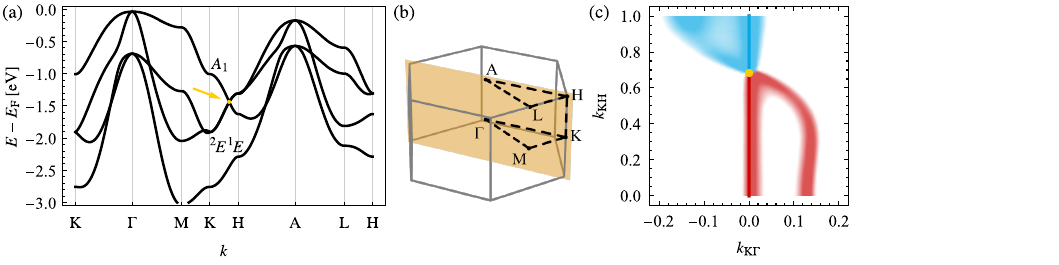}
    \caption{
    Nodal line structure of \ce{MgH2O2} near the type-$\mathsf{B}_l$ triple point on the $P$ line with little co-group $\bar{3}'$.
    The organization of the panels is in one-to-one correspondence with \cref{fig:TPmat:Na2LiN} with the difference that the shown plane is not a mirror plane. The cutoff on the gap size is $0.05\nunit{eV}$.
    Instead we have performed DFT calculations in planes perpendicular to the central line and determined the approximate planes in which nodal lines attaching to the triple point appear (see supplementary data and code~\cite{Lenggenhager:2022:TPClassif:SDC}). Here that plane is the $\mathrm{\Gamma K H}$-plane and its three mirror-symmetry-related copies, implying $n_a^\mathrm{nexus}=6$.
    }
    \label{fig:TPmat:MgH2O2}
\end{figure*}

\begin{figure*}
    \centering
    \includegraphics{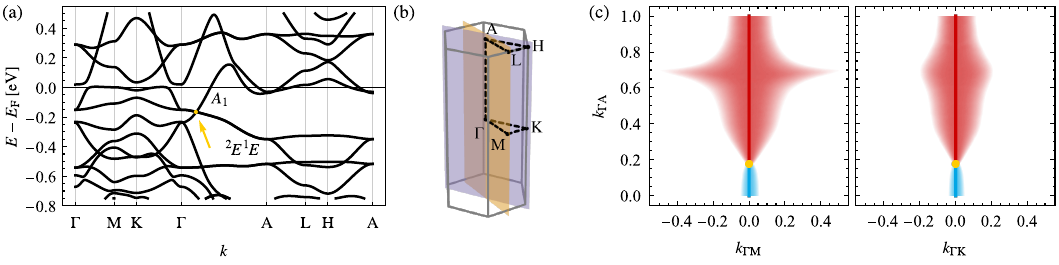}
    \caption{
    Nodal line structure of \ce{Li2Co12P7} near the type-$\mathsf{A}$ triple point on the $\Delta$ line with little co-group $\bar{6}'$.
    The organization of the panels is in one-to-one correspondence with \cref{fig:TPmat:Na2LiN} with the difference that the shown planes are \emph{not} mirror planes. The cutoff on the gap size is $0.01\nunit{eV}$.
    The little co-group $\bar{6}'$ contains neither a vertical mirror symmetry nor $\mcP\mcT$ symmetry, such that nodal lines away from the rotation axis cannot be stabilized.
    }
    \label{fig:TPmat:Li2Co12P7}
\end{figure*}

\begin{figure*}
    \centering
    \includegraphics{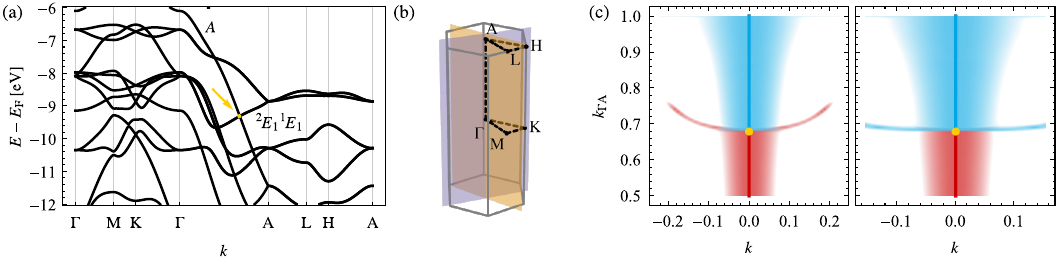}
    \caption{
    Nodal line structure of \ce{C3N4} (SG 176) near the type-$\mathsf{B}_q$ triple point at $E_\mathrm{TP}=-9.3\nunit{eV}$ on the $\Delta$ line with little co-group $6/m'$
    The organization of the panels is in one-to-one correspondence with \cref{fig:TPmat:Na2LiN} with the difference that the shown plane is not a mirror plane. The cutoff on the gap size is $0.05\nunit{eV}$.
    Using additional DFT calculations in planes perpendicular to the central nodal line (see supplementary data and code~\cite{Lenggenhager:2022:TPClassif:SDC}), we have determined that six nodal lines approximately lie in the planes indicated in panel (b) each, implying $n_a^\mathrm{nexus}=12$. The green plane is spanned by $\vec{k}_{\mathrm{\Gamma A}}$ and $\vec{k}=0.2\vec{k}_\mathrm{\Gamma M}+0.11\vec{k}_\mathrm{\Gamma M'}$, where $M'$ is the $M$-point rotated by $\tfrac{\pi}{3}$, and the red plane by $\vec{k}_{\mathrm{\Gamma A}}$ and $\vec{k}=0.02\vec{k}_\mathrm{\Gamma M}+0.2\vec{k}_\mathrm{\Gamma M'}$.
    }
    \label{fig:TPmat:C3N4_6_1}
\end{figure*}

\begin{figure*}
    \centering
    \includegraphics{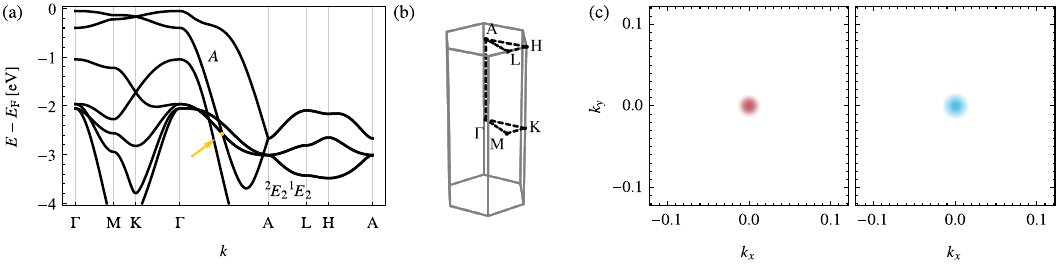}
    \caption{
    Nodal line structure of \ce{C3N4} (SG 176) near the type-$\mathsf{A}$ triple point at $E_\mathrm{TP}=-2.6\nunit{eV}$ on the $\Delta$ line with little co-group $6/m'$.
    The organization of the panels is in one-to-one correspondence with \cref{fig:TPmat:Li4HN} (but with a cutoff of $0.02\nunit{eV}$) and again there are no nodal-line arcs connecting to the triple point.
    }
    \label{fig:TPmat:C3N4_6_2}
\end{figure*}

\clearpage
\makeatletter\onecolumngrid@pop\makeatother


\bibliographystyle{apsrev4-2}

\end{document}